\documentclass[12pt]{article}
\usepackage{cite}
\usepackage{axodraw}
\usepackage{epsfig}

\makeatletter

\def\paragraph{\@startsection{paragraph}{4}{\z@}{+2.00ex plus
 +1ex minus +.2ex}{1.5ex plus .2ex}{\it\normalsize}}

\def\section{\@startsection {section}{1}{\z@}{+3.0ex plus +1ex minus
  +.2ex}{2.3ex plus .2ex}{\normalsize\bf\boldmath}}
\def\subsection{\@startsection{subsection}{2}{\z@}{+2.5ex plus +1ex
minus +.2ex}{1.5ex plus .2ex}{\normalsize\bf\boldmath}}
\def\subsubsection{\@startsection{subsubsection}{3}{\z@}{+3.25ex plus
 +1ex minus +.2ex}{1.5ex plus .2ex}{\normalsize\it}}

 \expandafter\ifx\csname mathrm\endcsname\relax\def\mathrm#1{{\rm #1}}\fi

\newcounter{saveeqn}

\@addtoreset{equation}{section}

\def\nl{\nonumber\\}

\newcommand{\lsim}
{\mathrel{\raisebox{-.3em}{$\stackrel{\displaystyle <}{\sim}$}}}
\newcommand{\gsim}
{\mathrel{\raisebox{-.3em}{$\stackrel{\displaystyle >}{\sim}$}}}
\def\asymp#1%
{\mathrel{\raisebox{-.4em}{$\widetilde{\scriptstyle #1}$}}}

\def\Nequal#1%
{\mathrel{\raisebox{-.5em}{$\stackrel{=}{\scriptstyle\rm#1}$}}}
\newcommand{\dsl}[1]{\not \hspace{-0.7mm}#1}
\def\dsl{\mathpalette\make@slash}
\def\make@slash#1#2{\setbox\z@\hbox{$#1#2$}%
  \hbox to 0pt{\hss$#1/$\hss\kern-\wd0}\box0}

\def\beq{\begin{equation}}
\def\eeq{\end{equation}}
\def\bit{\begin{itemize}}
\def\eit{\end{itemize}}
\def\beqar{\begin{eqnarray}}
\def\eeqar{\end{eqnarray}}
\def\barr#1{\begin{array}{#1}}
\def\earr{\end{array}}
\def\bfi{\begin{figure}}
\def\efi{\end{figure}}
\def\btab{\begin{table}}
\def\etab{\end{table}}
\def\bce{\begin{center}}
\def\ece{\end{center}}
\def\nn{\nonumber}

\def\text{\textstyle}

\def\Ga{\Gamma}
\def\ga{\gamma}
\def\de{\delta}

\def\eps{\epsilon}
\def\veps{\varepsilon}

\def\la{\lambda}

\def\si{\sigma}

\def\refeq#1{\mbox{(\ref{#1})}}

\def\reffi#1{\mbox{Figure~\ref{#1}}}
\def\reffis#1{\mbox{Figures~\ref{#1}}}
\def\refta#1{\mbox{Table~\ref{#1}}}

\def\refse#1{\mbox{Section~\ref{#1}}}

\def\refapp#1{\mbox{App.~\ref{#1}}}
\def\citere#1{\mbox{Ref.~\cite{#1}}}
\def\citeres#1{\mbox{Refs.~\cite{#1}}}

\newcommand{\TeV}{\unskip\,\mathrm{TeV}}
\newcommand{\GeV}{\unskip\,\mathrm{GeV}}
\newcommand{\MeV}{\unskip\,\mathrm{MeV}}

\newcommand{\fb}{\unskip\,\mathrm{fb}}

\newcommand{\ri}{{\mathrm{i}}}
\newcommand{\rI}{{\mathrm{I}}}
\newcommand{\rd}{{\mathrm{d}}}

\newcommand{\rT}{{\mathrm{T}}}

\newcommand{\A}{{\cal{A}}}

\newcommand{\M}{{\cal{M}}}

\def\mathswitchr#1{\relax\ifmmode{\mathrm{#1}}\else$\mathrm{#1}$\fi}

\newcommand{\PW}{\mathswitchr W}

\newcommand{\PZ}{\mathswitchr Z}

\newcommand{\Pg}{\mathswitchr g}

\newcommand{\PH}{\mathswitchr H}

\newcommand{\Pb}{\mathswitchr b}

\newcommand{\Pp}{\mathswitchr p}
\newcommand{\Pt}{\mathswitchr t}

\newcommand{\PWp}{\mathswitchr {W^+}}
\newcommand{\PWm}{\mathswitchr {W^-}}

\newcommand{\jet}{\mathswitchr {j}}

\def\mathswitch#1{\relax\ifmmode#1\else$#1$\fi}

\newcommand{\MH}{\mathswitch {M_\PH}}

\newcommand{\Mb}{\mathswitch {m_\Pb}}
\newcommand{\Mt}{\mathswitch {m_\Pt}}

\def\solid{\raise.9mm\hbox{\protect\rule{1.1cm}{.2mm}}}
\def\dash{\raise.9mm\hbox{\protect\rule{2mm}{.2mm}}\hspace*{1mm}}

\def\ie{i.e.\ }

\newcommand{\LO}{{\mathrm{LO}}}
\newcommand{\NLO}{{\mathrm{NLO}}}

\newcommand{\onel}{{\mbox{\scriptsize 1-loop}}}

\newcommand{\SU}{\mathrm{SU}}

\def\lra{\mathop{\mathrm{\leftrightarrow}}\nolimits}

\def\draftdate{\relax}
\def\mda{\relax}
\def\mua{\relax}
\def\mla{\relax}
\def\Mda{\relax}
\def\Mua{\relax}
\def\Mla{\relax}
\def\draft{
\def\thtystars{******************************}
\def\sixtystars{\thtystars\thtystars}
\typeout{}
\typeout{\sixtystars**}
\typeout{* Draft mode!
         For final version remove \protect\draft\space in source file *}
\typeout{\sixtystars**}
\typeout{}
\def\draftdate{\today}
\def\mua{\marginpar[\boldmath\hfil$\uparrow$]%
                   {\boldmath$\uparrow$\hfil}%
                    \typeout{marginpar: $\uparrow$}\ignorespaces}
\def\mda{\marginpar[\boldmath\hfil$\downarrow$]%
                   {\boldmath$\downarrow$\hfil}%
                    \typeout{marginpar: $\downarrow$}\ignorespaces}
\def\mla{\marginpar[\boldmath\hfil$\rightarrow$]%
                   {\boldmath$\leftarrow $\hfil}%
                    \typeout{marginpar: $\lra$}\ignorespaces}
\def\Mua{\marginpar[\boldmath\hfil$\Uparrow$]%
                   {\boldmath$\Uparrow$\hfil}%
                    \typeout{marginpar: $\uparrow$}\ignorespaces}
\def\Mda{\marginpar[\boldmath\hfil$\Downarrow$]%
                   {\boldmath$\Downarrow$\hfil}%
                    \typeout{marginpar: $\downarrow$}\ignorespaces}
\def\Mla{\marginpar[\boldmath\hfil$\Rightarrow$]%
                   {\boldmath$\Leftarrow $\hfil}%
                    \typeout{marginpar: $\lra$}\ignorespaces}
\overfullrule 5pt
\oddsidemargin -15mm
\marginparwidth 29mm
}

\def\stars{\strut\leaders\hbox{*}\hfill\strut}
\def\starline{\hfil\strut\hfil\hbox to \textwidth {\stars}\hfil}

\newcommand{\qqbremsdiagA}{
\begin{picture}(160,120) (-80,-60)
\Gluon(-50,35)(-20,65){4}{3.5}
\Vertex(-50,35){2}
\ArrowLine(-80,-35)(-20,-35)
\Vertex(-20,-35){2}
\ArrowLine(-20,-35)(-20,35)
\Vertex(-20,35){2}
\ArrowLine(-20,35)(-50,35)
\ArrowLine(-50,35)(-80,35)
\Gluon(-20,35)(30,35){4}{3.5}
\Gluon(-20,-35)(30,-35){4}{3.5}
\ArrowLine(60,15)(30,35)
\Vertex(30,35){2}
\ArrowLine(30,35)(60,55)
\ArrowLine(60,-15)(30,-35)
\Vertex(30,-35){2}
\ArrowLine(30,-35)(60,-55)
\end{picture}}

\newcommand{\gqbremsdiagA}{
\begin{picture}(160,120) (-80,-60)
\ArrowLine(-50,35)(-20,65)
\Vertex(-50,35){2}
\ArrowLine(-80,-35)(-20,-35)
\Vertex(-20,-35){2}
\ArrowLine(-20,-35)(-20,35)
\Vertex(-20,35){2}
\ArrowLine(-20,35)(-50,35)
\Gluon(-50,35)(-90,35){4}{3}
\Gluon(-20,35)(30,35){4}{3.5}
\Gluon(-20,-35)(30,-35){4}{3.5}
\ArrowLine(60,15)(30,35)
\Vertex(30,35){2}
\ArrowLine(30,35)(60,55)
\ArrowLine(60,-15)(30,-35)
\Vertex(30,-35){2}
\ArrowLine(30,-35)(60,-55)
\end{picture}}

\newcommand{\qqbremsdiagB}{
\begin{picture}(160,120) (-80,-60)
\ArrowLine(-60,-40)(-40,0)
\Vertex(-40,0){2}
\ArrowLine(-40,0)(-60,40)
\Vertex(10,0){2}
\Gluon(-40,0)(10,0){4}{3.5}
\Gluon(10,0)(40,30){4}{3.5}
\Gluon(10,0)(40,-30){4}{3.5}
\Gluon(40,30)(80,10){4}{3.5}
\Vertex(40,30){2}
\Gluon(40,30)(80,60){4}{3.5}
\ArrowLine(110,60)(80,60)
\Vertex(80,60){2}
\ArrowLine(80,60)(80,90)
\ArrowLine(70,-30)(40,-30)
\Vertex(40,-30){2}
\ArrowLine(40,-30)(40,-60)
\end{picture}
}

\newcommand{\qqbremsdiagC}{
\begin{picture}(160,120) (-80,-60)
\ArrowLine(-60,-40)(-40,0)
\Vertex(-40,0){2}
\ArrowLine(-40,0)(-60,40)
\Vertex(10,0){2}
\Gluon(-40,0)(10,0){4}{3.5}
\Gluon(10,0)(40,30){4}{3.5}
\Gluon(10,0)(40,-30){4}{3.5}
\Gluon(40,60)(80,60){4}{3.5}
\Vertex(40,60){2}
\ArrowLine(80,30)(40,30)
\Vertex(40,30){2}
\ArrowLine(40,30)(40,60)
\ArrowLine(40,60)(40,90)
\ArrowLine(70,-30)(40,-30)
\Vertex(40,-30){2}
\ArrowLine(40,-30)(40,-60)
\end{picture}
}

\newcommand{\ggbremsdiagA}{
\begin{picture}(160,120) (-80,-60)
\Gluon(-50,35)(-20,65){4}{3.5}
\Vertex(-50,35){2}
\Gluon(-80,-35)(-20,-35){4}{3.5}
\Vertex(-20,-35){2}
\Gluon(-20,-35)(-20,35){4}{3.5}
\Vertex(-20,35){2}
\Gluon(-20,35)(-50,35){4}{1.8}
\Gluon(-50,35)(-80,35){4}{1.8}
\Gluon(-20,35)(30,35){4}{3.5}
\Gluon(-20,-35)(30,-35){4}{3.5}
\ArrowLine(60,15)(30,35)
\Vertex(30,35){2}
\ArrowLine(30,35)(60,55)
\ArrowLine(60,-15)(30,-35)
\Vertex(30,-35){2}
\ArrowLine(30,-35)(60,-55)
\end{picture}}

\newcommand{\ggbremsdiagB}{
\begin{picture}(160,120) (-80,-60)
\Gluon(-60,-40)(-40,0){4}{3}
\Vertex(-40,0){2}
\Gluon(-40,0)(-60,40){4}{3}
\Vertex(10,0){2}
\Gluon(-40,0)(10,0){4}{3.5}
\Gluon(10,0)(40,30){4}{3.5}
\Gluon(10,0)(40,-30){4}{3.5}
\Gluon(40,30)(80,10){4}{3.5}
\Vertex(40,30){2}
\Gluon(40,30)(80,60){4}{3.5}
\ArrowLine(110,60)(80,60)
\Vertex(80,60){2}
\ArrowLine(80,60)(80,90)
\ArrowLine(70,-30)(40,-30)
\Vertex(40,-30){2}
\ArrowLine(40,-30)(40,-60)
\end{picture}
}

\newcommand{\ggbremsdiagC}{
\begin{picture}(160,120) (-80,-60)
\Gluon(-60,-40)(-40,0){4}{3}
\Vertex(-40,0){2}
\Gluon(-40,0)(-60,40){4}{3}
\Vertex(10,0){2}
\Gluon(-40,0)(10,0){4}{3.5}
\Gluon(10,0)(40,30){4}{3.5}
\Gluon(10,0)(40,-30){4}{3.5}
\Gluon(40,60)(80,60){4}{3.5}
\Vertex(40,60){2}
\ArrowLine(80,30)(40,30)
\Vertex(40,30){2}
\ArrowLine(40,30)(40,60)
\ArrowLine(40,60)(40,90)
\ArrowLine(70,-30)(40,-30)
\Vertex(40,-30){2}
\ArrowLine(40,-30)(40,-60)
\end{picture}
}

\newcommand{\ggbremsdiagD}{
\begin{picture}(160,120) (-80,-60)
\Gluon(-40,-40)(0,0){4}{3}
\Gluon(0,0)(-40,40){4}{3}
\Vertex(0,0){2}
\Gluon(0,0)(40,30){4}{3.5}
\Gluon(0,0)(40,-30){4}{3.5}
\Gluon(40,60)(80,60){4}{3.5}
\Vertex(40,60){2}
\ArrowLine(80,30)(40,30)
\Vertex(40,30){2}
\ArrowLine(40,30)(40,60)
\ArrowLine(40,60)(40,90)
\ArrowLine(70,-30)(40,-30)
\Vertex(40,-30){2}
\ArrowLine(40,-30)(40,-60)
\end{picture}
}

\newcommand{\qqtreediagA}{
\begin{picture}(160,120) (-80,-60)
\ArrowLine(-60,-35)(-20,-35)
\Vertex(-20,-35){2}
\ArrowLine(-20,-35)(-20,35)
\Vertex(-20,35){2}
\ArrowLine(-20,35)(-60,35)
\Gluon(-20,35)(30,35){4}{3.5}
\Gluon(-20,-35)(30,-35){4}{3.5}
\ArrowLine(60,15)(30,35)
\Vertex(30,35){2}
\ArrowLine(30,35)(60,55)
\ArrowLine(60,-15)(30,-35)
\Vertex(30,-35){2}
\ArrowLine(30,-35)(60,-55)
\end{picture}}

\newcommand{\ggtreediagA}{
\begin{picture}(160,120) (-80,-60)
\Gluon(-60,-35)(-20,-35){4}{3.5}
\Vertex(-20,-35){2}
\Gluon(-20,-35)(-20,35){4}{3.5}
\Vertex(-20,35){2}
\Gluon(-20,35)(-60,35){4}{3.5}
\Gluon(-20,35)(30,35){4}{3.5}
\Gluon(-20,-35)(30,-35){4}{3.5}
\ArrowLine(60,15)(30,35)
\Vertex(30,35){2}
\ArrowLine(30,35)(60,55)
\ArrowLine(60,-15)(30,-35)
\Vertex(30,-35){2}
\ArrowLine(30,-35)(60,-55)
\end{picture}}

\newcommand{\ggtreediagB}{
\begin{picture}(160,120) (-80,-60)
\Gluon(-60,-35)(-20,-35){4}{3.5}
\Vertex(-20,-35){2}
\ArrowLine(-20,-35)(-20,0)
\Vertex(-20,0){2}
\ArrowLine(-20,0)(-20,35)
\Vertex(-20,35){2}
\Gluon(-20,35)(-60,35){4}{3.5}
\ArrowLine(-20,35)(30,35)
\Gluon(-20,0)(30,0){4}{3.5}
\ArrowLine(30,-35)(-20,-35)
\ArrowLine(60,-20)(30,0)
\Vertex(30,0){2}
\ArrowLine(30,0)(60,20)
\end{picture}}

\newcommand{\ggtreediagC}{
\begin{picture}(160,120) (-80,-60)
\Gluon(-95,-40)(-75,0){4}{3}
\Vertex(-75,0){2}
\Gluon(-75,0)(-95,40){4}{3}
\Gluon(-75,0)(-40,0){4}{3.5}
\Vertex(-40,0){2}
\ArrowLine(-40,0)(-15,25)
\Vertex(-15,25){2}
\ArrowLine(-15,25)(10,50)
\ArrowLine(-40,0)(0,-40)
\Gluon(-15,25)(10,0){4}{3.5}
\ArrowLine(40,-30)(10,0)
\Vertex(10,0){2}
\ArrowLine(10,0)(40,30)
\end{picture}}

\newcommand{\qqtreediagB}{
\begin{picture}(160,120) (-80,-60)
\ArrowLine(-60,-40)(-40,0)
\Vertex(-40,0){2}
\ArrowLine(-40,0)(-60,40)
\Vertex(10,0){2}
\Gluon(-40,0)(10,0){4}{3.5}
\Gluon(10,0)(40,30){4}{3.5}
\Gluon(10,0)(40,-30){4}{3.5}
\ArrowLine(70,30)(40,30)
\Vertex(40,30){2}
\ArrowLine(40,30)(40,60)
\ArrowLine(70,-30)(40,-30)
\Vertex(40,-30){2}
\ArrowLine(40,-30)(40,-60)
\end{picture}
}

\newcommand{\qqhexagon}{
\begin{picture}(160,120) (-80,-60)
\ArrowLine(-40,-25)(-40,25)
\Vertex(-40,25){2}
\ArrowLine(-40,25)(-70,25)
\Vertex(-40,-25){2}
\ArrowLine(-70,-25)(-40,-25)
\Gluon(-40,25)(0,50){4}{3.5}
\Gluon(-40,-25)(0,-50){4}{3.5}
\Gluon(40,-25)(40,25){4}{3.5}
\ArrowLine(70,45)(40,25)
\Vertex(40,25){2}
\ArrowLine(40,25)(0,50)
\Vertex(0,50){2}
\ArrowLine(0,50)(30,70)
\ArrowLine(70,-45)(40,-25)
\Vertex(40,-25){2}
\ArrowLine(40,-25)(0,-50)
\Vertex(0,-50){2}
\ArrowLine(0,-50)(30,-70)
\end{picture}}

\newcommand{\gghexagon}{
\begin{picture}(160,120) (-80,-60)
\Gluon(-40,-25)(-40,25){4}{3.5}
\Vertex(-40,25){2}
\Gluon(-40,25)(-80,25){4}{3.5}
\Vertex(-40,-25){2}
\Gluon(-80,-25)(-40,-25){4}{3.5}
\Gluon(-40,25)(0,50){4}{3.5}
\Gluon(-40,-25)(0,-50){4}{3.5}
\Gluon(40,-25)(40,25){4}{3.5}
\ArrowLine(70,45)(40,25)
\Vertex(40,25){2}
\ArrowLine(40,25)(0,50)
\Vertex(0,50){2}
\ArrowLine(0,50)(30,70)
\ArrowLine(70,-45)(40,-25)
\Vertex(40,-25){2}
\ArrowLine(40,-25)(0,-50)
\Vertex(0,-50){2}
\ArrowLine(0,-50)(30,-70)
\end{picture}}

\newcommand{\qqpentagon}{
\begin{picture}(170,120)(-90,-60)
\ArrowLine(-95,-40)(-75,0)
\Vertex(-75,0){2}
\ArrowLine(-75,0)(-95,40)
\Gluon(-75,0)(-40,0){4}{3.5}
\Vertex(-40,0){2}
\Gluon(-40,0)(0,45){4}{3.5}
\Gluon(-40,0)(0,-45){4}{3.5}
\Gluon(40,-25)(40,25){4}{3.5}
\ArrowLine(70,45)(40,25)
\Vertex(40,25){2}
\ArrowLine(40,25)(0,45)
\Vertex(0,45){2}
\ArrowLine(0,45)(30,65)
\ArrowLine(70,-45)(40,-25)
\Vertex(40,-25){2}
\ArrowLine(40,-25)(0,-45)
\Vertex(0,-45){2}
\ArrowLine(0,-45)(30,-65)
\end{picture}}

\newcommand{\ggpentagon}{
\begin{picture}(170,120)(-90,-60)
\Gluon(-95,-40)(-75,0){4}{3}
\Vertex(-75,0){2}
\Gluon(-75,0)(-95,40){4}{3}
\Gluon(-75,0)(-40,0){4}{3.5}
\Vertex(-40,0){2}
\Gluon(-40,0)(0,45){4}{3.5}
\Gluon(-40,0)(0,-45){4}{3.5}
\Gluon(40,-25)(40,25){4}{3.5}
\ArrowLine(70,45)(40,25)
\Vertex(40,25){2}
\ArrowLine(40,25)(0,45)
\Vertex(0,45){2}
\ArrowLine(0,45)(30,65)
\ArrowLine(70,-45)(40,-25)
\Vertex(40,-25){2}
\ArrowLine(40,-25)(0,-45)
\Vertex(0,-45){2}
\ArrowLine(0,-45)(30,-65)
\end{picture}}

\begin{document}
\enlargethispage{2cm}
\thispagestyle{empty}
\def\thefootnote{\fnsymbol{footnote}}
\setcounter{footnote}{1}
\null
\draftdate
\hfill CERN-PH-TH/2009/008\\
\strut\hfill FR-PHENO-2010-003\\
\strut\hfill PSI-PR-10-02\\
\vspace{1.5cm}
\begin{center}
{\Large \bf\boldmath
NLO QCD corrections to 
$\Pt\bar\Pt\Pb\bar\Pb$ production 
\\[.5cm]
at the LHC:
2. full hadronic results
\par} 
\vspace{1.5cm}
{\large
{\sc A.\ Bredenstein$^1$, A.\ Denner$^2$, S.\ Dittmaier$^3$ 
and S.\ Pozzorini$^4$} } \\[.5cm]
$^1$ {\it High Energy Accelerator Research
                Organization (KEK),
\\
Tsukuba, Ibaraki 305-0801, Japan} \\[0.5cm]
$^2$ {\it Paul Scherrer Institut, W\"urenlingen und Villigen,
\\
CH-5232 Villigen PSI, Switzerland} \\[0.5cm]
$^3$ {\it Albert-Ludwigs-Universit\"at Freiburg, Physikalisches Institut, 
\\
D-79104 Freiburg, Germany}\\[0.5cm]
$^4$ {\it Physics Department, Theory Group, CERN, \\
CH-1211 Geneva 23, Switzerland}

\par \vskip 1em
\end{center}\par
\vfill {\bf Abstract:} 
\par 
We present predictions for $\Pt\bar\Pt\Pb\bar\Pb$ production at the
LHC in next-to-leading order QCD.  The precise description of this
background process is a prerequisite to observe associated
$\Pt\bar\Pt\PH$ production in the $\PH\to\Pb\bar\Pb$ decay channel and
to directly measure the top-quark Yukawa coupling at the LHC.  The
leading-order cross section is extremely sensitive to scale
variations. We observe that the traditional scale choice adopted in
ATLAS simulations underestimates the $\Pt\bar\Pt\Pb\bar\Pb$ background
by a factor two and introduce a new dynamical scale that stabilizes
the perturbative predictions.  We study various kinematic
distributions and observe that the corrections have little impact on
their shapes if standard cuts are applied.  In the regime of highly
boosted Higgs bosons, which offers better perspectives to observe the
$\Pt\bar\Pt\PH$ signal, we find significant distortions of the
kinematic distributions.  The one-loop amplitudes are computed using
process-independent algebraic manipulations of Feynman diagrams and
numerical tensor reduction.  We find that this approach provides very
high numerical stability and CPU efficiency.

\par
\vskip 2.5cm
\noindent
January 2010
\null
\setcounter{page}{0}
\clearpage
\def\thefootnote{\arabic{footnote}}
\setcounter{footnote}{0}

\section{Introduction}

The discovery of the Higgs boson and the measurement of its interactions
with massive quarks and vector bosons represent a central goal of the ATLAS
\cite{atlas-cms-tdrs,Aad:2009wy} and CMS \cite{Ball:2007zza} experiments at
the Large Hadron Collider (LHC).
The present limits from direct searches and electroweak precision
measurements favour a Higgs-mass range below the W-decay threshold. In this
light-Higgs scenario, the Higgs predominantly decays into bottom quarks, and
its observation is very challenging at the LHC.
In the dominant Higgs-production channel, \ie in gluon--gluon fusion, the
$\PH\to\Pb\bar\Pb$ signal is completely obscured by a huge QCD background.

Associated production mechanisms, where the Higgs boson is accompanied by a
massive gauge boson or top-quark pairs, feature more distinctive signatures
that can be exploited to reduce the background to the $\PH\to\Pb\bar\Pb$
final state.
These associated Higgs-production channels provide direct access to the
interactions of the Higgs boson with gauge bosons and heavy quarks. Their
observation would permit to test the electroweak
symmetry-breaking mechanism.
But the QCD background to associated $\PW\PH$, $\PZ\PH$, and
$\Pt\bar\Pt\PH$ production followed by $\PH\to\Pb\bar\Pb$ decay remains a
critical issue, which requires significant progress in two directions. The
low signal-to-background ratios must be increased by means of improved
selection strategies; and more precise descriptions of the background are
needed in order to reduce systematic uncertainties. In this paper we
address the latter issue by providing next-to-leading-order (NLO)
QCD predictions
for the irreducible QCD background to $\Pt\bar\Pt\PH(\PH\to\Pb\bar\Pb)$
production.

The strategies elaborated by ATLAS and CMS to identify the
$\Pt\bar\Pt\PH(\PH\to\Pb\bar\Pb)$ signal
\cite{Kersevan:2002vu,Cammin:2003,Aad:2009wy,Drollinger:2001ym,Cucciarelli:2006,Ball:2007zza}
are based on the full reconstruction of the $\Pt\bar\Pt\Pb\bar\Pb$
signature, starting from a final state with four b quarks and
additional light jets.  After imposing four b-taggings, a
reconstruction of the top quarks is performed. This permits to
identify two b quarks as top-decay products.  The remaining two b
quarks constitute a Higgs candidate, and their invariant-mass
distribution is the relevant observable to find the Higgs signal.
However, the presence of multiple b quarks and light jets in the final
state represents a serious obstacle to the correct identification of
the $\Pb\bar\Pb$ Higgs candidates.  Realistic simulations indicate
that only about 1/3 of the selected b-quark pairs have correct
combinatorics, while the other Higgs candidates contain b jets from
top decays or miss-tagged light jets.  This so-called combinatorial
background significantly dilutes the Higgs signal and increases its
background contamination.  The QCD processes
$\Pp\Pp\to\Pt\bar\Pt\Pb\bar\Pb$ and $\Pt\bar\Pt\jet\jet$ are the main
background components.  The latest ATLAS and CMS simulations
\cite{Aad:2009wy,Ball:2007zza}, for $30\fb^{-1}$ and $60\fb^{-1}$,
respectively, anticipate a statistical significance around $2\sigma$
(ignoring systematic uncertainties) and a fairly low
signal-to-background ratio of order 1/10.  This calls for better than
10\% precision in the background description, a very demanding
requirement both from the experimental and theoretical point of view.

More recently, alternative search strategies based on the selection of
highly boosted Higgs bosons, which decay into ``fat jets'' containing
two $\Pb$~quarks, have opened new and very promising perspectives,
both for $V\PH(\PH\to\Pb\bar\Pb)$ \cite{Butterworth:2008iy} and
$\Pt\bar\Pt\PH(\PH\to\Pb\bar\Pb)$ \cite{Plehn:2009rk}.  In the case of
$\Pt\bar\Pt\PH$, this novel approach might enable a better background
suppression and increase the signal-to-background ratio beyond $1/3$.
Moreover, three b-taggings would be sufficient to strongly suppress
the $\Pt\bar\Pt\jet\jet$ contamination. In this case the background
would be completely dominated by $\Pt\bar\Pt\Pb\bar\Pb$ production.

The calculation of the NLO QCD corrections to the process
$\Pp\Pp\to\Pt\bar\Pt\Pb\bar\Pb$, first presented in
\citeres{Bredenstein:2008zb, Bredenstein:2009aj} and subsequently
confirmed in \citere{Bevilacqua:2009zn}, constitutes another important
step towards the observability of the
$\Pt\bar\Pt\PH(\PH\to\Pb\bar\Pb)$ signal at the LHC.
The ATLAS simulations of the $\Pt\bar\Pt\Pb\bar\Pb$ background
\cite{Kersevan:2002vu,Cammin:2003,Aad:2009wy} are based on
leading-order (LO) matrix elements and the scale choice
$\mu_\mathrm{R}=\mu_\mathrm{F}=E_{\mathrm{thr}}/2$, where
$E_{\mathrm{thr}}=2 \Mt+m_{\Pb\bar\Pb}$ is the partonic threshold
energy.\footnote{The scale choice adopted in the CMS simulations
  \cite{Drollinger:2001ym,Cucciarelli:2006,Ball:2007zza} is not
  documented.}  These predictions suffer from huge scale
uncertainties: the LO $\Pt\bar\Pt\Pb\bar\Pb$ cross section can vary up
to a factor four if the renormalization and factorization scales are
identified with different kinematic parameters \cite{Kersevan:2002vu}.
However, in the case of the $\Pt\bar\Pt\PH$ signal
\cite{Beenakker:2001rj,Beenakker:2002nc} it was found that setting the
QCD scale equal to half the threshold energy leads to fairly moderate
NLO corrections ($K\simeq 1.2$).  The same behaviour was subsequently
observed in two other processes that feature a final state similar to
$\Pt\bar\Pt\Pb\bar\Pb$.  At the scale
$\mu_\mathrm{R,F}=E_{\mathrm{thr}}/2$, the NLO QCD corrections to
$\Pp\Pp\to\Pt\bar\Pt\jet$ 
\cite{Dittmaier:2007wz} and
$\Pt\bar\Pt\PZ$ 
\cite{Lazopoulos:2008de} at the LHC amount to $K\simeq1.0{-}1.15$ (for
$p_{\rT,\mathrm{jet}}\gsim20{-}50\GeV$) and $K\simeq 1.35$,
respectively.  On the basis of these observations one might have
expected that LO $\Pt\bar\Pt\Pb\bar\Pb$ predictions obtained with the
same scale choice might have a decent level of precision, say at the
20--30\% level.  However, it turned out that the NLO corrections to
$\Pt\bar\Pt\Pb\bar\Pb$ production, for
$\mu_\mathrm{R,F}=E_{\mathrm{thr}}/2$, are much larger and correspond
to a $K$ factor of about 1.8
\cite{Bredenstein:2009aj,Bevilacqua:2009zn}.  Apart from the sizable
impact on the $\Pt\bar\Pt\PH$ analysis, this big $K$ factor suggests
the presence of large logarithms that tend to spoil the convergence of
the perturbative expansion. As we argue, this is mainly due to the
fact that the scale $\mu_\mathrm{R,F}=E_{\mathrm{thr}}/2$ does not
provide an adequate description of the QCD dynamics of
$\Pt\bar\Pt\Pb\bar\Pb$ production.  To cure this problem we introduce
a new and more natural scale choice. This leads to a much smaller $K$
factor and also reduces the residual scale dependence at NLO.  Besides
the issue of scale dependences, in this paper we also investigate NLO
effects on various differential distributions and selection cuts that
are relevant for the $\Pt\bar\Pt\PH$ analysis, both within the
traditional ATLAS/CMS approach and in the boosted-Higgs framework.

In addition to its phenomenological relevance, the calculation of the
NLO corrections to $\Pp\Pp\to\Pt\bar\Pt\Pb\bar\Pb$ constitutes also an
important technical benchmark.  The description of many-particle
processes at NLO plays an important role for the LHC physics programme
\cite{Buttar:2006zd, Bern:2008ef}.  Numerous Higgs and new-physics
signals, as well as their background, are characterized by
multi-particle final states.  These processes often involve high
powers of $\alpha_{\mathrm{s}}$ that give rise to very large scale
uncertainties if NLO effects are not taken into account.  The most
prominent multi-particle reactions that require NLO predictions are
summarised in the so-called Les Houches priority list
\cite{Buttar:2006zd,Bern:2008ef}.
In the recent years, the technical challenges raised by these
calculations have triggered an impressive amount of conceptual and
technical developments. In particular, in order to treat one-loop
amplitudes with particle multiplicities higher than five, various
methods based on tensor integrals
\cite{Ferroglia:2002mz,Denner:2002ii,Denner:2005nn,delAguila:2004nf,Giele:2004iy,Giele:2004ub,Binoth:2005ff,Binoth:2006hk,Diakonidis:2008ij,Diakonidis:2009fx}
or on-shell reductions \cite{Bern:1994cg,Berger:2006ci,Bern:2007dw,
  Berger:2008sj,Britto:2004nc,Brandhuber:2005jw,
  Britto:2006sj,Forde:2007mi,Badger:2007si,Anastasiou:2006gt,Giele:2008ve,
  Ellis:2008ir,Ossola:2006us,Mastrolia:2008jb,
  Draggiotis:2009yb,vanHameren:2009dr} have been developed.  Very
recently, these different techniques have lead to the first NLO
results for six-particle processes at the LHC, namely for
$\Pp\Pp\to\Pt\bar\Pt\Pb\bar\Pb$\cite{Bredenstein:2009aj,Bevilacqua:2009zn},
the leading- \cite{KeithEllis:2009bu} and the full-colour
contributions \cite{Berger:2009ep} to $\Pp\Pp\to\PW\jet\jet\jet$, and
for the $q\bar q$ contribution to
$\Pp\Pp\to\Pb\bar\Pb\Pb\bar\Pb$\cite{Binoth:2009rv}.

To compute the virtual corrections to $\Pt\bar\Pt\Pb\bar\Pb$
production we employ explicit diagrammatic representations of the
one-loop amplitudes.  A key feature of our approach is the
factorization of colour structures at the level of individual
diagrams.  This permits to reduce the CPU cost of colour sums
essentially to zero.  Helicity-dependent structures are algebraically
reduced to a common set of so-called standard matrix elements.  In
this way the sums over physical helicities are strongly boosted.
Tensor loop integrals are related to scalar integrals by means of
numerical algorithms that systematically avoid numerical instabilities
from inverse Gram determinants and other spurious singularities
\cite{Denner:2002ii,Denner:2005nn}.  The efficiency of the calculation
is strongly increased by recycling a multitude of common
subexpressions, which occur both inside individual diagrams and in
tensor integrals of different diagrams that share common
sub-topologies.  As demonstrated by the remarkably high CPU speed of
the numerical code, these procedures strongly mitigate the factorial
complexity that is inherent in Feynman diagrams.  The real corrections
are handled with the dipole subtraction
method~\cite{Catani:1996vz,Dittmaier:1999mb,Phaf:2001gc,Catani:2002hc},
and the phase-space integration is performed with adaptive
multi-channel methods
\cite{Berends:1994pv,Denner:1999gp,Dittmaier:2002ap}.  Our results
have been confirmed with the OPP
method~\cite{Ossola:2006us,Mastrolia:2008jb,Draggiotis:2009yb,vanHameren:2009dr}
and {\sc HELAC-1LOOP}~\cite{Cafarella:2007pc,Czakon:2009ss} within the
statistical Monte Carlo error of 0.2\% \cite{Bevilacqua:2009zn}.

The paper is organised as follows.  \refse{se:calculation} is devoted
to technical aspects of the calculation of the virtual and real
corrections.  In \refse{se:numres} we present predictions for the LHC.
In particular, we discuss the scale dependence and investigate NLO
effects on the shape of several distributions.  Our results are
summarised in \refse{se:conclusion}.  In \refapp{app:smes} we outline
the algebraic reduction of helicity structures, and in
\refapp{app:benchmark} we provide benchmark results for the matrix
element squared in lowest order and including virtual corrections for
one phase-space point.

\section{Details of the calculation}
\label{se:calculation}
\begin{figure}
\centerline{
{\unitlength 0.45pt
\SetScale{0.45}
\SetWidth{1.3}
\qqtreediagA\hspace*{1em}
\qqtreediagB\hspace*{1em}
\ggtreediagA\hspace*{1em}
\ggtreediagB\hspace*{1em}
\ggtreediagC
}}
\caption{Sample tree diagrams contributing to the 
$q\bar q\to\Pt\bar\Pt\Pb\bar\Pb$ and $\Pg\Pg\to\Pt\bar\Pt\Pb\bar\Pb$
channels.}
\label{fig:trees}
\end{figure}
In LO, the hadronic production of $\Pt\bar\Pt\Pb\bar\Pb$ proceeds via
the partonic processes $q\bar q\to\Pt\bar\Pt\Pb\bar\Pb$ and
$\Pg\Pg\to\Pt\bar\Pt\Pb\bar\Pb$, which are described by 7 and 36 tree
diagrams, respectively (see \reffi{fig:trees}).
\begin{figure}
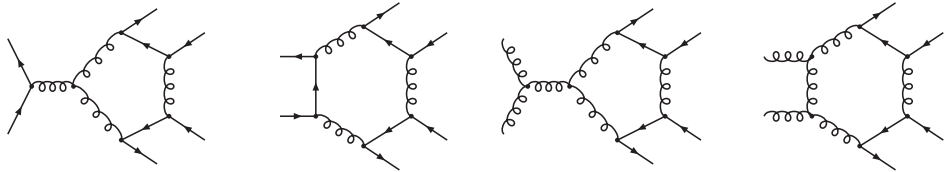

\centerline{
{\unitlength 0.45pt
\SetScale{0.45}
\SetWidth{1.3}
\qqpentagon\hspace*{1em}
\qqhexagon\hspace*{1em}
\ggpentagon\hspace*{1em}
\gghexagon
}}
\caption{
  Sample pentagon and hexagon graphs contributing to $q\bar
  q\to\Pt\bar\Pt\Pb\bar\Pb$ and $\Pg\Pg\to\Pt\bar\Pt\Pb\bar\Pb$.  The $\Pg\Pg$
  ($q\bar q$) channel comprises in total 1003 (188) graphs, including
  40 (8) hexagons and 114 (24) pentagons.  }
\label{fig:loops}
\end{figure}
The virtual NLO QCD corrections to these channels involve 188 and 1003
one-loop diagrams, respectively.
A few examples of pentagon and hexagon graphs are
illustrated in \reffi{fig:loops}.  The real emission contributions are
induced by the partonic processes $q\bar q\to\Pt\bar\Pt\Pb\bar\Pb \Pg$,
$\Pg\Pg\to\Pt\bar\Pt\Pb\bar\Pb \Pg$, $q\Pg\to\Pt\bar\Pt\Pb\bar\Pb q$, and
$\Pg\bar q\to\Pt\bar\Pt\Pb\bar\Pb \bar q$.  The $\Pg\Pg$ channel involves
341 tree diagrams. The $q\bar q$, $q\Pg$ and $\Pg\bar q$ channels, which
are related by crossing transformations, are described by 64 tree
diagrams each (see \reffi{fig:nlotrees}).
\begin{figure}
\centerline{
{\unitlength 0.45pt
\SetScale{0.45}
\SetWidth{1.3}
\qqbremsdiagA\hspace*{1em}
\qqbremsdiagB\hspace*{1em}
\qqbremsdiagC\hspace*{1em}
\gqbremsdiagA
}}
\vspace{7mm}
\centerline{
{\unitlength 0.45pt
\SetScale{0.45}
\SetWidth{1.3}
\ggbremsdiagA\hspace*{1em}
\ggbremsdiagB\hspace*{1em}
\ggbremsdiagC\hspace*{1em}
\ggbremsdiagD
}}
\caption{
Sample real-emission diagrams
contributing to the channels 
$q\bar q\to\Pt\bar\Pt\Pb\bar\Pb \Pg$, \mbox{$q\Pg\to\Pt\bar\Pt\Pb\bar\Pb q$},
and $\Pg\Pg\to\Pt\bar\Pt\Pb\bar\Pb \Pg$.
}
\label{fig:nlotrees}
\end{figure}

In the following we describe the calculation of the virtual
and real NLO corrections. Each of these contributions has been 
worked out twice and independently, resulting in two completely
independent computer codes. Top quarks are treated fully inclusively, \ie 
we do not include top decays. Moreover we handle bottom quarks in the
massless approximation, corresponding to the five-flavour scheme. 
However, we do not take into account the suppressed contribution 
from initial-state bottom quarks.

\subsection{Virtual corrections}
\label{se:virtcor}
The virtual corrections are generated with two in-house {\sc
  Mathematica} programs that reduce Feynman diagrams and generate {\sc
  Fortran77} code in a fully automatized way.  One of the two programs
relies on {\sc FormCalc}~5.2~\cite{Hahn:1998yk} for preliminary
algebraic manipulations.  Here we outline the underlying structure of
the calculation, with emphasis on colour/helicity structures and
tensor integrals. In this respect, both programs are organised in a
fairly similar way.  Since the treatment of the $q\bar q$ channel is
already documented in \citere{Bredenstein:2008zb}, we focus on the
$\Pg\Pg$ channel.

\subsubsection*{Diagram-by-diagram approach}
The virtual corrections are obtained from the 
interference of the one-loop and LO matrix elements
summed over external-state colours and helicities.
This quantity is computed 
on a diagram-by-diagram basis,
\beqar
\sum_{\mathrm{col}}
\sum_{\mathrm{hel}}
\M^{(\onel)} \left(\M^{(\LO)}\right)^*
&=&
\sum_{\Gamma}
\left[\sum_{\mathrm{col}}
\sum_{\mathrm{hel}}
\M^{(\Gamma)} \left(\M^{(\LO)}\right)^*
\right].
\eeqar
The contributions of individual loop diagrams $(\Gamma)$ are evaluated
by separate numerical routines and summed explicitly.
The Feynman diagrams are generated with two independent tools, {\sc
FeynArts}~1.0~\cite{Kublbeck:1990xc} and {\sc
FeynArts}~3.2~\cite{Hahn:2000kx}.

\subsubsection*{Colour factorization}
One of the key features of the diagram-by-diagram approach is that the
cost related to the large number of diagrams is compensated by the
possibility to perform colour sums very efficiently.  This is a
consequence of colour factorization: individual (sub)diagrams consist
of a single colour-stripped amplitude ${\cal A}^{(\Gamma)}$ multiplied
by a trivial colour factor ${\cal C}^{(\Gamma)}$,
\beqar\label{eq:colfact}
\M^{(\Gamma)}
&=&
\A^{(\Gamma)}
{\cal C}^{(\Gamma)}.
\eeqar
More precisely, each diagram gives rise to $3^{n_4}$ colour-factorized
contributions of type \refeq{eq:colfact}, where $n_4$ is the number of
quartic gluon vertices in the diagram.  These terms are handled as
separate subdiagrams.  However, most diagrams do not involve quartic
couplings, and their colour structures factorize completely.  For
instance, the last diagram in \reffi{fig:loops} involves a single
colour structure
\beqar
{\cal C}^{(\mathrm{hex})}
&=&
\sum_{b,c,d,e}f^{a_1bc}f^{a_2cd}
\left(T^bT^e\right)_{i_3i_4}
\left(T^dT^e\right)_{i_5i_6},
\eeqar
where $a_1,a_2$ and $i_3,i_4,i_5,i_6$ are the colour indices of the
$\Pg\Pg$ and $\Pt\bar\Pt\Pb\bar\Pb$ external states, numbered in this
order.  All colour structures can be easily reduced to Kronecker
symbols and Gell-Mann matrices $T^a=\la^a/2$ by using
\beq
f^{abc}T^c=-\ri[T^a,T^b],\qquad  
T^a_{ij}T^a_{kl}=
\frac{1}{2}\left(\delta_{il}\delta_{kj}-
\frac{1}{N_{\mathrm{c}}}\delta_{ij}\delta_{kl}\right), 
\eeq
and other well-known $\SU(N_{\mathrm{c}})$ relations.
In the $\Pg\Pg$ channel, this reduction leads to a colour basis
of 14 elements,
\newcommand{\csA}[6]{T^{a_#1}_{i_#2i_#3}T^{a_#4}_{i_#5i_#6}}
\newcommand{\csB}[6]{\de_{i_#1i_#2} \left(T^{a_{#3}}T^{a_{#4}}\right)_{i_#5i_#6}}
\newcommand{\csC}[6]{\de^{a_#1a_#2}\de_{i_#3i_#4}\de_{i_#5i_#6}}
\beqar
{\cal C}_1&=&\csA{1}{3}{4}{2}{5}{6},\quad
{\cal C}_2=\csB{3}{4}{1}{2}{5}{6},\quad
{\cal C}_3=\csB{3}{4}{2}{1}{5}{6},\nl
{\cal C}_4&=&\csA{1}{5}{6}{2}{3}{4},\quad
{\cal C}_5=\csB{5}{6}{1}{2}{3}{4},\quad
{\cal C}_6=\csB{5}{6}{2}{1}{3}{4},\nl
{\cal C}_7&=&\csA{1}{3}{6}{2}{5}{4},\quad
{\cal C}_8=\csB{3}{6}{1}{2}{5}{4},\quad
{\cal C}_{9}=\csB{3}{6}{2}{1}{5}{4},\nl
{\cal C}_{10}&=&\csA{1}{5}{4}{2}{3}{6},\quad
{\cal C}_{11}=\csB{5}{4}{1}{2}{3}{6},\quad
{\cal C}_{12}=\csB{5}{4}{2}{1}{3}{6},\nl
{\cal C}_{13}&=&\csC{1}{2}{3}{4}{5}{6},\quad
{\cal C}_{14}=\csC{1}{2}{3}{6}{5}{4}.
\eeqar
For $N_{\mathrm{c}}=3$, only 13 of these colour operators are independent
owing to the relation
\beq
{\cal C}_{14} = -2\sum_{i=1}^6 {\cal C}_i
+2\sum_{i=7}^{12} {\cal C}_i + {\cal C}_{13}.
\eeq
The summation over external colours is performed once and for all at
the level of the colour basis and the LO matrix element.  To this end,
we compute the colour-interference matrix
\beqar
I_{kl}=
\sum_{\mathrm{col}} 
{\cal C}_k
{\cal C}^*_{l},
\eeqar
and reducing the tree matrix element in colour space,
\beqar
\M^{(\LO)}
&=&
\sum_{l}
\M_{l}^{(\LO)}
{\cal C}_{l},
\eeqar
we build the interference of $\M^{(\LO)}$ with the elements of the
colour basis as
\beqar\label{coltreeint}
\tilde\M_{k}^{(\LO)}
&=&
\sum_{\mathrm{col}} 
{\cal C}_k
\left(\M^{(\LO)}\right)^*
=
\sum_{l}{I}_{kl}\left(\M_{l}^{(\LO)}\right)^*.
\eeqar
Then, upon reduction of the factorized colour structure of the 
loop diagrams, 
\beqar\label{colofactb}
\M^{(\Gamma)}
&=&
\A^{(\Gamma)}
{\cal C}^{(\Gamma)}
=
\A^{(\Gamma)}
\left(\sum_k c_k^{(\Gamma)}{\cal C}_{k}\right),
\eeqar
we obtain the colour-summed loop--tree interference as
\beqar\label{colofactc}
\sum_{\mathrm{col}}
\M^{(\Gamma)}\left(\M^{(\LO)}\right)^*
&=&
\A^{(\Gamma)}
\left(\sum_{k} c_k^{(\Gamma)}
\tilde\M_{k}^{(\LO)}\right),
\eeqar
where the coefficients $c_k^{(\Gamma)}$ are simple numbers.  The
colour-summed result is given by a combination of previously computed
colour--Born interference terms \refeq{coltreeint}.  This requires
{\em a single evaluation} of the non-trivial colour-stripped
amplitude ${\cal A}^{(\Gamma)}$ of each (sub)diagram.

Helicity structures are handled in a very similar way.  The
helicity-dependent parts of all diagrams are reduced to a common basis
of so-called Standard Matrix Elements (SMEs), and helicity sums are
performed once and for all at the level of the SMEs--Born interference
(see below).  The diagram-independent treatment of the
helicity-dependent parts of loop graphs is made possible by the
covariant decomposition of tensor integrals, \ie by replacing loop
momenta (in the numerator) with external momenta and metric tensors.

\subsubsection*{Covariant decomposition and numerical reduction of
  tensor integrals} 
Tensor one-loop integrals with $N$ propagators and $P$ Lorentz indices
are expressed in terms of totally symmetric covariant structures
$\{g\dots g p\dots p\}^{\mu_1\dots\mu_P}_{j_1\dots j_P}$ involving
$g^{\mu\nu}$ and the external momenta $p_1,\dots,p_{N-1}$,
\beq
\frac{(2\pi\mu)^{4-D}}{\ri\pi^{2}}\int \rd^{D}q\,
\frac{q^{\mu_1}\dots q^{\mu_P}}
{\prod_{i=0}^{N-1}\left[(q+p_{i})^2-m_i^2+\ri 0\right]}
=
\sum_{j_1,\dots,j_P=0}^{N-1} 
{T^{N}_{j_1\dots j_P}}\; 
{\{g\dots g p\dots p\}^{\mu_1\dots\mu_P}_{j_1\dots j_P}},
\eeq
with $D$ denoting the number of space--time dimensions.  For details
of the notation we refer to \citere{Denner:2005nn}.  To describe
$N$-point integrals with $N\ge 5$, tensor structures with only four
external momenta would be sufficient.  However, in order to avoid
potential instabilities due to inverse Gram determinants we use a
redundant set of structures, including the metric tensor and $N-1$
momenta.

The virtual corrections to $q\bar q\to\Pt\bar\Pt\Pb\bar\Pb$ and
$\Pg\Pg\to\Pt\bar\Pt\Pb\bar\Pb$ involve tensor integrals up to rank
$P=3$ and $P=4$, respectively.  The one-loop amplitudes are expressed
as linear combinations of tensor-integral coefficients
$T^{N}_{j_1,\dots,j_P}$.  The latter are evaluated by {\em
  numerical\/} libraries that recursively reduce them to master
integrals using
the methods of \citeres{Denner:2002ii,Denner:2005nn}%
\footnote{We note in passing that the reduction methods of
  \citeres{Denner:2002ii,Denner:2005nn} have also been used in the
  related calculation~\cite{Lei:2007rv} of NLO QCD corrections to the
  $2\to4$ particle process $\ga\ga\to\Pt\bar\Pt\Pb\bar\Pb$ at a
  $\ga\ga$ collider.}.  Avoiding an explicit reduction of analytic
expressions to master integrals, this numerical approach prevents
prohibitively large expressions and permits to adapt the reduction
strategy to the specific numerical problems that appear in different
phase-space regions.

Tensor $N$-point integrals with $N\ge 5$ are expressed in terms of
lower-rank and lower-point integrals exploiting the
four-dimensionality of
space--time~\cite{Denner:2002ii,Denner:2005nn}.%
\footnote{Similar reductions are described in \citere{Binoth:2005ff}.}
The tensor rank and the number of propagators are simultaneously
reduced without introducing inverse Gram determinants.  Consequently,
the maximal power of inverse Gram determinants resulting from the
entire reduction is given by the maximal rank of four-point integrals,
which never exceeds four in renormalizable gauges.  Scalar hexagons
and pentagons are reduced to boxes using Melrose's method~\cite{Me65}.
Tensor 4-point and 3-point integrals are reduced to scalar integrals
with the Passarino--Veltman algorithm \cite{Passarino:1979jh} as long
as no small Gram determinant appears in the reduction.  
If small Gram determinants occur, alternative
schemes are applied \cite{Denner:2005nn}.%
\footnote{Similar procedures based on numerical evaluations of
  specific one-loop integrals~\cite{Binoth:2005ff,Ferroglia:2002mz} or
  expansions in small determinants~\cite{Giele:2004ub} have also been
  proposed by other authors.}  More precisely, we make use of expansions of
the tensor coefficients about the limit of vanishing Gram determinants
and possibly other kinematical determinants.  
 One- and two-point tensor integrals 
are obtained with numerically stable analytic expressions.

Ultraviolet (UV) divergences are regularized dimensionally throughout,
but infrared (IR) divergences are treated in different variants, which
comprise pure dimensional regularization with strictly massless light
quarks and a hybrid scheme with small quark masses.  The corresponding
scalar integrals are evaluated using the methods and results of
\citeres{'tHooft:1979xw,Beenakker:1990jr}, and different
regularization schemes are translated into each other as described in
\citere{Dittmaier:2003bc}.

The calculation of tensor integrals is implemented in two independent
{\sc Fortran} libraries. This permits to perform detailed cross
checks, which confirm the excellent numerical stability of the
reduction procedure.  An automatic cache system is implemented that
strongly boosts the reduction by recycling a multitude of tensor
integrals among Feynman diagrams with common sub-topologies.  The
virtual corrections to $\Pg\Pg\to\Pt\bar\Pt\Pb\bar\Pb$ comprise about
350 different scalar integrals, which require roughly $10\,$ms CPU
time per phase-space point on a 3 GHz Intel Xeon processor.  
The calculation of the complete set of
scalar and tensor integrals with and without cache system takes
approximately $40\,$ms and $200\,$ms, respectively.

\subsubsection*{Rational parts}
In $D=4-2\epsilon$ dimensions, UV-singular tensor integrals
give rise to $1/\epsilon_{\mathrm{UV}}$ poles,
\beqar\label{rationala}
T^{N}_{j_1\dots j_P}
&=&
\hat{T}^{N}_{j_1\dots j_P}
+
\frac{R^{N}_{j_1\dots j_P}}{\epsilon_{\mathrm{UV}}}.
\eeqar
Consequently, their $D$-dimensional coefficients 
need to be expanded in $D-4$, 
\beqar\label{rationalb}
f(D) T^{N}_{j_1\dots j_P}
&=&
f(4) T^{N}_{j_1\dots j_P}
-2 f'(4) R^{N}_{j_1\dots j_P},
\eeqar
resulting in so-called rational terms that are
proportional to the pole residues $R^{N}_{j_1\dots j_P}$.  Rational
contributions originate from $D$-dependent terms in tensor-reduction
identities and in the loop-momentum-independent part of the diagram
numerators.  The relevant expansions are automatically performed by
means of a catalogue of residues of UV poles.

Note that in \refeq{rationalb} we have implicitly assumed that
rational terms resulting from $1/\epsilon$ and $1/\epsilon^2$ poles of
IR kind vanish.  This is a non-trivial and general property
of one-loop QCD amplitudes. More precisely, while rational terms of IR
origin can be present in the wave-function renormalization factors, in
truncated one-loop amplitudes they cancel.  This holds within the 't
Hooft--Feynman gauge and similar gauge fixings, as was explicitly
proven in Appendix~A of \citere{Bredenstein:2008zb}.

\subsubsection*{Algebraic reduction of helicity structures and helicity sums}
The helicity structures encountered in the explicit evaluation of all
Feynman diagrams are algebraically reduced to a common basis of SMEs
as described below.  The general form of SMEs for the
$\Pg\Pg\to\Pt\bar\Pt\Pb\bar\Pb$ channel is\footnote{For convenience we
  consider the crossed process $\Pg\Pg\bar\Pt\Pt\bar\Pb\Pb\to 0$, \ie
  we treat all particles and momenta as incoming.}
\beq\label{SMEs}
{\hat\M}_m =
Q_m^{\mu_1\mu_2\rho_1\dots \rho_l}
\varepsilon_{\mu_1}(p_1)
\varepsilon_{\mu_2}(p_2)
\left[\bar v(p_3) \gamma_{\rho_1}\dots \gamma_{\rho_k}u(p_4)\right]
\left[\bar v(p_5) \gamma_{\rho_{k+1}}\dots
\gamma_{\rho_l}u(p_6)\right],\nn\nl
\eeq
where $Q_m^{\mu_1\mu_2\rho_1\dots \rho_l}$ consists of combinations of
metric tensors and external momenta.  These compact spinor chains
permit to decouple helicity information from the remnant parts of the
diagrams, so that helicity sums can be performed in a
diagram-independent and efficient way.  In practice, the
colour-stripped part of each loop diagram [see \refeq{colofactb}] is
expressed as a linear combination of SMEs and tensor integrals,
\beqar\label{SMEdecomp}
\A^{(\Gamma)}
&=&
\sum_m {\cal F}^{(\Gamma)}_m {\hat \M}_m,
\nn\\
{\cal F}^{(\Gamma)}_m &=& \sum_{P} 
\sum_{j_1,\ldots,j_P=0}^{N-1} 
{\cal K}^{(\Gamma)}_{m;j_1\dots j_P}
{T^{N}_{j_1\dots j_P}} 
\;+\; \mbox{rational parts}.
\eeqar
The coefficients ${\cal K}^{(\Gamma)}_{m;i_1\dots i_P}$ are rational
functions of the kinematic invariants.  These functions involve only
denominators from intermediate-particle propagators and are free
of spurious poles that might generate numerical instabilities.

Helicity sums are performed at the level of the interference of the
diagram-independent SMEs with the colour-projected Born amplitude
\refeq{coltreeint},
\beqar
M_{km}
&=&
\sum_{\mathrm{hel}}
{\hat\M}_m \tilde\M_k^{(\LO)}
=
\sum_{l}
I_{kl}
\sum_{\mathrm{hel}}
 {\hat\M}_m \left(\M^{(\LO)}_{l}\right)^*.
\eeqar
This matrix is computed only once per phase-space point employing the
Weyl--van der Waerden spinor formalism of \citere{Dittmaier:1998nn}.
Using $M_{km}$ one can directly obtain the colour- and helicity-summed
contributions of each loop diagram in terms of its colour- and
helicity-independent form factors ${\cal F}^{(\Gamma)}_m$ and the
coefficients ${c}^{(\Gamma)}_k$ of its factorized colour structure
\refeq{colofactb},
\beqar\label{colhelsum}
\sum_{\mathrm{col}}
\sum_{\mathrm{hel}}
\M^{(\Gamma)} \left(\M^{(\LO)}\right)^*
&=&
\sum_m {\cal F}_m^{(\Gamma)}
\left(\sum_{k}
c_k^{(\Gamma)} 
M_{km}\right).
\eeqar
Owing to the high number of SMEs ${\hat \M_m}$ and the complexity of
the corresponding form factors ${\cal F}_m^{(\Gamma)}$, the
representation \refeq{colhelsum} yields fairly large expressions.  For
instance, the size of the numerical routines describing individual
hexagon diagrams in the $\Pg\Pg$ channel
is of the order of $0.5{-}1\,$MB.
The reduction of helicity structures to SMEs is one of the key aspects
that determine the size and the speed of the code.  In order to avoid
possible numerical cancellations, this procedure is entirely based on
algebraic identities that are free of denominators.  The reduction
algorithm consists of two main steps.  The first step is based on
process-independent identities in $D$ dimensions.  To reduce helicity
structures of type \refeq{SMEs} we employ: momentum conservation;
Dirac algebra; ordering of Dirac matrices inside Dirac chains; Dirac
equation; transversality and gauge-fixing conditions for the
gluon-polarization vectors, $p_i^{\mu}\varepsilon_{\mu}(p_j)=0$ for
$i,j=1,2$.  The basis of SMEs obtained with these identities contains
more than thousand elements for the $\Pg\Pg$ channel.

After these manipulations in $D$ dimensions we extract all rational
terms performing the relevant expansions in $D-4$.  We then proceed
with a second reduction step, based on four-dimensional relations.
Specifically, we apply two alternative reduction algorithms that are
based on relations derived from Chisholm's identity.

The first algorithm is constructed along the lines of the reduction
described in \citere{Bredenstein:2008zb} for the $q\bar q$ channel.
Each fermion chain is split into two contributions via insertion of
chiral projectors $\omega_\pm=(1\pm\gamma^5)/2$,
\beq
\left[\bar{v}(p_3)
\Gamma_{a}
u(p_4)\right]
\left[\bar{v}(p_5)
\Gamma_{b}
u(p_6)\right]
=
\sum_{\lambda,\rho=\pm}
\left[\bar{v}(p_3)
\Gamma_{a}
\omega_\lambda
u(p_4)\right]
\left[\bar{v}(p_5)
\Gamma_{b}
\omega_\rho
u(p_6)\right].
\eeq
This permits to employ various relations of the
type~\cite{Bredenstein:2008zb,Denner:2005fg}
\beqar\label{eq:chisholm1}
{
\gamma^\mu}
{\gamma^\alpha
\gamma^\beta}
\omega_\pm
\otimes
{
\gamma_{\mu}}
&=&
{
\gamma^\mu}
\omega_\pm
\otimes
\left(
\gamma_{\mu}
\gamma^\beta
\gamma^\alpha
\omega_\pm
+
\gamma^\alpha
\gamma^\beta
\gamma_{\mu}
\omega_\mp
\right),
\eeqar
where the tensor product connects Dirac matrices that belong to
different fermion chains.  By means of such identities one can
exchange Dirac matrices between chains that are connected by
$\gamma^\mu\otimes\gamma_\mu$ contractions.  As described in
\citere{Bredenstein:2008zb}, using identities of type
\refeq{eq:chisholm1} in combination with the above-mentioned
$D$-dimensional relations (Dirac equation, etc.)  one can obtain a
rich variety of non-trivial reduction identities.  In this way we have
constructed a fairly sophisticated reduction algorithm 
(see \refapp{app:smes})
that relates
all helicity structures present in the $\Pg\Pg$ channel to 502 SMEs.
In spite of its efficiency, this reduction procedure has the
disadvantage of depending on process-specific aspects, like the number
of massive and massless fermion chains and the number of external
momenta.  It is thus important to investigate the trade-off between
the obtained efficiency and the time-consuming task of designing the
reduction on a process-by-process basis.  To this end, we have
implemented an alternative and much simpler reduction method.  This
procedure is entirely process-independent.  It does not make use of
chiral projectors, and consists of a single four-dimensional identity,
\newcommand{\QG}[5]{
\gamma^{\mu_{#1}}
\gamma^{\mu_{#2}}
\gamma^{\mu_{#3}}
\gamma^{\mu_{#4}}
\gamma^{\mu_{#5}}
}
\newcommand{\TG}[3]{
\gamma^{\mu_{#1}}
\gamma^{\mu_{#2}}
\gamma^{\mu_{#3}}
}
\newcommand{\SG}[1]{
\gamma^{\mu_{#1}}
}
\newcommand{\GTE}[2]{
g^{\mu_{#1}\mu_{#2}}
}
\beqar\label{fivegammaid}
\lefteqn{\QG{1}{2}{3}{4}{5} 
=
\GTE{1}{2}\TG{3}{4}{5}
-\GTE{1}{3}\TG{2}{4}{5}
+\GTE{1}{4}\TG{2}{3}{5}
-\GTE{1}{5}\TG{2}{3}{4}
}\quad&&\nl&&{}
+\GTE{2}{3}\TG{1}{4}{5}
-\GTE{2}{4}\TG{1}{3}{5}
+\GTE{2}{5}\TG{1}{3}{4}
+\GTE{3}{4}\TG{1}{2}{5}
\nl&&{}
-\GTE{3}{5}\TG{1}{2}{4}
+\GTE{4}{5}\TG{1}{2}{3}
-\left(
\GTE{1}{2}\GTE{3}{4}
-\GTE{1}{3}\GTE{2}{4}
+\GTE{1}{4}\GTE{2}{3}
\right)\SG{5}
\nl&&{}
+\left(
\GTE{1}{2}\GTE{3}{5}
-\GTE{1}{3}\GTE{2}{5}
+\GTE{1}{5}\GTE{2}{3}
\right)\SG{4}
-\left(
\GTE{1}{2}\GTE{4}{5}
-\GTE{1}{4}\GTE{2}{5}
\right.\nl&&{}\left.
+\GTE{1}{5}\GTE{2}{4}
\right)\SG{3}
+\left(
\GTE{1}{3}\GTE{4}{5}
-\GTE{1}{4}\GTE{3}{5}
+\GTE{1}{5}\GTE{3}{4}
\right)\SG{2}
-\left(
\GTE{2}{3}\GTE{4}{5}
\right.\nl&&{}\left.
-\GTE{2}{4}\GTE{3}{5}
+\GTE{2}{5}\GTE{3}{4}
\right)\SG{1},
\eeqar
which can be derived from Chisholm's identity and permits to eliminate
any spinor chain involving more than three Dirac matrices without
introducing $\gamma_5$ and $\eps$-tensors.\footnote{ Products of four
  Dirac matrices can occur only inside the massive top-quark chain.
  In this case one can introduce a fifth Dirac matrix by rewriting the
  massive spinor as $u(p)={\dsl p}u(p)/m$, and then use
  \refeq{fivegammaid}.}  In this case we could reduce all
$\Pg\Pg$-channel helicity structures to 970 SMEs.

Comparing the number of SMEs obtained with the process-dependent and
process-independent algorithms, we observe that the former is superior
by roughly a factor two. Thus, if we naively assume that the CPU
efficiency scales with the number of SMEs, we would expect a
factor-two difference in the speed of the numerical code. In contrast,
we find that the CPU efficiency obtained with the two reductions is
almost identical. This suggests that the reduction of the number of
SMEs is compensated by an increase in the size of the form factors.
This unexpected result means that the obtained CPU performance---at
least for this process---does
not depend on sophisticated and process-dependent optimisations.

\subsection{Real corrections}
\label{se:realcor}

\newcommand{\qparbar}{\raisebox{.6em}{\tiny $(-)$}\hspace{-.83em}q}

The calculation of the $q\bar q$ channel has been described in
\citere{Bredenstein:2009aj}. The evaluation of the $\qparbar\,\Pg$
channels is done in the same way. In the following we sketch the
calculation for the $\Pg\Pg$ channel. We have again performed two
independent calculations of all building blocks.

In both evaluations of the real corrections the amplitudes are
calculated as helicity matrix elements which have been generated with
{\sc Madgraph 4.1.33} \cite{Stelzer:1994ta}. While the amplitudes for
$q\bar q\to\Pt\bar\Pt\Pb\bar\Pb\Pg$ and $\raisebox{.6em}{\tiny
  $(-)$}\hspace{-.83em}q\,\Pg \to\Pt\bar\Pt\Pb\bar\Pb \qparbar\,$ have
been checked with the Weyl--van der Waerden spinor formalism of
\citere{Dittmaier:1998nn}, those for
$\Pg\Pg\to\Pt\bar\Pt\Pb\bar\Pb\Pg$ have been verified with an
implementation of off-shell recursion relations
\cite{Berends:1987me,Caravaglios:1995cd,Draggiotis:1998gr}. The
singularities for soft and collinear gluon emission are isolated via
dipole
subtraction~\cite{Catani:1996vz,Dittmaier:1999mb,Phaf:2001gc,Catani:2002hc}
for NLO QCD calculations using the formulation~\cite{Catani:2002hc}
for massive quarks.
Soft and collinear singularities in the ``endpoint part'' of the
subtraction function (the $I$ operator of
\citeres{Catani:1996vz,Catani:2002hc}), \ie the part of the
subtraction terms that has to be combined with the virtual
corrections, are regularized using the same regularization
prescription (dimensional or with small quark masses) as the
corresponding virtual corrections.  No regularization is needed in the
subtraction terms for the real corrections. For both the $q\bar q$ and
$\Pg\Pg$ channels 30 different dipole subtraction terms need to be
included while each $\qparbar\,\Pg$ channel requires only 10, since we
demand b~quarks with finite transverse momentum in the final state.
After combining virtual and real corrections,
singularities connected to collinear configurations in the final state
cancel for ``collinear-safe'' observables after applying a jet
algorithm. Singularities connected to collinear initial-state
splittings are removed via $\overline{\mathrm{MS}}$ QCD factorization
by PDF redefinitions.  
In both evaluations the phase-space integration is performed with
multi-channel Monte Carlo generators~\cite{Berends:1994pv} and adaptive
weight optimisation similar to the one implemented in {\sc
  Lusifer}~\cite{Dittmaier:2002ap}.

\begin{sloppypar}
  {\it Version~1} of the real corrections employs the {\sc MadDipole}
  implementation of dipole subtraction~\cite{Frederix:2008hu}.  The
  phase-space integration, implemented in {\sc C++}, is based on {\sc
    RacoonWW}, but the phase-space mappings are built up in a more
  generic way very similar to the approach of {\sc
    Lusifer}~\cite{Dittmaier:2002ap}.  For each of the 341
  bremsstrahlung Feynman diagrams a corresponding channel is taken into
  account in the Monte Carlo integration.
\end{sloppypar}

In {\it version~2} all dipole subtraction terms have been implemented
directly into the Monte Carlo generator.  The Monte Carlo generator is
a further development of the one used in {\sc COFFER$\ga\ga$}
\cite{Bredenstein:2005zk} and for the calculation of the NLO
corrections to $\Pp\Pp\to\PH\jet\jet+X$ \cite{Ciccolini:2007jr}.  In
addition to the Monte Carlo channels for the bremsstrahlung diagrams
$30\times 36=1080$ channels are used to map the dipole subtraction
terms in the $\Pg\Pg$ channel. These additional channels lead to some
improvement in the convergence of the Monte Carlo integration.

All real bremsstrahlung amplitudes of Madgraph have been checked
against independent calculations for several phase-space points.  The
cancellation between real matrix elements and dipole subtraction terms
has been verified numerically in all soft and collinear regions.  The
individual dipole subtraction terms, the subtracted real matrix
elements, and the integrated subtraction terms ($P$ and $K$ terms of
\citeres{Catani:1996vz,Catani:2002hc}) have been compared point-wise
between the two independent calculations. The agreement was generally
at the level of 13 digits. The integrated LO cross section has been
verified with {\tt SHERPA}\cite{Gleisberg:2003xi} at the level of the
integration errors of 0.2\%. From the point of view of Monte Carlo
integration the most complicated and time-consuming part is the
integration of the real corrections in the $\Pg\Pg$ channel.
For the complete NLO cross section we found agreement between the two
versions of our
code within 1--$3\sigma$, where $1\si$ corresponds to 0.1--0.2\%. The
results for the distributions coincide within 1--$3\sigma$ for each
bin.

In order to give an impression on the statistics and the required CPU
time we show in \refta{Tab:cpu} some numbers for $2\times10^7$
generated phase-space points before cuts.
This yields an accuracy for the NLO cross section of about 0.5\%. The
contributions of the $\raisebox{.6em}{\tiny
  $(-)$}\hspace{-.83em}q\,\Pg$ channel were calculated for every 4th
event and those of the virtual corrections in the $\Pg\Pg$ channel and
the contributions of the $q\bar q$ channel for every 20th event. The
bulk of the runtime is taken by the $\Pg\Pg$ channel.
For the virtual corrections the CPU time is dominated by the $\Pg\Pg$
channel and amounts to $180\,$ms per event. The virtual correction in
the  $q\bar q$ channel are by a factor 20 faster.
In order to produce the plots for the scale variations we
generated $2\times10^7$ phase-space points for the NLO predictions and
$2\times10^8$ for the LO cross section.  To generate the 
distributions we used about 20 times more events for the NLO results.

\begin{table}
$$
\begin{array}{c@{\qquad}c@{\qquad}c@{\qquad}c@{\qquad}c@{\qquad}c}
\hline
& \sigma/{\sigma_\mathrm{LO}}
& \mbox{\# events}  &  
{(\Delta \sigma)_{\mathrm{stat}}/{\sigma}} &  \mbox{runtime}  &
{\mbox{time/event}}  \\
&& \mbox{(after cuts)}  &&  \\
\hline 
\mathrm{tree~level} & \phantom{-}86\% & 5.3\times 10^{6}&  0.4 \times 10^{-3} & 38\,\mathrm{min} &
0.4\,\mathrm{ms} \\
\mathrm{virtual} & {-11\%} &  0.26\times 10^{6} &  {0.6 \times
10^{-3}} &
13\,\mathrm{h} &\,180\,\mathrm{ms}  \\
\mathrm{real+dipoles} & \phantom{-}49\% &  10\times 10^{6} &
3 \times 10^{-3}& 40\,\mathrm{h} &  14\,\mathrm{ms}\\
\hline
\mathrm{total} & 124\% & & 4 \times 10^{-3}& 53\,\mathrm{h}\\
\hline
\end{array}
$$
\caption{Statistics and speed of various parts of the calculation 
based on $2\times10^7$ events before applying 
cuts generated on a 3\,GHz Intel Xeon
processor using the pgf77 compiler with standard options.}
\label{Tab:cpu}
\end{table}

\section{Predictions for the LHC}
\label{se:numres}
The thorough description of a complex signature like
$\Pt\bar\Pt\Pb\bar\Pb$ involves numerous possible observables. Here we
investigate distributions and cuts that are relevant for the search of
$\Pt\bar\Pt\PH (\PH\to\Pb\bar\Pb)$ at the
LHC\cite{Cammin:2003,Cucciarelli:2006}, where $\Pt\bar\Pt\Pb\bar \Pb$
contributes to the irreducible background.
In our previous work \cite{Bredenstein:2009aj} we found a $K$ factor
of about $1.8$ for the integrated $\Pt\bar\Pt\Pb\bar\Pb$ cross section
at the LHC.  This unexpectedly large NLO effect raises two important
issues that we address in the present analysis.
Firstly, we discuss the relation between the large $K$ factor and the
scale choice.  This leads us to a new and more appropriate scale
choice, which improves the convergence of the perturbative expansion.
Secondly, we consider possible strategies to reduce the
$\Pt\bar\Pt\Pb\bar\Pb$ cross section in order to facilitate the
extraction of the $\Pt\bar\Pt\PH$ signal.  In particular, we study the
influence of a jet veto on the NLO cross section and its perturbative
stability.
We also explore the kinematic region of highly-boosted $\Pb\bar\Pb$
pairs, which helps to separate the Higgs signal from its QCD
background, as first suggested for associated WH and ZH production
\cite{Butterworth:2008iy} and, very recently, also for $\Pt\bar\Pt\PH$
production \cite{Plehn:2009rk}.

Let us remind that top-quark decays are not included in our
calculation.  In practice we treat top quarks in a completely
inclusive way, and we restrict our analysis to the kinematic
properties of those two b quarks that do not result from top decays.
From the experimental view-point, the presented distributions
correspond to the unrealistic situation of perfect top-quark
reconstruction.  The detailed description of top-decay products and
the related issue of b-quark combinatorics are left for future
studies.

\subsection{Input parameters, jet definition, and  cuts}

We study the process $\Pp\Pp\to\Pt\bar\Pt\Pb\bar\Pb+X$ at
$\sqrt{s}=14\TeV$.  For the top-quark mass, renormalized in the
on-shell scheme, we take the numerical value $\Mt=172.6\GeV$
\cite{Group:2008nq}. All other QCD partons, including $\Pb$~quarks,
are treated as massless particles.  Collinear final-state
configurations, which give rise to singularities, are recombined into
IR-safe jets using a \mbox{$k_{\rT}$-algorithm} \cite{Catani:1992zp}.
Specifically, we adopt the \mbox{$k_\rT$-algorithm} of
\citere{Blazey:2000qt} and recombine all final-state b quarks and
gluons with pseudorapidity $|\eta| < 5$ into jets with separation
$\sqrt{\Delta\phi^2+\Delta y^2}>D=0.4$ in the
rapidity--azimuthal-angle plane.  Requiring two b jets, this also
avoids collinear singularities resulting from massless
$\Pg\to\Pb\bar\Pb$ splittings.\footnote{ Note that, as compared to our
  previous analysis\cite{Bredenstein:2008zb,Bredenstein:2009aj}, we
  have reduced the jet-algorithm parameter from $D=0.8$ to $D=0.4$.
  This is particularly important for highly boosted b-quark pairs with
  $m_{\Pb\bar\Pb}\sim \MH$, since $D=0.8$ would lead to their
  recombination into a single jet and, consequently, to their
  rejection.}

After recombination, we impose the following cuts on the 
transverse momenta and rapidities of the b jets:
\beq\label{standardcuts}
p_{\rT,\Pb}>p_{\rT,\Pb,\mathrm{cut}}=20\GeV, \qquad
|y_\Pb|<y_{\Pb,\mathrm{cut}}=2.5.
\eeq
This choice is dictated by the detector geometry and the search for a
$\Pt\bar\Pt\PH(\PH\to\Pb\bar\Pb)$ signal at the
LHC\cite{Cammin:2003,Cucciarelli:2006}.  The outgoing (anti)top quarks
are neither affected by the jet algorithm nor by phase-space cuts.
For what concerns b~quarks, the jet algorithm and the requirement of
having two b~jets with $p_{\rT,\Pb}>20\GeV$ sets an effective lower
limit on the $\Pb\bar\Pb$ invariant mass of roughly $10\GeV$.  But the
$m_{\Pb\bar\Pb}$-range that is relevant for the Higgs-boson search is
actually much higher.  In this kinematic region, with $p_{\rT,\Pb}\gg
\Mb$ and $m_{\Pb\bar\Pb}\gg \Mb$, we expect that the $m_b=0$
approximation works fairly well.  To asses its precision we compared
the LO cross section for $m_\Pb=0$ and $m_\Pb=4.2\GeV$ using {\tt
  SHERPA}\cite{Gleisberg:2003xi}.  For the integrated cross section,
which is dominated by $m_{\Pb\bar\Pb}$ values well below $100\GeV$, we
found that the finite-$\Mb$ effect is smaller than 3\%.

We consistently use the CTEQ6~\cite{Pumplin:2002vw} set of PDFs, \ie
we take CTEQ6L1 PDFs with a one-loop running $\alpha_{\mathrm{s}}$ in
LO and CTEQ6M PDFs with a two-loop running $\alpha_{\mathrm{s}}$ in
NLO, but neglect the suppressed contributions from b~quarks in the
initial state.  The number of active flavours is $N_{\mathrm{F}}=5$,
and the respective QCD parameters are
$\Lambda_5^{\mathrm{LO}}=165\MeV$ and
$\Lambda_5^{\overline{\mathrm{MS}}}=226\MeV$.  In the renormalization
of the strong coupling constant the top-quark loop in the gluon
self-energy is subtracted at zero momentum. In this scheme, the
running of $\alpha_{\mathrm{s}}$ is generated solely by the
contributions of the light-quark and gluon loops.

\subsection{Renormalization and factorization scales}
The perturbative expansion of the $\Pp\Pp\to\Pt\bar\Pt\Pb\bar\Pb$
cross section starts with the fourth power of $\alpha_{\mathrm{s}}$.
Consequently the LO predictions, and also the $K$ factor, are
extremely sensitive to variations of the renormalization scale.  In
\citere{Kersevan:2002vu} it was pointed out that the LO
$\Pt\bar\Pt\Pb\bar\Pb$ cross section can vary by up to a factor four
if the QCD scale is identified with different kinematic parameters.
In all recent ATLAS studies of $\Pt\bar\Pt\PH (\PH\to\Pb\bar\Pb)$
\cite{Kersevan:2002vu,Cammin:2003,Aad:2009wy} the signal and its
$\Pt\bar\Pt\Pb\bar\Pb$ background were simulated by setting the
renormalization and factorization scales equal to half the threshold
energy, $E_{\mathrm{thr}}=2 \Mt+m_{\Pb\bar\Pb}$.  For $\Pp\Pp\to
\Pt\bar\Pt\PH+X$, this choice was well supported by the existing NLO
analysis \cite{Beenakker:2001rj,Beenakker:2002nc}.  But, in the
absence of NLO predictions for $\Pt\bar\Pt\Pb\bar\Pb$, the choice of
the same scale for signal and background was motivated solely by the
assumption that the two processes have similar kinematics.  However,
in \citere{Bredenstein:2009aj} we found that, if both processes are
evaluated at $\mu_\mathrm{R}=\mu_\mathrm{F}=E_{\mathrm{thr}}/2$,
$\Pp\Pp\to\Pt\bar\Pt\Pb\bar\Pb$ receives much larger NLO corrections
($K\simeq 1.8$) than $\Pp\Pp\to\Pt\bar\Pt\PH$ ($K\simeq 1.2$).  This
is mainly due to the fact that the scale $E_{\mathrm{thr}}/2$ does not
provide an adequate description of the QCD dynamics that governs
$\Pt\bar\Pt\Pb\bar\Pb$ production.

The main difference between $\Pp\Pp\to
\Pt\bar\Pt\PH(\PH\to\Pb\bar\Pb)$ and its irreducible QCD background is
that the former process involves only two powers of
$\alpha_{\mathrm{s}}$ at LO.  Moreover, the part of the signal process
that is mediated by strong interactions does not involve any scale
significantly smaller than $E_{\mathrm{thr}}/2$.  In contrast,
$\Pp\Pp\to \Pt\bar\Pt\Pb\bar\Pb$ is entirely driven by QCD and is
proportional to $\alpha_{\mathrm{s}}^4$ at LO.  In the
$m_{\Pb\bar\Pb}\to 0$ limit, the dominant $\Pt\bar\Pt\Pb\bar\Pb$
production mechanism is $\Pp\Pp\to \Pt\bar\Pt g(g\to\Pb\bar\Pb)$,
where a gluon with small virtuality plays a role analogous to the
intermediate Higgs boson in the signal process.  In this regime, the
factorization of $\Pt\bar\Pt\Pg$ production and $\Pg\to\Pb\bar\Pb$
splitting provides two well-defined and widely separated scales:
$E_{\mathrm{thr}}/2$ and $m_{\Pb\bar\Pb}$, respectively.  This simple
picture is, however, not applicable to the kinematic region of
interest, where \mbox{$m_{\Pb\bar\Pb} \gsim 100\GeV$}.  Here,
$\Pp\Pp\to \Pt\bar\Pt\Pb\bar\Pb$ involves various other mechanisms.
For instance, the radiation of one or both b quarks off initial-state
gluons can play an important role due to collinear enhancements.  In
order to find an optimal QCD scale, we have tried to identify a
dominant production mechanism.  To this end, we have inspected the
relative weights of the channels corresponding to various
Feynman-diagram topologies in our adaptive Monte Carlo generator.
However, we found that none of these channels is strongly enhanced
with respect to the others.
Similarly we were not able to reproduce the bulk of the cross section
in terms of effective approximations based on collinear
$\Pg\to\Pb\bar\Pb$ splittings of the incoming gluons. 
This suggests that $\Pp\Pp\to
\Pt\bar\Pt\Pb\bar\Pb$ is a genuinely multi-scale and multi-channel
reaction.

We thus decided to adopt a pragmatic scale choice, 
based on the kinematic properties of the $\Pt\bar\Pt\Pb\bar\Pb$ final
state. While $\Mt$ sets a clear scale for the couplings to the top quarks, 
the inspection of differential distributions reveals that the cross
section is saturated by b quarks with $p_{\rT,\Pb}\ll \Mt$ (see
Figs.~\ref{fig:pTbhard_dist_1} and \ref{fig:pTbsoft_dist_1}).  In
order to account for these different scales we have adopted a
dynamical QCD scale corresponding to their geometric average,
\beq\label{centralscale}
\mu^2_0=\Mt\sqrt{p_{\rT,\Pb}p_{\rT,\bar\Pb}}.
\eeq
Our LO and NLO predictions are obtained by varying the renormalization
($\mu_{\mathrm{R}}$) and factorization ($\mu_{\mathrm{F}}$) scales around
the central value \refeq{centralscale},
\beq
\mu_{\mathrm{R}}=\xi_{\mathrm{R}}\mu_0,\qquad
\mu_{\mathrm{F}}=\xi_{\mathrm{F}}\mu_0.
\eeq
In the following sections we investigate the dependence of the LO and
NLO integrated cross section with respect to uniform
($\xi_{\mathrm{F}}=\xi_{\mathrm{R}}$) and antipodal
($\xi_{\mathrm{F}}=\xi_{\mathrm{R}}^{-1}$) scale variations in the
range $1/8\le \xi_{\mathrm{F}},\xi_{\mathrm{R}} \le 8$.  We find that
uniform variations have a larger impact on the cross section as
compared to antipodal variations.  For all distributions we provide LO
and NLO predictions with uncertainty bands corresponding to factor-two
uniform scale variations. More precisely, all observables are
evaluated at three different scales:
$\xi_{\mathrm{F}}=\xi_{\mathrm{R}}=0.5,1,2$.  As we will see from the
reduction of the $K$ factor and the scale uncertainties, the scale
choice \refeq{centralscale} clearly improves the convergence of the
perturbative expansion as compared to \citere{Bredenstein:2009aj}.

\subsection{Additional cuts}
Besides the standard cuts \refeq{standardcuts},
we have imposed the following kinematic restrictions to the
$\Pb\bar\Pb$ system and the extra jet that is radiated at NLO: 
\beq\label{extracuts}
m_{\Pb\bar\Pb}>m_{\Pb\bar\Pb,\mathrm{cut}},\qquad
p_{\rT,\Pb\bar\Pb}>p_{\rT,\Pb\bar\Pb,\mathrm{cut}},\qquad
p_{\rT,\mathrm{jet}}<p_{\mathrm{jet,veto}}.
\eeq

\btab
\bce
\begin{tabular}{c|ccc|cc|ccc}
Setup& $m_{\Pb\bar\Pb,\mathrm{cut}}$ &
$p_{\rT,\Pb\bar\Pb,\mathrm{cut}}$ &
$p_{\mathrm{jet,veto}}$&
$p_{\rT,\Pb,\mathrm{cut}}$ &
$y_{\Pb,\mathrm{cut}}$ &
$\sigma_\mathrm{LO}$&
$\sigma_\mathrm{NLO}$&
$K$
\\\hline
I &100&-&-&20&2.5& 
$786.3(2){\strut+78\%\atop -41\%}$ & 
$978(3){\strut+13\%\atop-21\%}$ &
1.24\\
II&-&200&-&20&2.5&
$451.8(2){\strut+79\% \atop-41\%}$ &
$592(4){\strut+13\% \atop-22\%}$ &
1.31 \\
III&100&-&100&20&2.5&
$786.1(6){\strut+78\% \atop-41\%}$ &
$700(3){\strut+0.4\%\atop-19\%}$ &
0.89\\
IV&100&-&-&50&2.5&
$419.4(1){\strut+77\% \atop-40\%}$ &
$526(2){\strut+13\% \atop-21\%}$ &
1.25
\end{tabular}
\caption{Cut parameters (in GeV), integrated LO and NLO cross section (in fb) 
with statistical errors and scale variations by factors 2 up and down 
as well as  $K$ factors for the four different setups.}
\label{tab:setups}
\ece
\etab

In order to investigate the individual effect of these extra cuts and
correlations with other observables, we have generated differential
distributions in four different setups described in
\refta{tab:setups}.  The setups I--III implement the standard cuts
\refeq{standardcuts} and explore the individual impact of the extra
cuts \refeq{extracuts}.  Setup IV is a variant of setup I, where the
cut \refeq{standardcuts} on the b-jet $p_\rT$ is increased from
$20\GeV$ to $50\GeV$.

\subsection{Setup I}
In this setup we impose the cut $m_{\Pb\bar\Pb}> 100\GeV$. This
removes a large part of the cross section
and selects the kinematic region of interest 
for the $\Pt\bar\Pt\PH(\PH\to\Pb\bar\Pb)$ signal.

\subsubsection*{Scale dependence}
\begin{figure}
\includegraphics[bb= 95 445 280 655, width=.45\textwidth]
{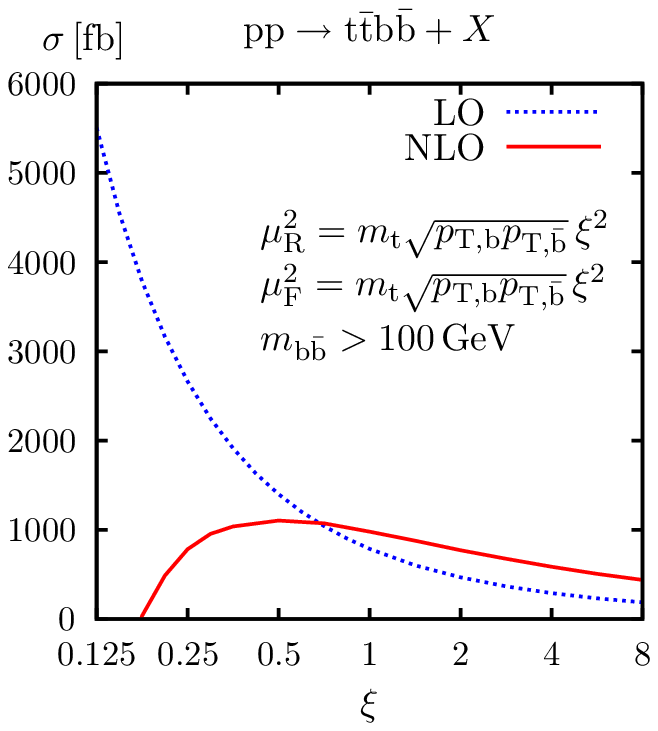}
\hfill
\includegraphics[bb= 95 445 280 655, width=.45\textwidth]
{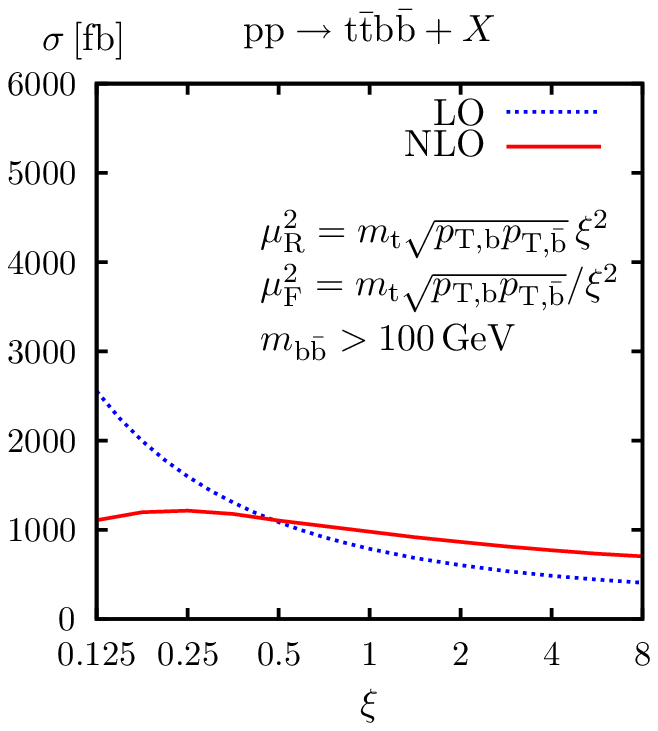}
\vspace*{-.8em}
\caption{Scale dependence of the LO and NLO 
$\Pp\Pp\to\Pt\bar\Pt\Pb\bar\Pb+X$ cross section at \mbox{$\sqrt{s}=14\TeV$}
in setup~I.
The left and the right plots describe  
uniform ($\xi_{\mathrm{R}}=\xi_{\mathrm{F}}=\xi$)
and antipodal ($\xi_{\mathrm{R}}=\xi_{\mathrm{F}}^{-1}=\xi$)
scale variations, respectively. 
}
\label{fig:sigma_tot_1}
\end{figure}

The LO and NLO integrated cross sections and their dependence with
respect to uniform (left plot) and antipodal (right plot) scale
variations are displayed in \reffi{fig:sigma_tot_1}.  At the central
scale we obtain $\sigma_{\mathrm{LO}}=786.3(2)\fb$ and
$\sigma_{\mathrm{NLO}}=978(3)\fb$, where the numbers in parentheses
are the errors of the Monte Carlo integration for $2\times10^8$ LO
events and $2\times10^7$ NLO events before applying cuts.  These
predictions are not directly comparable to those of
\citere{Bredenstein:2009aj}, where we did not apply any cut to
$m_{\Pb\bar\Pb}$.  Still we can compare the $K$ factors, which are
rather insensitive to $m_{\Pb\bar\Pb}$. We observe that the new scale
choice \refeq{centralscale} reduces the NLO corrections from $K\simeq
1.77$ \cite{Bredenstein:2009aj} to $K\simeq 1.24$.  
We note that, in
spite of the smaller $K$ factor, the new scale choice yields larger LO
and NLO cross sections as compared to the scale $E_{\mathrm{thr}}/2$
used in \citere{Bredenstein:2009aj}.  This can be easily seen from
\reffi{fig:sigma_tot_1}, where the new and the previous central scales
correspond to $\mu_\mathrm{R}/\mu_0 = 1$ and
$\mu_\mathrm{R}/\mu_0=E_{\mathrm{thr}}/(2\mu_0)\gg 1$,
respectively.\footnote{ Using setup-I cuts and the ATLAS scale choice
  $\mu_{\mathrm{R,F}}=E_{\mathrm{thr}}/2$ we obtain
  $\sigma_{\mathrm{LO}}(E_{\mathrm{thr}}/2)=448.7(1)\fb$ and
  $\sigma_{\mathrm{NLO}}(E_{\mathrm{thr}}/2)=751(2)\fb$.  The increase
  of the cross section due to the combined effect of the new scale
  choice and the NLO correction factor is thus
  $\sigma_{\mathrm{NLO}}(\mu_0)/\sigma_{\mathrm{LO}}(E_{\mathrm{thr}}/2)\simeq
  2.18$ while the $K$ factor for setup-I cuts and the ATLAS scale
  choice results in $K\simeq 1.67$.}  In addition to the $K$ factor,
the new scale choice reduces also the NLO scale uncertainty.  Varying
the scale up or down by a factor 2 changes the LO and NLO cross
section by 78\% in LO and by 21\% in NLO.  The improvement with
respect to \citere{Bredenstein:2009aj}, where we had a 33\% NLO
uncertainty, is evident also from the shape of the NLO curves in
\reffi{fig:sigma_tot_1}.  Now we observe a stable point in the
vicinity of the central scale.

\subsubsection*{Jet veto}
\begin{figure}
\includegraphics[bb= 95 445 280 655, width=.45\textwidth]
{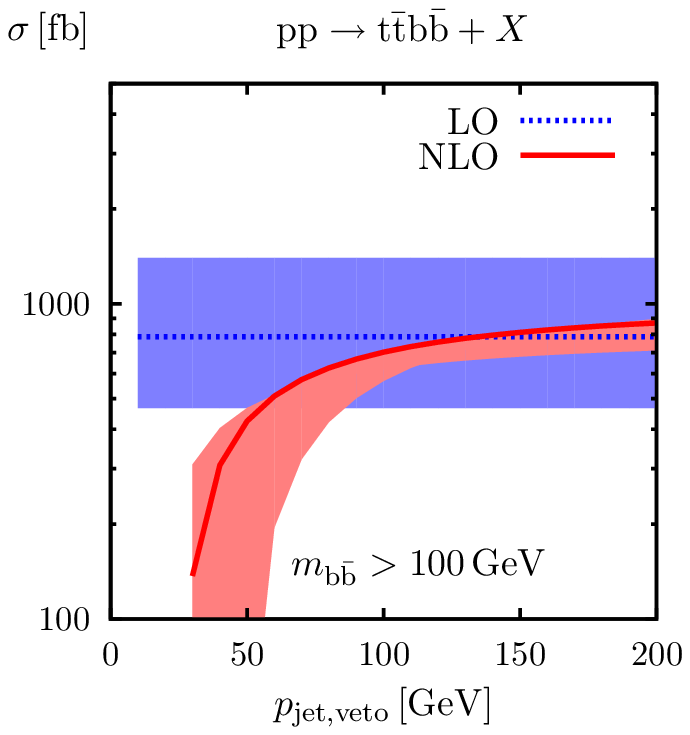}
\hfill
\includegraphics[bb= 95 445 280 655, width=.45\textwidth]
{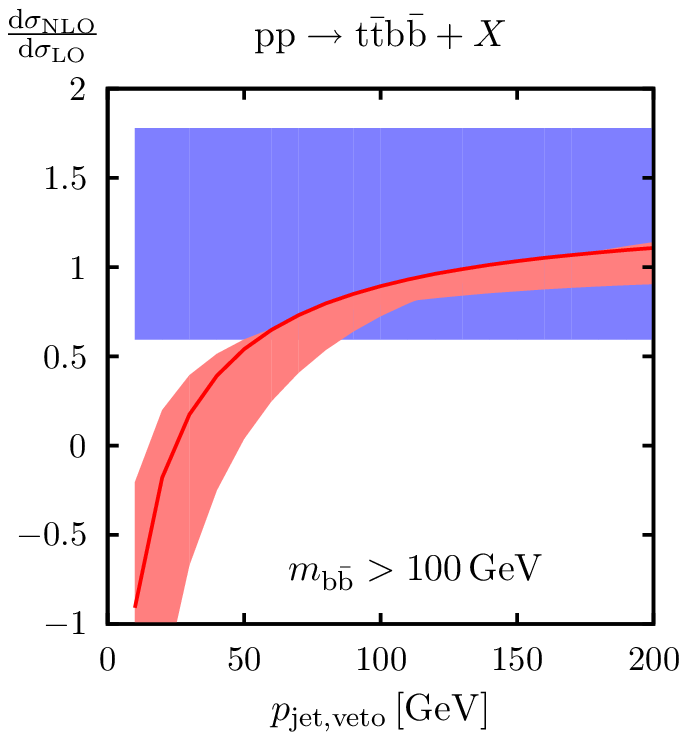}
\vspace*{-.8em}
\caption{Dependence of the
$\Pp\Pp\to\Pt\bar\Pt\Pb\bar\Pb+X$ cross section 
on a jet veto  
($p_{\mathrm{T,jet}}<p_{\mathrm{jet,veto}}$) in setup~I:
absolute
LO and NLO predictions (left) and NLO  $K$ factor (right).
The uncertainty bands correspond to factor-two scale variations.
}
\label{fig:jet_veto_1}
\end{figure}

As anticipated in \citere{Bredenstein:2009aj}, a jet veto can
significantly reduce the large cross section of the
$\Pt\bar\Pt\Pb\bar\Pb$ background.  This could facilitate the
extraction of the $\Pt\bar\Pt\PH$ signal at the LHC.  In
\reffi{fig:jet_veto_1}, the integrated $\Pt\bar\Pt\Pb\bar\Pb$ cross
section is plotted versus the upper bound, $p_{\mathrm{jet,veto}}$,
imposed to the jet transverse momentum.  Here, as well as in the
following figures, the left plot shows the absolute LO (blue) and NLO
(red) predictions. The curves and their uncertainty bands represent
factor-two (uniform) scale variations around the central value
\refeq{centralscale}.  The right plot displays the LO and NLO bands
normalized to the central value of the LO prediction.  There the blue
band illustrates the relative uncertainty of the
LO cross section, \ie $\si_{\LO}(\xi)/\si_{\LO}(\xi=1)$,
the red curve corresponds to the $K$ factor,
$K=\si_{\NLO}(\xi=1)/\si_{\LO}(\xi=1)$, and the red band shows the
variation of the $K$ factor when varying the scales in the NLO cross
section but keeping them fixed in the LO cross section, $K_\xi=\si_{\NLO}(\xi)/\si_{\LO}(\xi=1)$.
In \reffi{fig:jet_veto_1} the red (NLO) curve tends to saturate the
upper bound of its uncertainty band, a feature that can be observed in
various other distributions.  This is consistent with the shape of the
NLO curve in \reffi{fig:sigma_tot_1},
which develops a maximum in the vicinity of the central scale.

The NLO curve in \reffi{fig:jet_veto_1} shows that a sizable reduction
of the cross section requires a jet veto well below $200\GeV$. For
$p_{\mathrm{jet,veto}}=150,100$, and $50\GeV$, the $K$ factor is
reduced to $1.03$, $0.89$, and $0.54$, respectively.  However, there
is a trade-off between suppressing the NLO cross section and
increasing its perturbative uncertainty.  The jet veto tends to
destroy the cancellation between IR logarithms of virtual and real
origin and its effect grows as
$-\alpha_{\mathrm{s}}^5\ln^2(E_{\mathrm{thr}}/p_{\mathrm{jet,veto}})$
when $p_{\mathrm{jet,veto}}$ becomes small.
Since they are accompanied by an $\alpha_{\mathrm{s}}^5$ coefficient,
these logarithms can give rise to huge scale uncertainties already for
moderate values of $p_{\mathrm{jet,veto}}$.

This is reflected by the dramatic amplification of the NLO uncertainty
band in \reffi{fig:jet_veto_1}.  Its lower bound enters the pathologic
regime of negative cross sections around
$p_{\mathrm{jet,veto}}=50\GeV$.  Here, besides the NLO cross section
itself, also its uncertainty estimate becomes completely unreliable.
The region of small $p_{\mathrm{jet,veto}}$ would require a
resummation. This would stabilize the perturbative calculation and
compensate for its divergent behaviour. As a result, the unphysical
suppression of the NLO cross section for $p_{\mathrm{jet,veto}}\ll
100\GeV$ would be washed out.  If we restrict ourselves to the
fixed-order NLO result, the plot tells us that jet-veto values around
$100\GeV$ provide a good compromise: the reduction of the $K$ factor
is already significant ($K\simeq 0.89$) and the NLO scale uncertainty
(19\%) is at the same level as for the total cross section (21\%).

\subsubsection*{Invariant-mass distributions}
\begin{figure}
\includegraphics[bb= 95 445 280 655, width=.45\textwidth]
{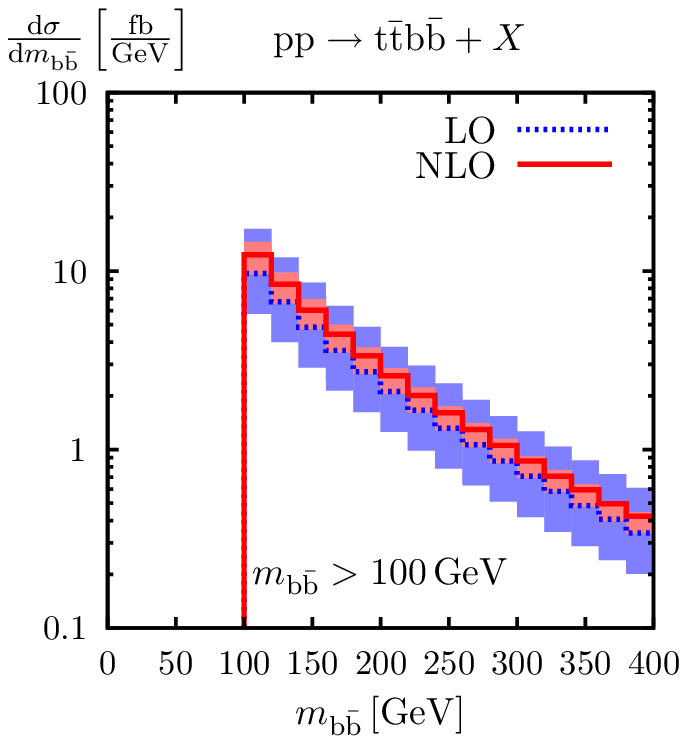}
\hfill
\includegraphics[bb= 95 445 280 655, width=.45\textwidth]
{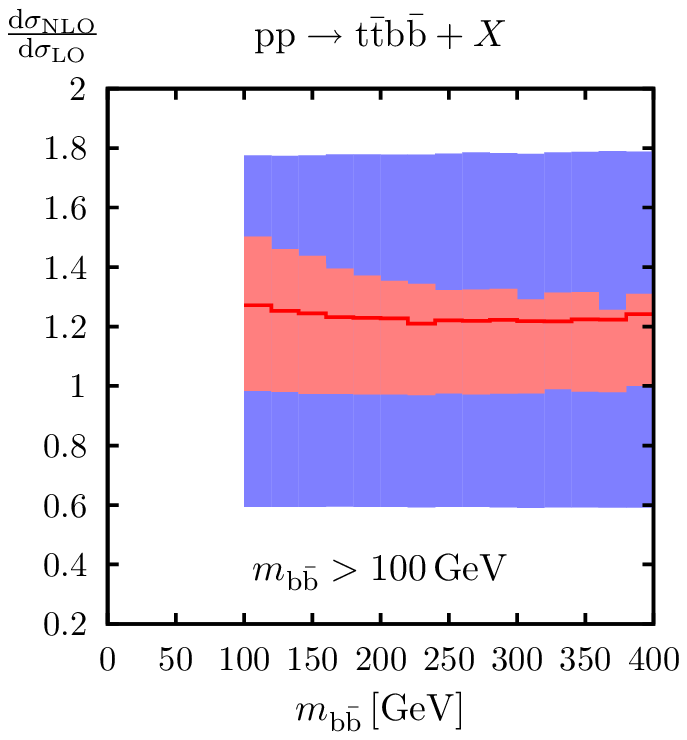}
\vspace*{-.8em}
\caption{Invariant-mass distribution of the
  $\Pb\bar\Pb$ pair in setup~I: absolute LO and NLO predictions (left) and NLO
  $K$ factor (right).  The uncertainty bands correspond to factor-two
  scale variations.  }
\label{fig:mbb_dist_1}
\vspace*{4em}
\includegraphics[bb= 95 445 280 655, width=.45\textwidth]
{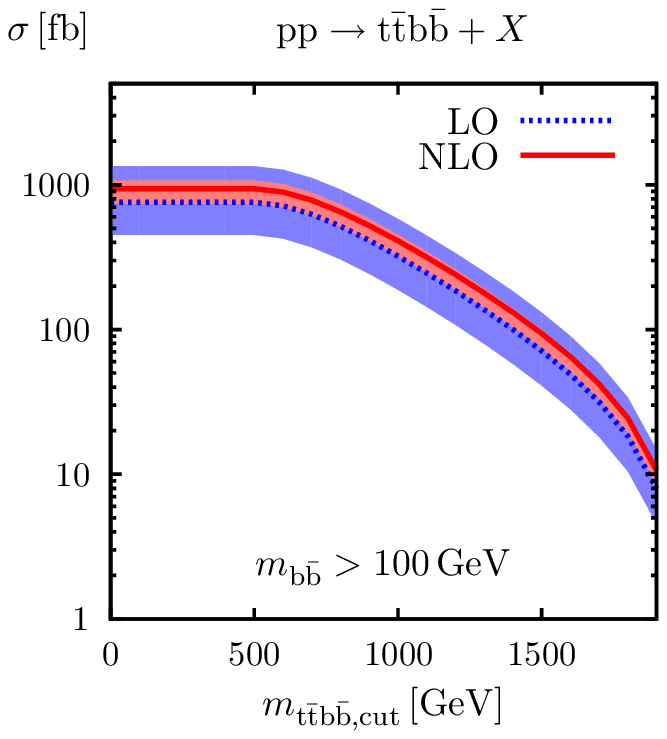}
\hfill
\includegraphics[bb= 95 445 280 655, width=.45\textwidth]
{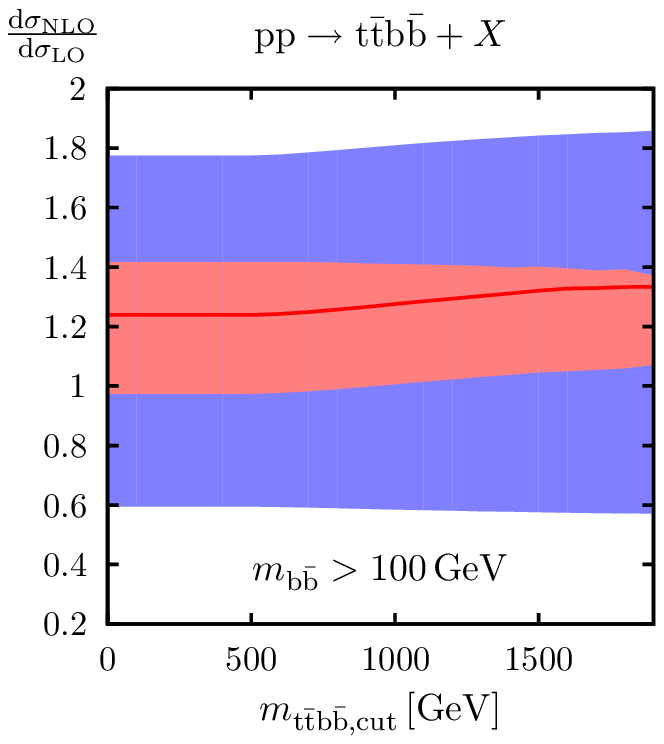}
\vspace*{-.8em}
\caption{Dependence of the
cross section with respect to a cut on the 
$\Pt\bar\Pt\Pb\bar\Pb$ invariant mass
  ($m_{\Pt\bar\Pt\Pb\bar\Pb}>m_{\Pt\bar\Pt\Pb\bar\Pb,\mathrm{cut}}$) in setup~I:
absolute LO and NLO predictions (left) and NLO  $K$ factor (right).
The uncertainty bands correspond to factor-two scale variations.
}
\label{fig:mttbb_cut_1}
\end{figure}

The invariant-mass distribution of the $\Pb\bar\Pb$ pair, shown in
\reffi{fig:mbb_dist_1}, constitutes a key observable for the search of
the $\Pt\bar\Pt\PH$ signal.  Because of limited resolution and b-quark
combinatorial problems, the Higgs boson would appear as a relatively
broad and small peak on top of this distribution.  The subtraction of
the $\Pt\bar\Pt\Pb\bar\Pb$ background requires an accurate
determination of its normalization, possibly by direct measurement in
a signal-free region, and a precise theoretical description of its
shape.

In \reffi{fig:mbb_dist_1} we observe that the NLO predictions
perfectly fit within the LO band and significantly reduce the QCD
uncertainty over the entire invariant-mass range.  The numerical
impact of the corrections is moderate and almost constant ($1.21<K<1.27$).  
This favourable behaviour is ensured by the dynamical
scale choice \refeq{centralscale}.  At high invariant masses the upper
bound of the NLO uncertainty band slightly decreases and approaches
the central NLO prediction.  The same trend appears in the high-energy
tail of other distributions.

\reffi{fig:mttbb_cut_1} displays the dependence of the cross section 
with respect to a cut on the total invariant mass of the 
$\Pt\bar\Pt\Pb\bar\Pb$ system
($m_{\Pt\bar\Pt\Pb\bar\Pb}>m_{\mathrm{cut},\Pt\bar\Pt\Pb\bar\Pb}$).
Since it corresponds to the 
invariant mass of the $\PWp\PWm\Pb\bar\Pb\Pb\bar\Pb$ final state,
this quantity is independent of the b-jet combinatorics.
It can thus be measured with better resolution as compared 
to observables that involve only a particular subset 
of the four b jets.
This property might be exploited in order to improve 
the signal-to-background ratio.
Apart from a slight increase in the high invariant-mass tail,
the NLO corrections behave similarly as for
the $m_{\Pb\bar\Pb}$~distribution. 

\subsubsection*{Transverse-momentum distributions}
\begin{figure}
\includegraphics[bb= 95 445 280 655, width=.45\textwidth]
{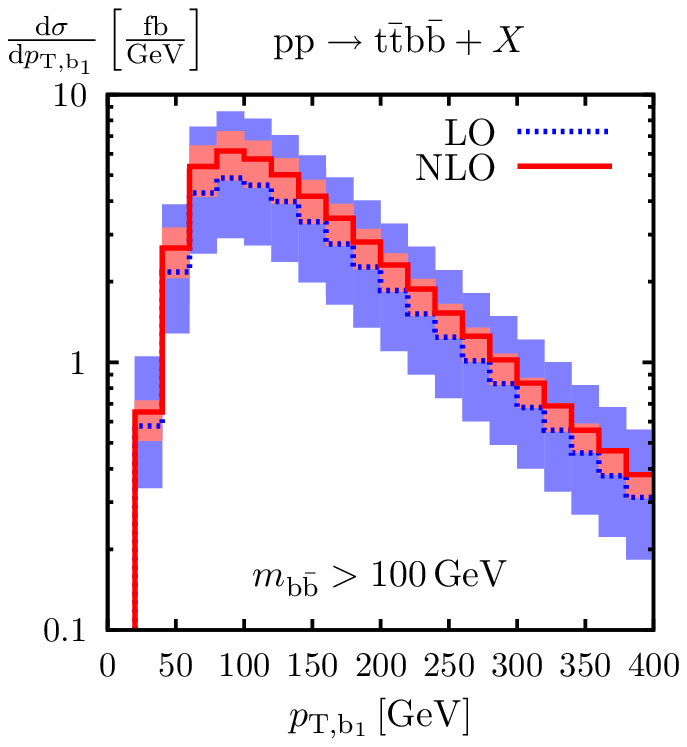}
\hfill
\includegraphics[bb= 95 445 280 655, width=.45\textwidth]
{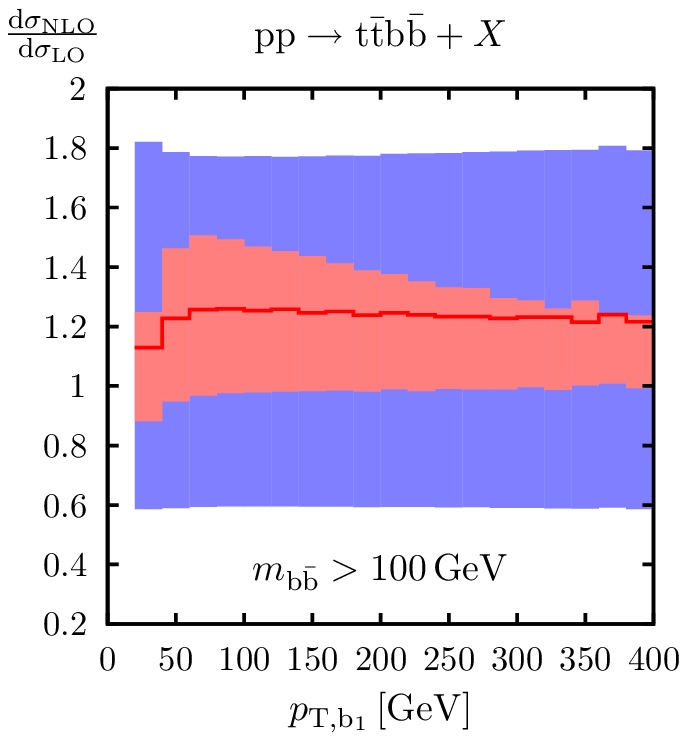}
\vspace*{-.8em}
\caption{Transverse-momentum distribution of the
harder b jet in setup~I: absolute
LO and NLO predictions (left) and NLO  $K$ factor (right).
The uncertainty bands correspond to factor-two scale variations.
}
\label{fig:pTbhard_dist_1}
\vspace*{4em}
\includegraphics[bb= 95 445 280 655, width=.45\textwidth]
{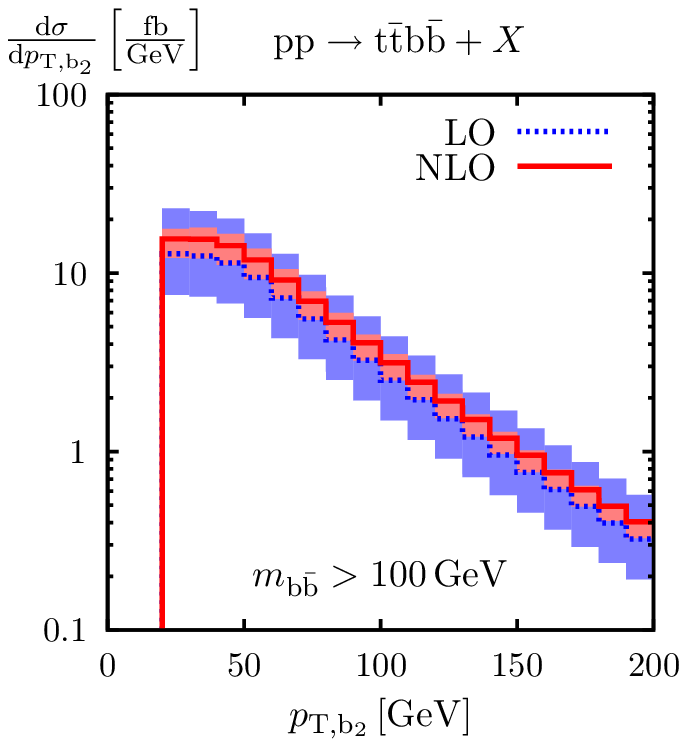}
\hfill
\includegraphics[bb= 95 445 280 655, width=.45\textwidth]
{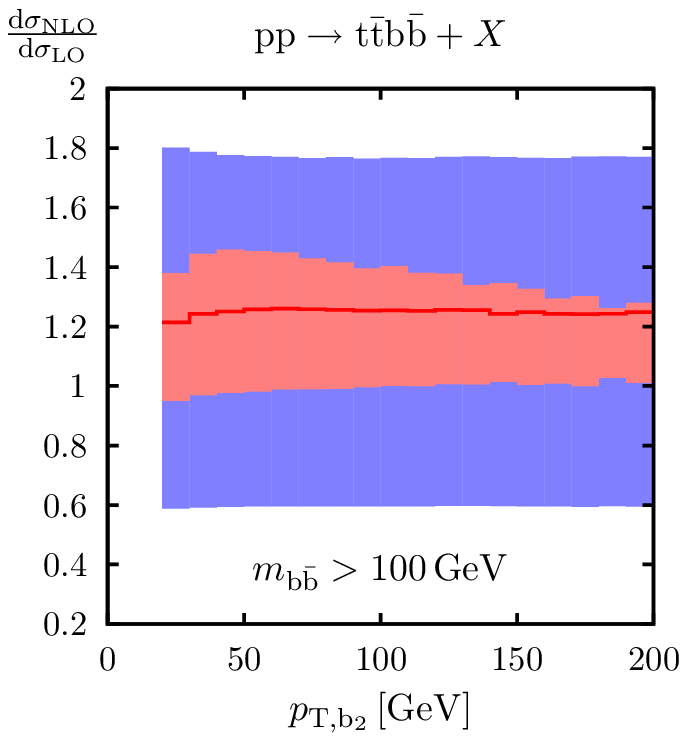}
\vspace*{-.8em}
\caption{Transverse-momentum distribution of the
softer b jet in setup~I: absolute
LO and NLO predictions (left) and NLO  $K$ factor (right).
The uncertainty bands correspond to factor-two scale variations.
}
\label{fig:pTbsoft_dist_1}
\end{figure}

The transverse-momentum distributions of the individual b jets,
ordered according to their $p_\rT$, are presented in
\reffi{fig:pTbhard_dist_1} (harder b jet) and
\reffi{fig:pTbsoft_dist_1} (softer b jet).  While the softer b jet
tends to saturate the lower bound of $20\GeV$ \refeq{standardcuts},
the harder behaves rather
differently.  Its distribution has a maximum around $100\GeV$ and a
tail that extends up to fairly high transverse momenta.  These shapes
suggest that one of the two quarks is often emitted from initial-state
gluons, while the other one participates to the hard scattering.  In
contrast, none of the b quarks resulting from $\Pt\bar\Pt\PH$
originates from initial-state radiation.  This feature, which renders
the cross section quite sensitive to $p_{\rT,\Pb}$, might be exploited
to improve the separation of the $\Pt\bar\Pt\PH$ signal.

\begin{figure}
\includegraphics[bb= 95 445 280 655, width=.45\textwidth]
{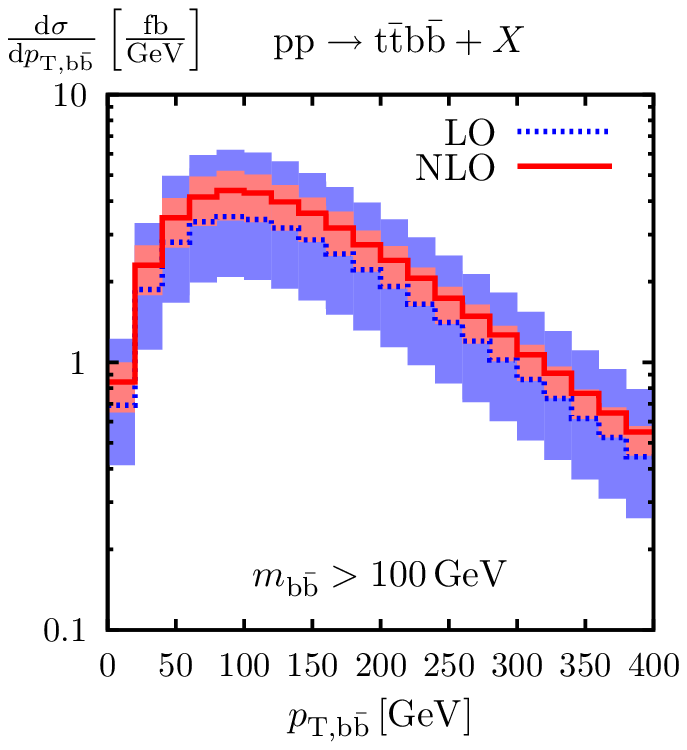}
\hfill
\includegraphics[bb= 95 445 280 655, width=.45\textwidth]
{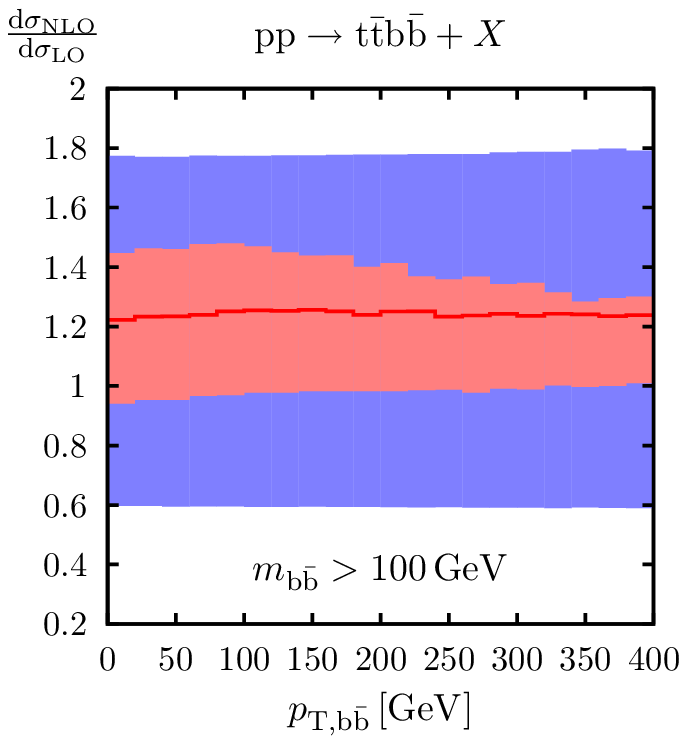}
\vspace*{-.8em}
\caption{Transverse-momentum distribution of the
$\Pb\bar\Pb$ system in setup~I: absolute
LO and NLO predictions (left) and NLO  $K$ factor (right).
The uncertainty bands correspond to factor-two scale variations.
}
\label{fig:pTbb_dist_1}
\end{figure}

The dynamical scale introduced in \refeq{centralscale} accounts for
the different kinematics of the two b jets and the extension of their
transverse momenta over a wide $p_\rT$~range.  The goodness of this
choice is confirmed by the stability of the $K$ factor over the entire
$p_\rT$-spectrum.
The transverse-momentum distribution of the $\Pb\bar\Pb$ pair is shown
in \reffi{fig:pTbb_dist_1}. Its shape resembles fairly closely that of
the harder b-jet distribution.  Also the $K$ factor and the scale
uncertainties behave similarly.

\subsubsection*{Rapidity and azimuthal distributions}
\begin{figure}
\includegraphics[bb= 95 445 280 655, width=.45\textwidth]
{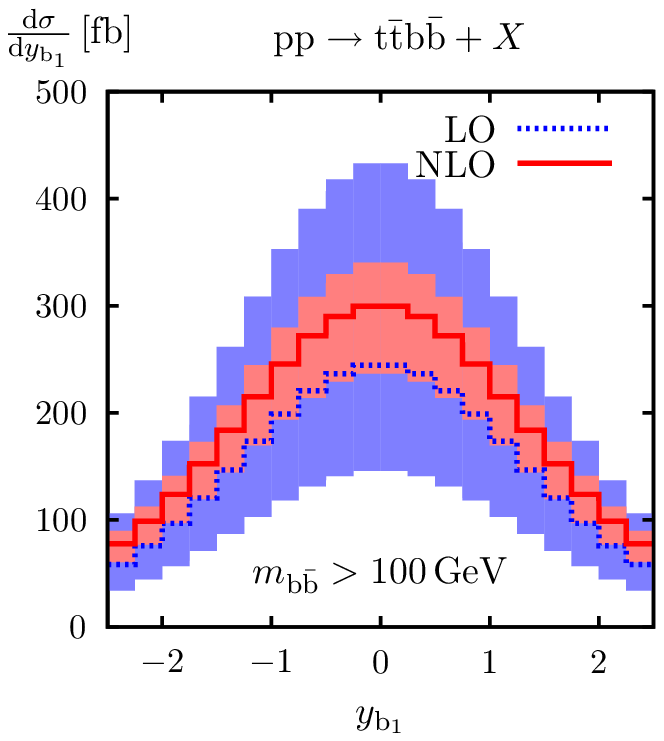}
\hfill
\includegraphics[bb= 95 445 280 655, width=.45\textwidth]
{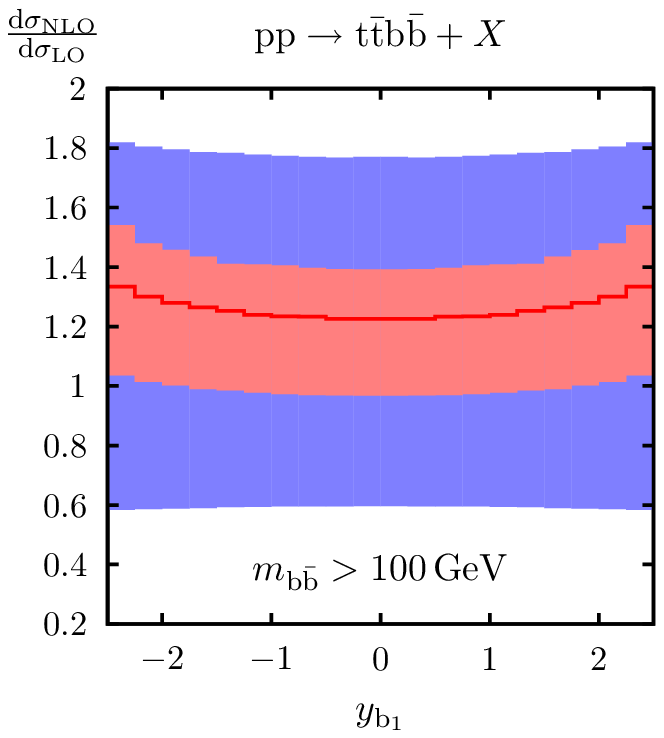}
\vspace*{-.8em}
\caption{Rapidity distribution of the
harder b jet in setup~I: absolute
LO and NLO predictions (left) and NLO  $K$ factor (right).
The uncertainty bands correspond to factor-two scale variations.
}
\label{fig:ybhard_dist_1}
\vspace*{4em}
\includegraphics[bb= 95 445 280 655, width=.45\textwidth]
{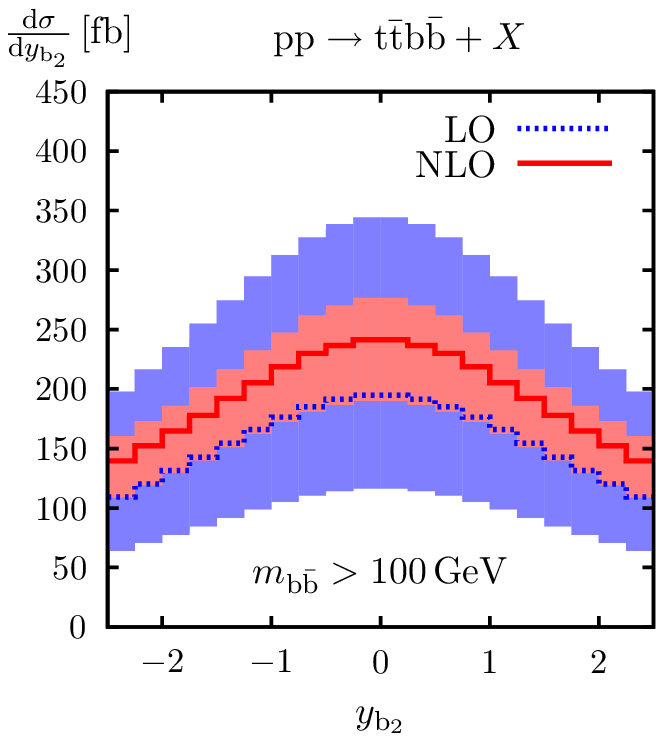}
\hfill
\includegraphics[bb= 95 445 280 655, width=.45\textwidth]
{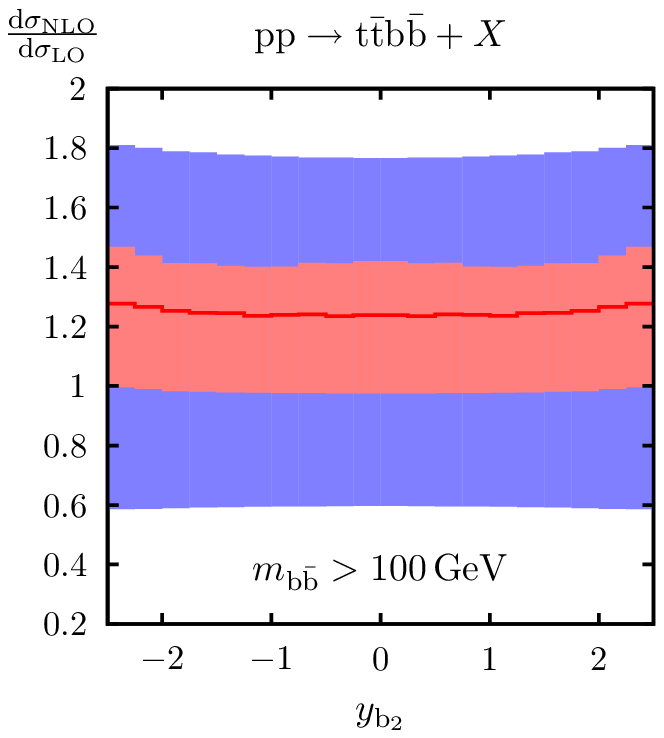}
\vspace*{-.8em}
\caption{Rapidity distribution of the
softer b jet in setup~I: absolute
LO and NLO predictions (left) and NLO  $K$ factor (right).
The uncertainty bands correspond to factor-two scale variations.
}
\label{fig:ybsoft_dist_1}
\end{figure}

The rapidities of the individual b jets, ordered in $p_\rT$, are shown
in \reffi{fig:ybhard_dist_1} \mbox{(harder b jet)} and
\reffi{fig:ybsoft_dist_1} (softer b jet).  Both b jets tend to
populate the central region.  But this feature is much more pronounced
for the harder b jet, while the softer one has a significant
probability to be emitted also in the forward and backward directions.
\begin{figure}
\includegraphics[bb= 95 445 280 655, width=.45\textwidth]
{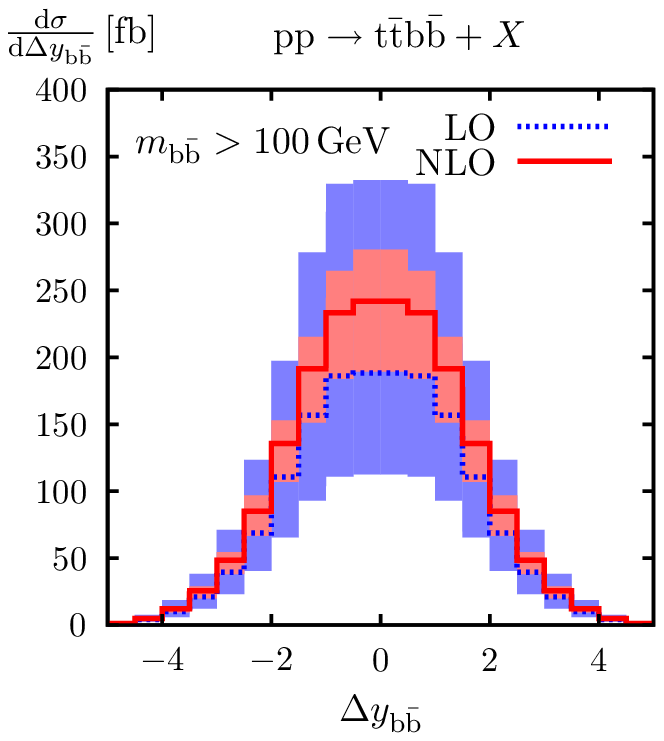}
\hfill
\includegraphics[bb= 95 445 280 655, width=.45\textwidth]
{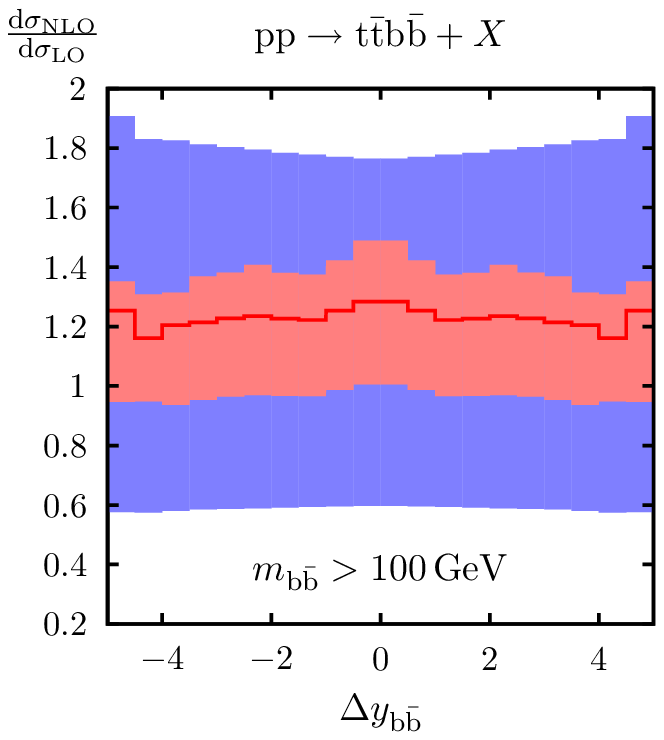}
\vspace*{-.8em}
\caption{Rapidity separation of the 
two b jets in setup~I: absolute
LO and NLO predictions (left) and NLO  $K$ factor (right).
The uncertainty bands correspond to factor-two scale variations.
}
\label{fig:dybb_dist_1}
\vspace*{4em}
\includegraphics[bb= 95 445 280 655, width=.45\textwidth]
{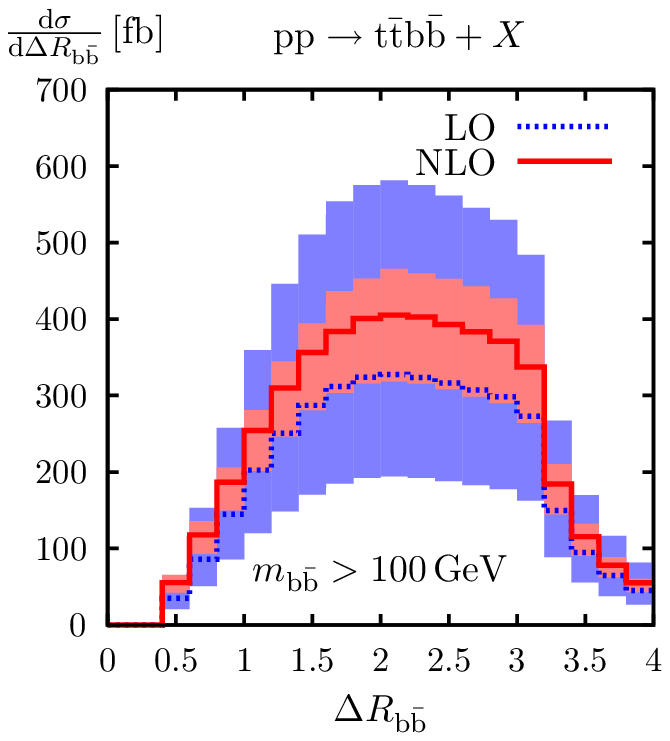}
\hfill
\includegraphics[bb= 95 445 280 655, width=.45\textwidth]
{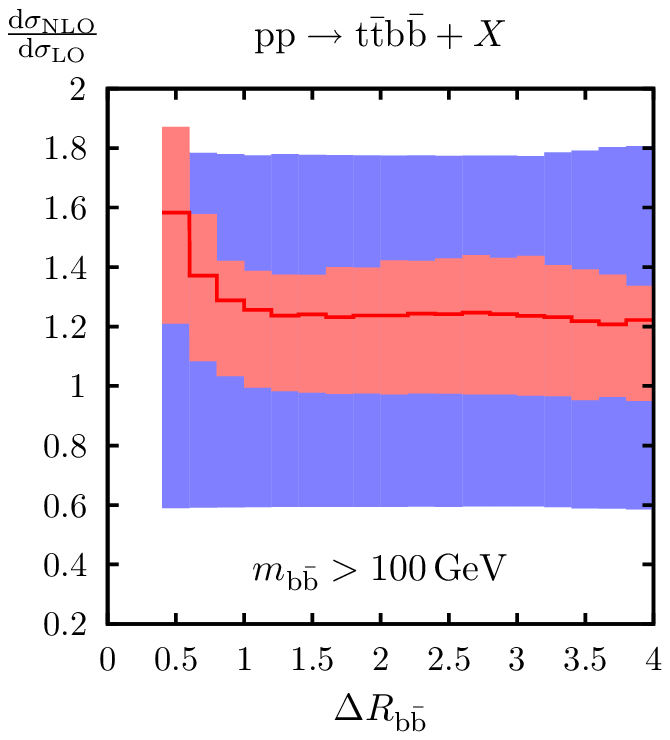}
\vspace*{-.8em}
\caption{Rapidity--azimuthal-angle separation of the 
two b jets in setup~I: absolute
LO and NLO predictions (left) and NLO  $K$ factor (right).
The uncertainty bands correspond to factor-two scale variations.
}
\label{fig:dRbb_dist_1}
\end{figure}
The rapidity distribution of the $\Pb\bar\Pb$~system (not plotted)
resembles that of the harder b jet, and the rapidity-separation
distribution (\reffi{fig:dybb_dist_1}) does not suggest any strong
correlation between the two b jets.  All rapidity distributions
receive moderate and almost constant NLO corrections.

The rapidity--azimuthal-angle separation, $\Delta R_{\Pb\bar\Pb}=
\sqrt{(y_{\Pb}-y_{\bar\Pb})^2+(\phi_{\Pb}-\phi_{\bar\Pb})^2}$, of the
b jets is displayed in \reffi{fig:dRbb_dist_1}.  The shape of this
distribution is determined by three kinematic constraints: the
rapidity cut \refeq{standardcuts} is responsible for the suppression
at high $\Delta R_{\Pb\bar\Pb}$; the sharp lower bound at $\Delta
R_{\Pb\bar\Pb}=0.4$ results from the jet algorithm; and the
invariant-mass cut $m_{\Pb\bar\Pb}>100\GeV$ keeps the two b jets at
intermediate $\Delta R$-separations.  The NLO corrections induce a
30--40\% distortion of the shape of this distribution in the region
$0.4< \Delta R_{\Pb\bar\Pb} \lsim 1$.  This effect can be attributed
to the recombination of b quarks and \mbox{non-b partons}, which can
turn b-quark pairs with $\Delta R <0.4$ into b-jet pairs with $\Delta
R >0.4$.

\begin{figure}
\includegraphics[bb= 95 445 280 655, width=.45\textwidth]
{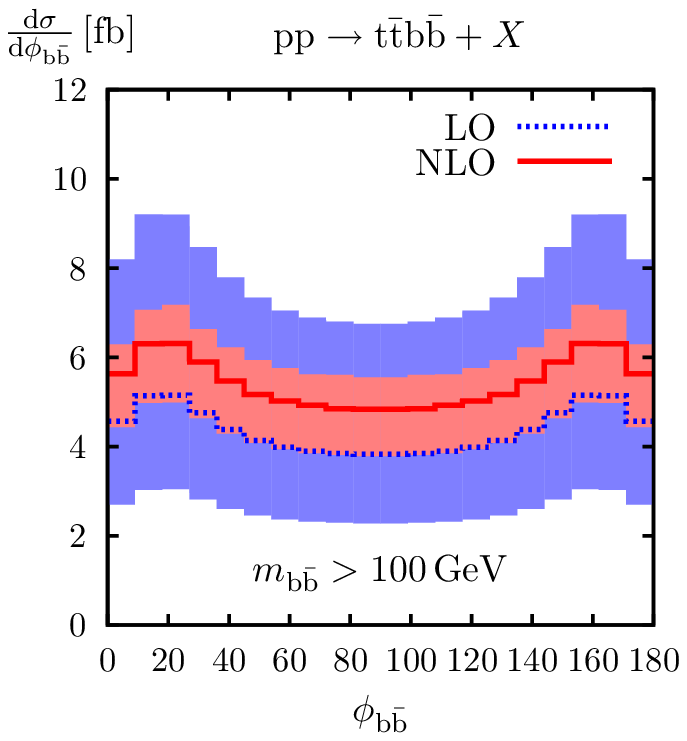}
\hfill
\includegraphics[bb= 95 445 280 655, width=.45\textwidth]
{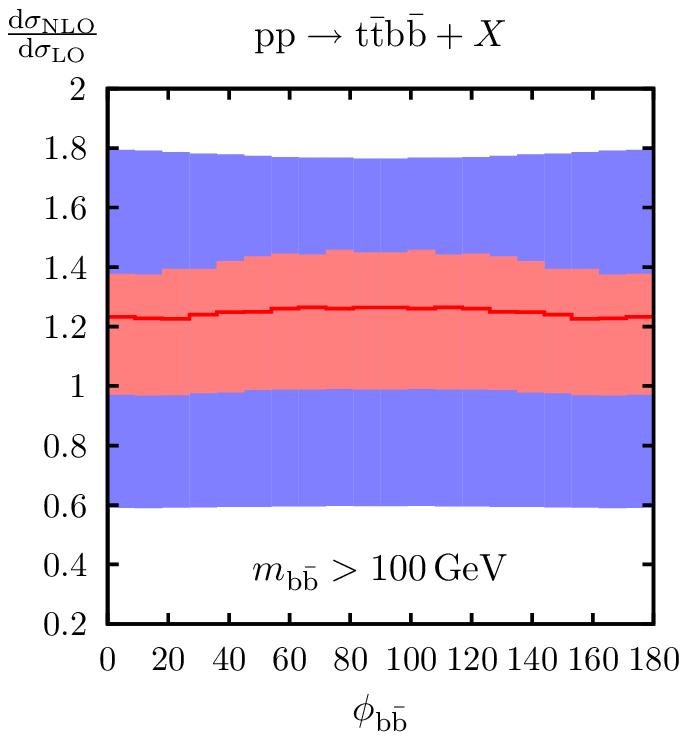}
\vspace*{-.8em}
\caption{Azimuthal orientation of the 
  b jets in the plane perpendicular to the $\Pb\bar\Pb$ system (see
  text) in setup~I: absolute LO and NLO predictions (left) and NLO $K$
  factor (right).  The uncertainty bands correspond to factor-two
  scale variations.  }
\label{fig:phibb_dist_1}
\end{figure}

Finally, in \reffi{fig:phibb_dist_1}, we plot the angular variable
$\phi_{\Pb\bar\Pb}$, which describes the azimuthal orientation of the
$\Pb$ jets.  This observable is defined as the opening angle between
two planes: the ``production'' plane spanned by the beam axis and the
total momentum of the $\Pb\bar\Pb$ system, and the ``decay'' plane,
which contains the momenta of the individual b jets,
defined by
\beqar
\cos{\phi_{\Pb\bar\Pb}} &=& 
\frac{[{\bf p}_{\mathrm{beam}}\times({\bf p}_{\Pb}+{\bf p}_{\bar\Pb})]
\cdot({\bf p}_{\Pb}\times{\bf p}_{\bar\Pb})}
{|{\bf p}_{\mathrm{beam}}\times({\bf p}_{\Pb}+{\bf p}_{\bar\Pb})|
|{\bf p}_{\Pb}\times{\bf p}_{\bar\Pb}|}.
\label{eq:phipr}
\eeqar
Equivalently $\phi_{\Pb\bar\Pb}$ represents the azimuthal orientation
of the b jets with respect to the beam direction in the plane
perpendicular to the $\Pb\bar\Pb$ momentum.  In the case where the
$\Pb\bar\Pb$ pair results from the decay of an intermediate particle,
like the Higgs boson in $\Pt\bar\Pt\PH$ production, the spin of the
latter can be determined from the $\phi_{\Pb\bar\Pb}$~distribution.  
Since the Higgs boson has spin~0, the $\phi_{\Pb\bar\Pb}$~distribution
is expected to be isotropic, \ie $\phi_{\Pb\bar\Pb}$-independent.
As we see from \reffi{fig:phibb_dist_1}, the NLO
corrections have a negligible influence on the shape of this
observable for the $\Pt\bar\Pt\Pb\bar\Pb$ background.

\subsection{Setup II}
As discussed in the introduction, the selection of $\Pt\bar\Pt\Pb\bar\Pb$
signatures with highly boosted b-quark pairs may help
to separate the $\Pt\bar\Pt\PH(\PH\to\Pb\bar\Pb)$ signal from
its backgrounds. This motivates us to study the irreducible 
$\Pt\bar\Pt\Pb\bar\Pb$ background in this particular phase-space region.
Specifically, in setup~II, we select highly boosted $\Pb\bar\Pb$ pairs 
with $p_{\rT,\Pb\bar\Pb}> 200\GeV$, as proposed in \citere{Plehn:2009rk}.
In contrast to setup~I, here we do not impose any cut 
on the ${\Pb\bar\Pb}$ invariant mass. Nevertheless the cuts on
$p_{\rT,\Pb\bar\Pb}$, $p_{\rT,\Pb}$, and $p_{\rT,\bar\Pb}$, together
with the bound $\Delta R_{\Pb\bar\Pb}>0.4$ resulting from the jet algorithm,
impose an effective lower bound
\beq\label{mbbbound}
m_{\Pb\bar\Pb}
\simeq
\Delta R_{\Pb\bar\Pb}\,
p_{\rT,\Pb\bar\Pb}\,
\sqrt{z(1-z)}
>
\Delta R_{\Pb\bar\Pb}\,
p_{\rT,\Pb,\mathrm{cut}}\,
\sqrt{\frac{p_{\rT,\Pb\bar\Pb,\mathrm{cut}}}{p_{\rT,\Pb,\mathrm{cut}}}-1}
= 24 \GeV,
\eeq
where $z$ and $1-z$ are the (transverse) momentum fractions of the two
b jets, and the first equation holds for small $\Delta
R_{\Pb\bar\Pb}$.

\begin{figure}
\includegraphics[bb= 95 445 280 655, width=.45\textwidth]
{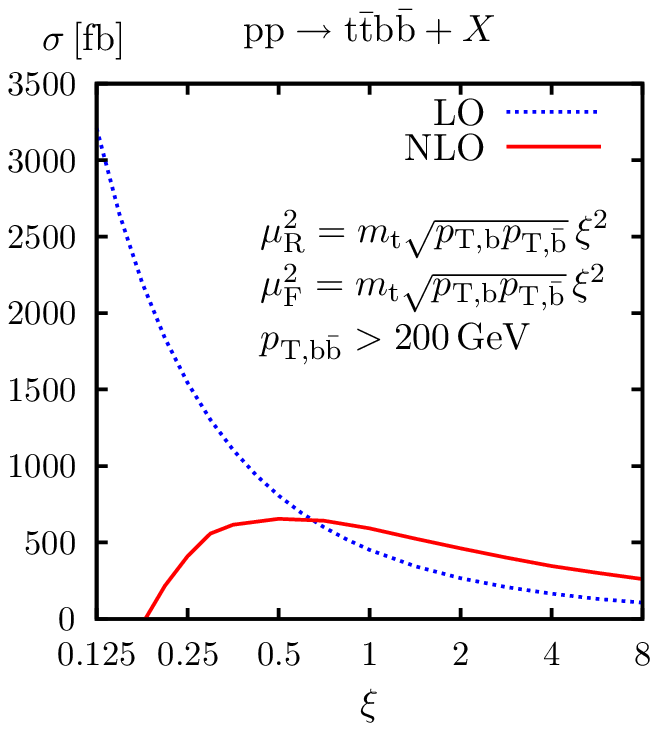}
\hfill
\includegraphics[bb= 95 445 280 655, width=.45\textwidth]
{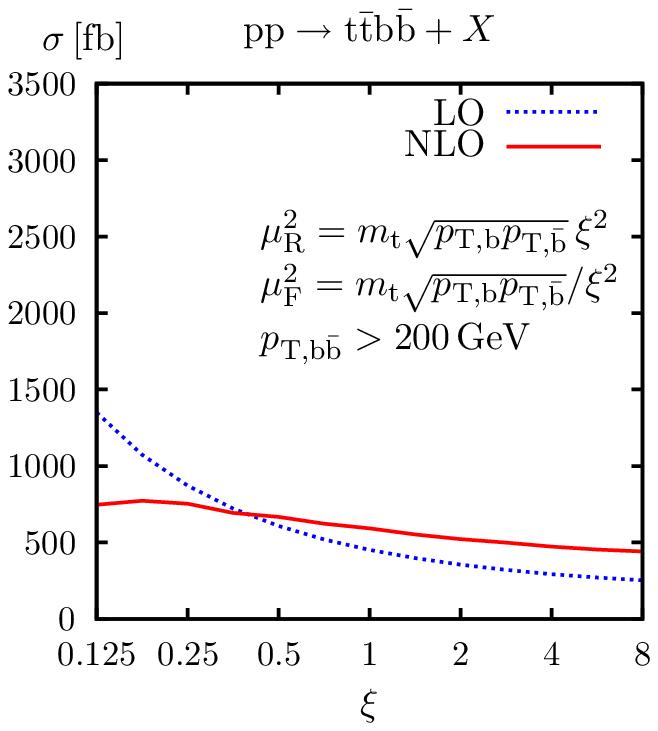}
\vspace*{-.8em}
\caption{Scale dependence of the LO and NLO 
  $\Pp\Pp\to\Pt\bar\Pt\Pb\bar\Pb+X$ cross section at
  \mbox{$\sqrt{s}=14\TeV$} in setup~II.  The left and the
  right plots describe uniform
  ($\xi_{\mathrm{R}}=\xi_{\mathrm{F}}=\xi$) and antipodal
  ($\xi_{\mathrm{R}}=\xi_{\mathrm{F}}^{-1}=\xi$) scale variations,
  respectively.  }
\label{fig:sigma_tot_2}
\end{figure}

\subsubsection*{Scale dependence}
The scale dependence of the LO and NLO integrated cross sections is
shown in \reffi{fig:sigma_tot_2}.  At the central scale we obtain
$\sigma_{\mathrm{LO}}=451.8(2)\fb$ and
$\sigma_{\mathrm{NLO}}=592(4)\fb$. This corresponds to an NLO
correction factor $K=1.31$.  The absolute NLO cross section is reduced
by about $40\%$ as compared to setup I.  The shape of the
scale-dependence curves is quite similar as in \reffi{fig:sigma_tot_1}
and indicates good convergence and stability of the perturbative
expansion.  The shifts induced by factor-two variations of the QCD
scales amount to 79\% in LO and 22\% in NLO.

Investigating the sensitivity of the NLO cross section to a jet veto
we found similar results as in setup I.  For a jet veto of 100 GeV the
$K$ factor and the NLO uncertainty amount to 0.84 and 33\%,
respectively.

\begin{figure}
\includegraphics[bb= 95 445 280 655, width=.45\textwidth]
{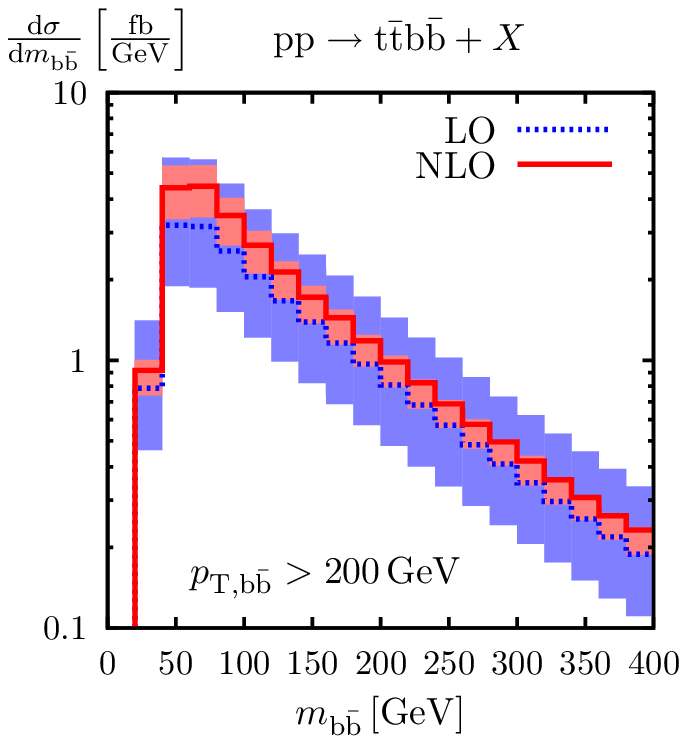}
\hfill
\includegraphics[bb= 95 445 280 655, width=.45\textwidth]
{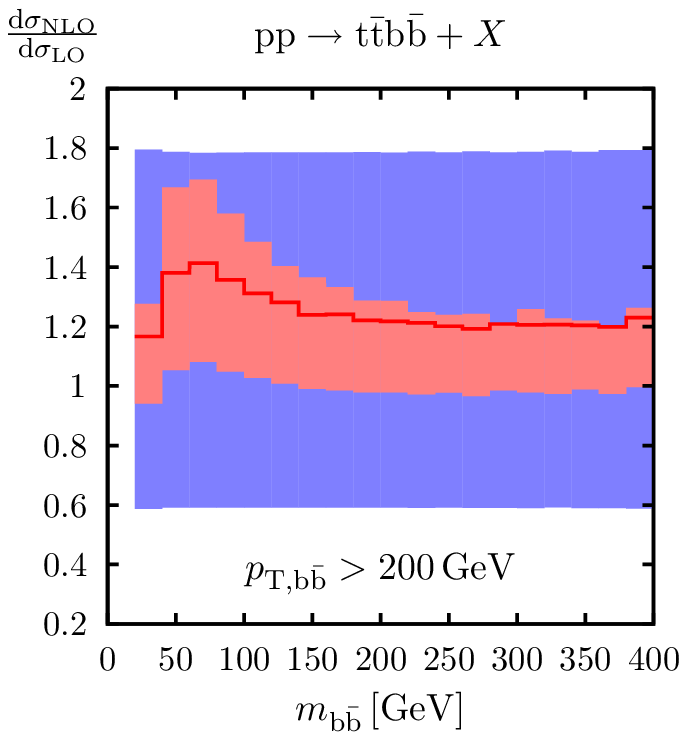}
\vspace*{-.8em}
\caption{Invariant-mass distribution of the
$\Pb\bar\Pb$ pair in setup~II: absolute
LO and NLO predictions (left) and NLO  $K$ factor (right).
The uncertainty bands correspond to factor-two scale variations.
}
\label{fig:mbb_dist_2}
\end{figure}

\subsubsection*{Invariant-mass and transverse-momentum distributions}
The $\Pb\bar\Pb$ invariant-mass distribution is displayed in
\reffi{fig:mbb_dist_2}. Its behaviour in the region
$m_{\Pb\bar\Pb}\lsim 50\GeV$ reflects the effective lower bound
\refeq{mbbbound}.  We find that the NLO corrections induce an
appreciable shape distortion of about 20\%, in particular near the
physically interesting region of $m_{\Pb\bar\Pb}\sim 100\GeV$.  Such
effect tends to mimic a Higgs signal and should be carefully taken
into account in the $\Pt\bar\Pt\PH(\PH\to\Pb\bar\Pb)$ analysis.

\begin{figure}
\includegraphics[bb= 95 445 280 655, width=.45\textwidth]
{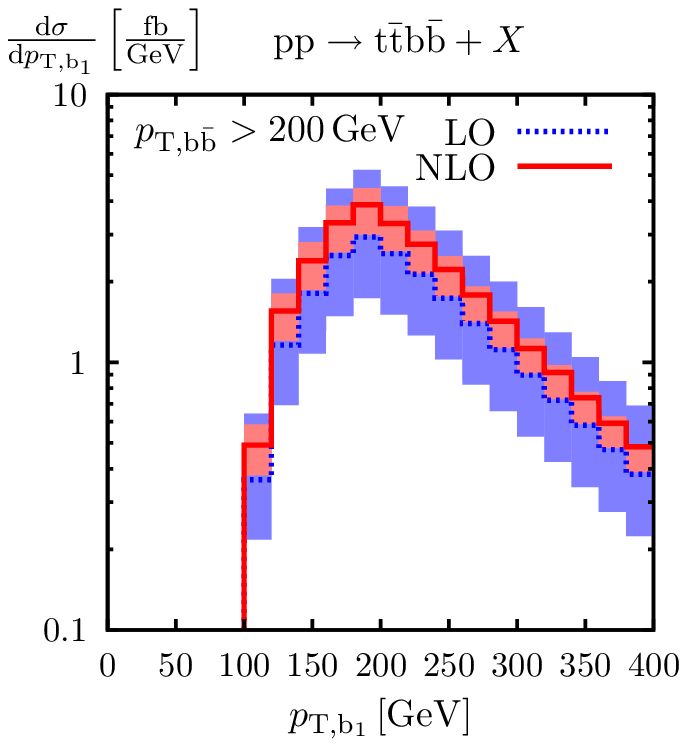}
\hfill
\includegraphics[bb= 95 445 280 655, width=.45\textwidth]
{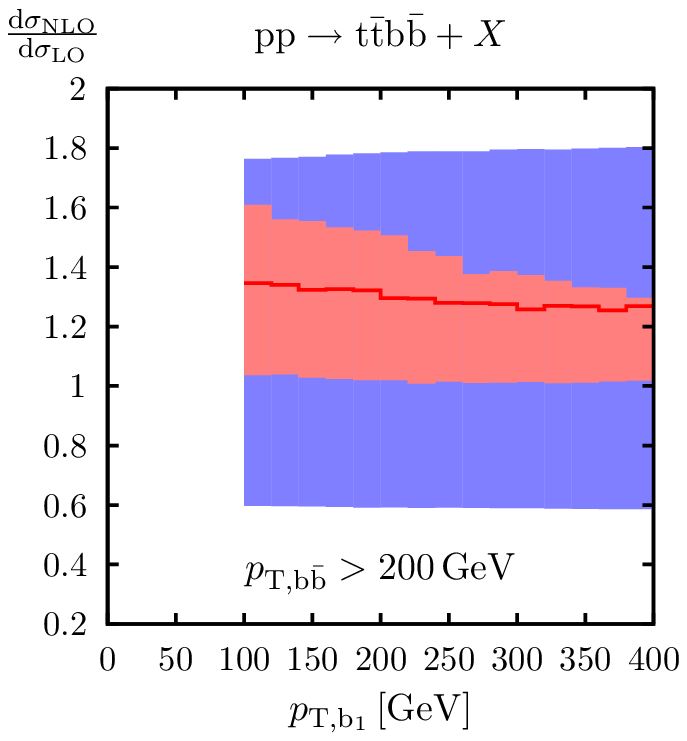}
\vspace*{-.8em}
\caption{Transverse-momentum distribution of the
harder b jet in setup~II: absolute
LO and NLO predictions (left) and NLO  $K$ factor (right).
The uncertainty bands correspond to {factor-two} scale variations.
}
\label{fig:pTbhard_dist_2}
\vspace*{4em}
\includegraphics[bb= 95 445 280 655, width=.45\textwidth]
{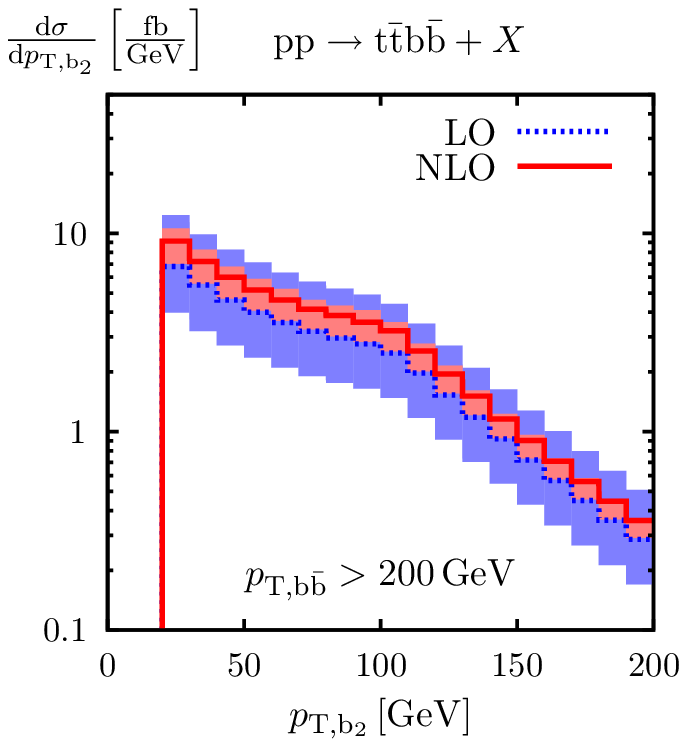}
\hfill
\includegraphics[bb= 95 445 280 655, width=.45\textwidth]
{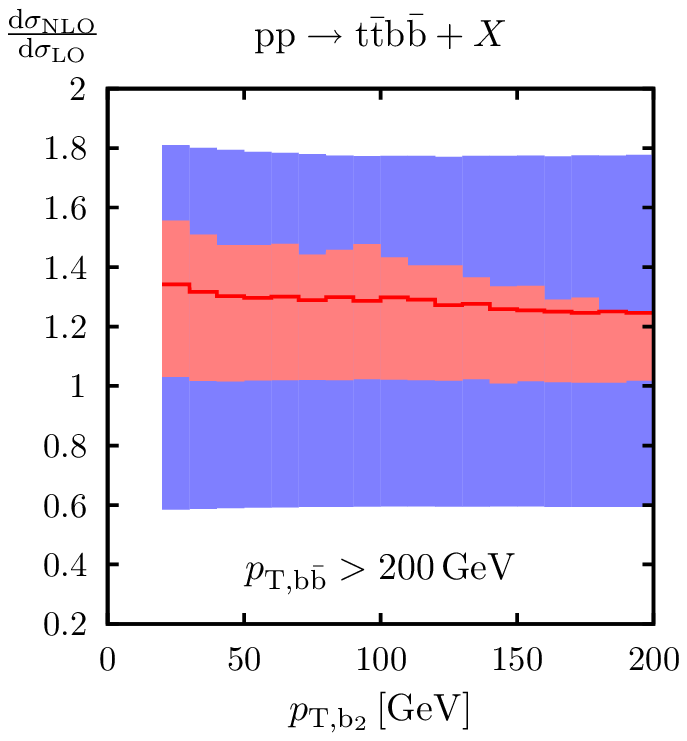}
\vspace*{-.8em}
\caption{Transverse-momentum distribution of the
softer b jet in setup~II: absolute
LO and NLO predictions (left) and NLO  $K$ factor (right).
The uncertainty bands correspond to factor-two scale variations.
}
\label{fig:pTbsoft_dist_2}
\end{figure}

The transverse-momentum distributions of the harder and softer b jets
are presented in \reffi{fig:pTbhard_dist_2} and
\reffi{fig:pTbsoft_dist_2}, respectively.  As a consequence of the cut
imposed on the transverse momentum of the b~pair, the harder b jet is
pushed to much higher $p_\rT$ values as compared to setup I.  The
maximum of its distribution is located around 200 GeV.  In contrast,
the softer b jet is much less sensitive to the $p_{\rT,\Pb\bar\Pb}$
cut and is predominantly produced in the region $20\GeV< p_\rT < 100
\GeV$.  This different kinematic behaviour of the two b jets might be
exploited to separate the $\Pt\bar\Pt\Pb\bar\Pb$ background from the
$\Pt\bar\Pt\PH$ signal, where both b jets are produced by the Higgs
boson and should thus have more similar $p_\rT$-values.  The NLO
corrections to both $p_{\rT,\Pb}$ distributions feature a slight
transverse-momentum dependence, with 10\% variations of the $K$ factor
within the plotted range.

\begin{figure}
\includegraphics[bb= 95 445 280 655, width=.45\textwidth]
{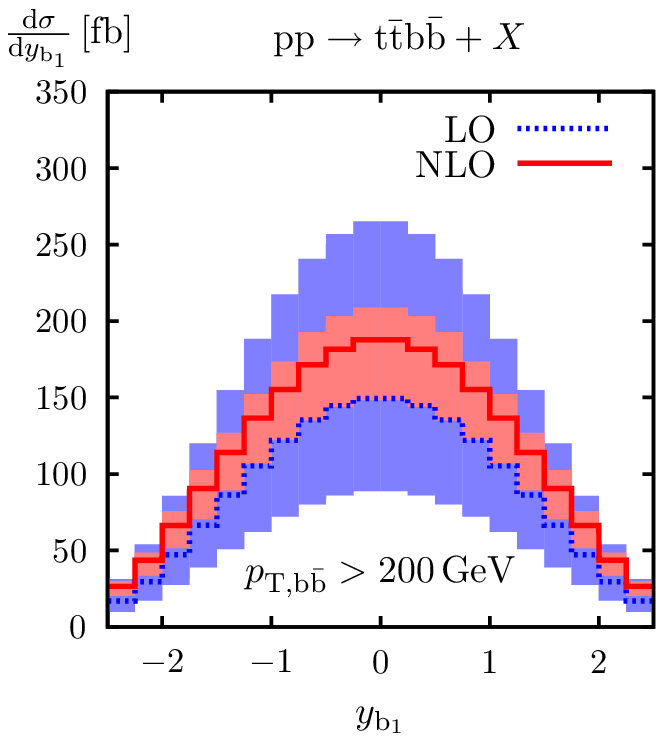}
\hfill
\includegraphics[bb= 95 445 280 655, width=.45\textwidth]
{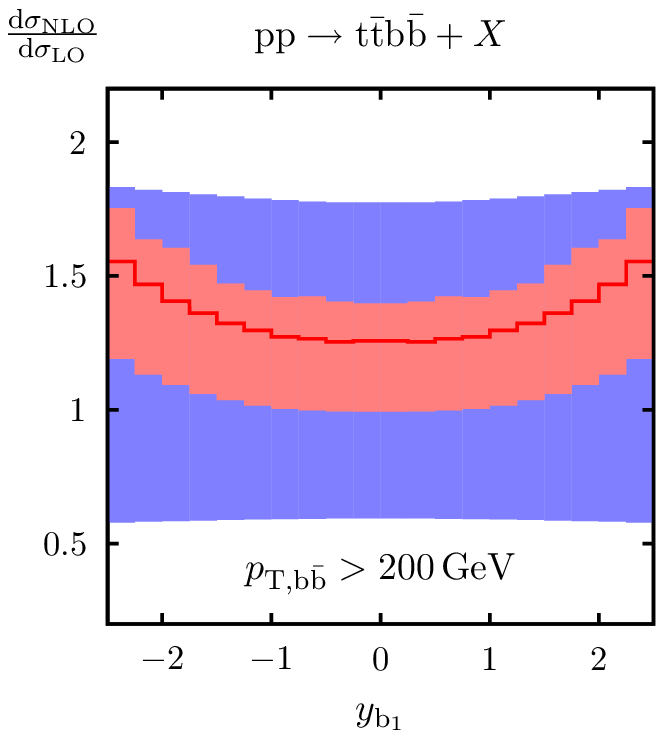}
\vspace*{-.8em}
\caption{Rapidity distribution of the
harder b jet in setup~II: absolute
LO and NLO predictions (left) and NLO  $K$ factor (right).
The uncertainty bands correspond to factor-two scale variations.
}
\label{fig:ybhard_dist_2}
\vspace*{4em}
\includegraphics[bb= 95 445 280 655, width=.45\textwidth]
{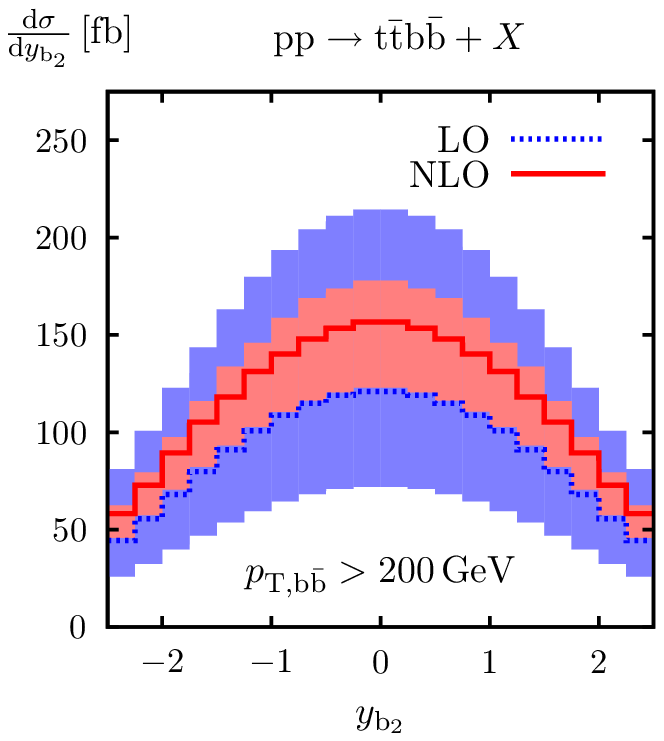}
\hfill
\includegraphics[bb= 95 445 280 655, width=.45\textwidth]
{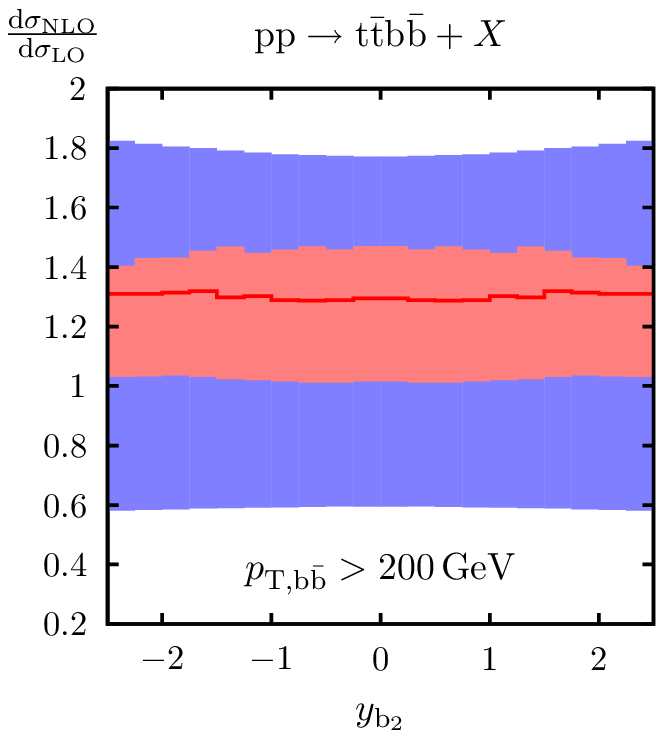}
\vspace*{-.8em}
\caption{Rapidity distribution of the
softer b jet in setup~II: absolute
LO and NLO predictions (left) and NLO  $K$ factor (right).
The uncertainty bands correspond to factor-two scale variations.
}
\label{fig:ybsoft_dist_2}
\end{figure}

\subsubsection*{Rapidity and azimuthal distributions}
The rapidities of the harder and softer b jets are shown in
\reffis{fig:ybhard_dist_2} and \ref{fig:ybsoft_dist_2}, respectively.
Due to the high $p_\rT$ of the b-pair system, both b jets tend to be
more central as compared to setup I.  While the $K$ factor is almost
insensitive to the rapidity of the soft b jet, the NLO corrections
have a non-negligible influence on the shape of the hard-b-jet
distribution for $|y_\Pb|>1$.  The rapidity distribution of the
$\Pb\bar\Pb$~system (not plotted) behaves similarly to the one of the
harder b jet.

\begin{figure}
\includegraphics[bb= 95 445 280 655, width=.45\textwidth]
{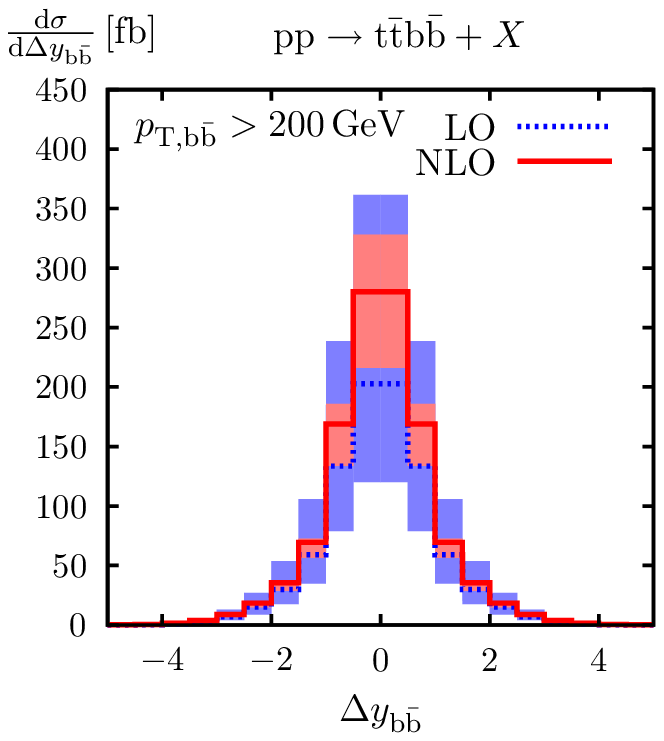}
\hfill
\includegraphics[bb= 95 445 280 655, width=.45\textwidth]
{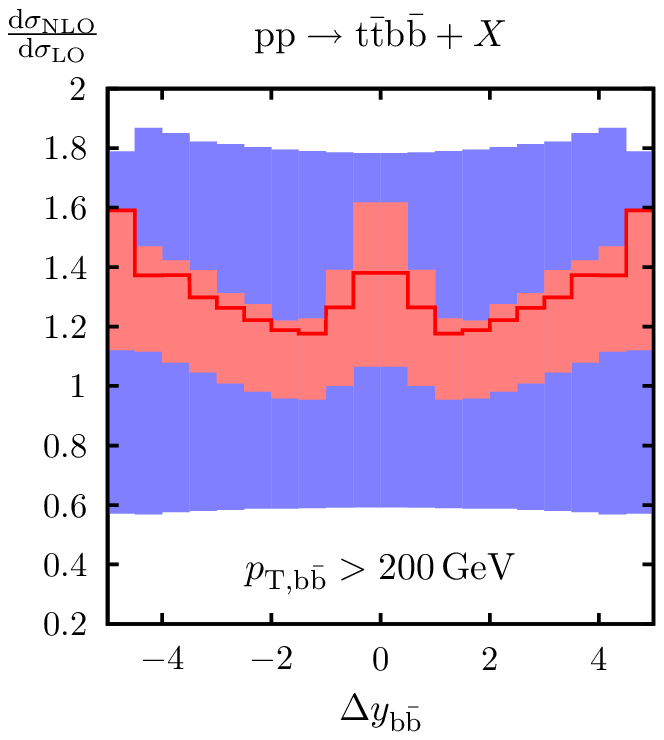}
\vspace*{-.8em}
\caption{Rapidity separation of the 
two b jets in setup~II: absolute
LO and NLO predictions (left) and NLO  $K$ factor (right).
The uncertainty bands correspond to factor-two scale variations.
}
\label{fig:dybb_dist_2}
\end{figure}

The rapidity-separation distribution (\reffi{fig:dybb_dist_2}) is
strongly peaked at small $\Delta y_{\Pb\bar\Pb}$ and the NLO
corrections have an appreciable influence on its shape. In the region
$\Delta y_{\Pb\bar\Pb}< 2$ the $K$ factor varies between 1.17 and
1.38.

\begin{figure}
\includegraphics[bb= 95 445 280 655, width=.45\textwidth]
{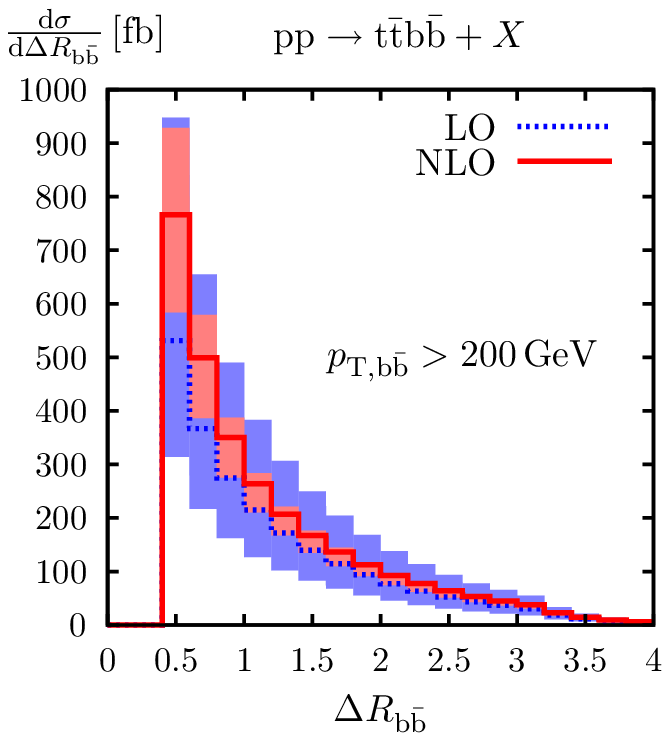}
\hfill
\includegraphics[bb= 95 445 280 655, width=.45\textwidth]
{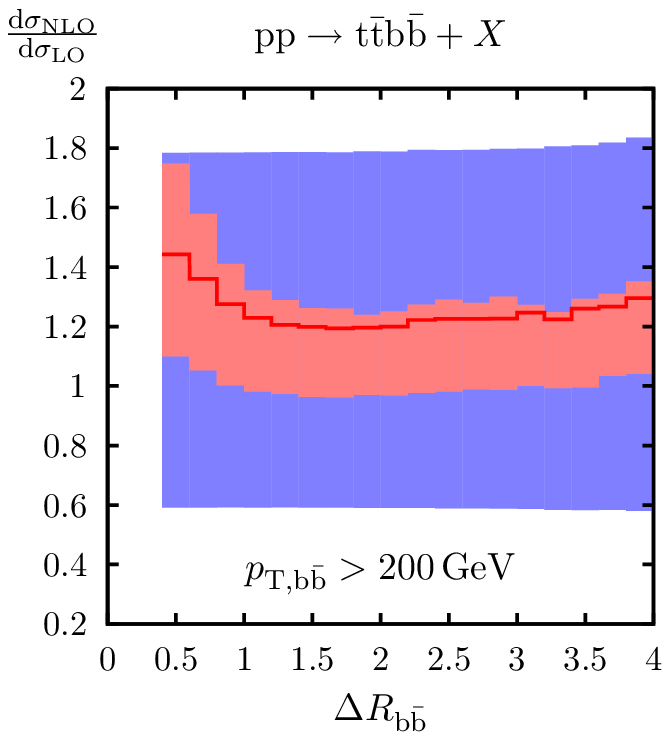}
\vspace*{-.8em}
\caption{Rapidity-azimuthal separation of the 
two b jets in setup~II: absolute
LO and NLO predictions (left) and NLO  $K$ factor (right).
The uncertainty bands correspond to factor-two scale variations.
}
\label{fig:dRbb_dist_2}
\end{figure}

Finally, in \reffi{fig:dRbb_dist_2} we show the
rapidity--azimuthal-angle separation of the b jets, which is strongly
peaked at small $\Delta R_{\Pb\bar\Pb}$.  Also in this distribution we
observe a significant shape distortion.  In the region $\Delta
R_{\Pb\bar\Pb}< 1$, which corresponds to invariant masses of the order
of 100 GeV, the $K$ factor increases by about 20\%.  This NLO effect
might have an important impact on the measurement of $\Pt\bar\Pt\PH$
in the highly-boosted Higgs regime.

\begin{figure}
\includegraphics[bb= 95 445 280 655, width=.45\textwidth]
{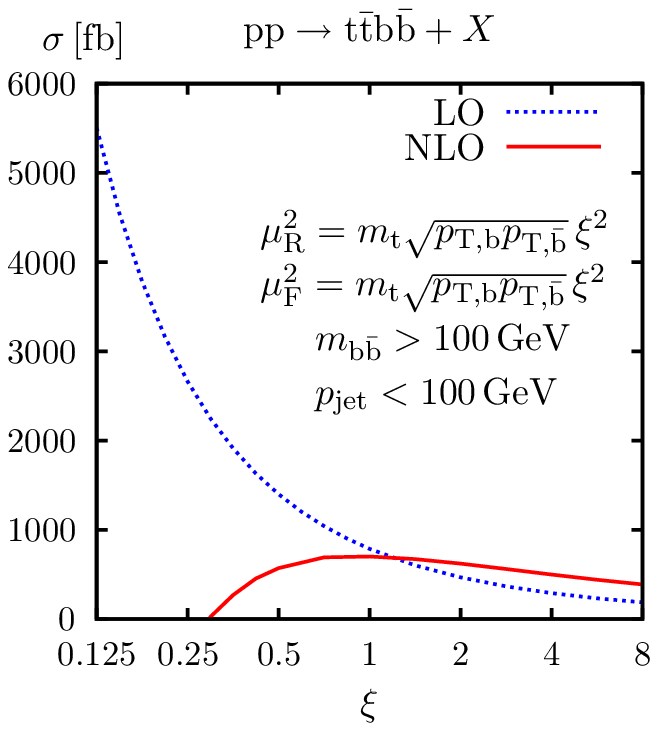}
\hfill
\includegraphics[bb= 95 445 280 655, width=.45\textwidth]
{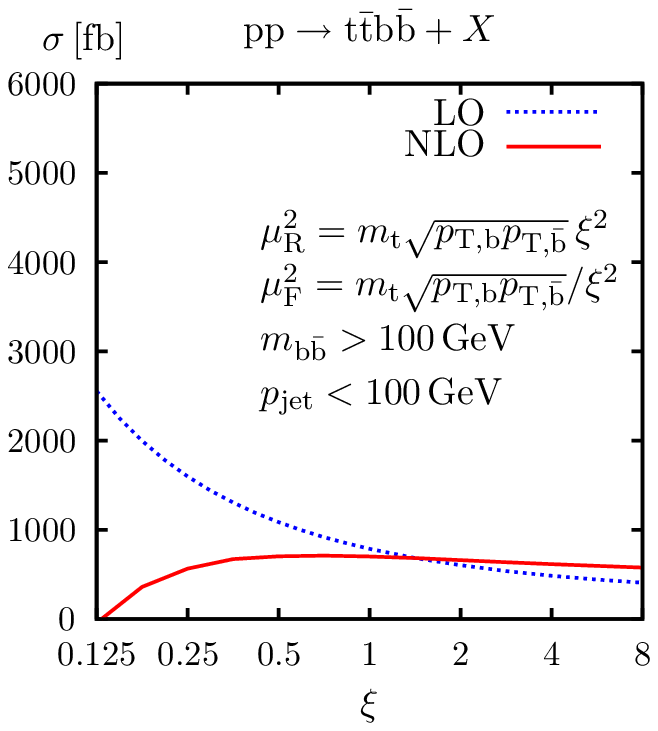}
\vspace*{-.8em}
\caption{Scale dependence of the LO and NLO 
$\Pp\Pp\to\Pt\bar\Pt\Pb\bar\Pb+X$ cross section at \mbox{$\sqrt{s}=14\TeV$}
in setup~III.
The left and the right plots describe  
uniform ($\xi_{\mathrm{R}}=\xi_{\mathrm{F}}=\xi$)
and antipodal ($\xi_{\mathrm{R}}=\xi_{\mathrm{F}}^{-1}=\xi$)
scale variations, respectively. 
}
\label{fig:sigma_tot_3}
\end{figure}

\subsection{Setup III}
In order to explore the effect of a jet veto and its possible
correlation with other observables, we have generated events with
$m_{\Pb\bar\Pb}>100 \GeV$ and $p_{\rT,\mathrm{jet}}<100 \GeV$.  The LO
and NLO cross sections and their scale dependence are shown in
\reffi{fig:sigma_tot_3}.  At the central scale we obtain
$\sigma_{\mathrm{LO}}=786.1(6)\fb$ and
$\sigma_{\mathrm{NLO}}=700(3)\fb$, corresponding to a correction
factor $K=0.89$.  It is evident from \reffi{fig:sigma_tot_3} that the
central scale is very close to a stable point.  This demonstrates that
a jet-veto value of 100 GeV is sufficiently large to avoid
perturbative instabilities.  Varying the QCD scales up and down by a
factor two shifts the NLO cross section by only 0.4\% and $-19\%$,
respectively.

Inspecting various kinematic distributions we find that the NLO
corrections have a much bigger impact on shapes as compared to setup
I. In particular, we observe that the suppression effect resulting
from the jet veto is rather sensitive to the transverse momentum of
the b jets. For instance, the NLO correction to the
$p_\rT$~distribution of the harder b jet varies by about 15\% in the
range $20 \GeV < p_{\rT,\Pb_1} < 200\GeV$.

\begin{figure}
\includegraphics[bb= 95 445 280 655, width=.45\textwidth]
{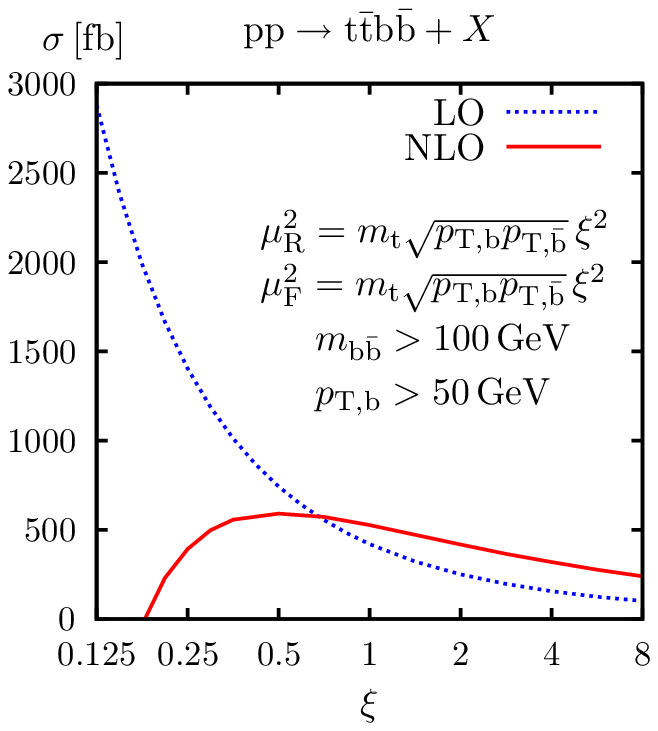}
\hfill
\includegraphics[bb= 95 445 280 655, width=.45\textwidth]
{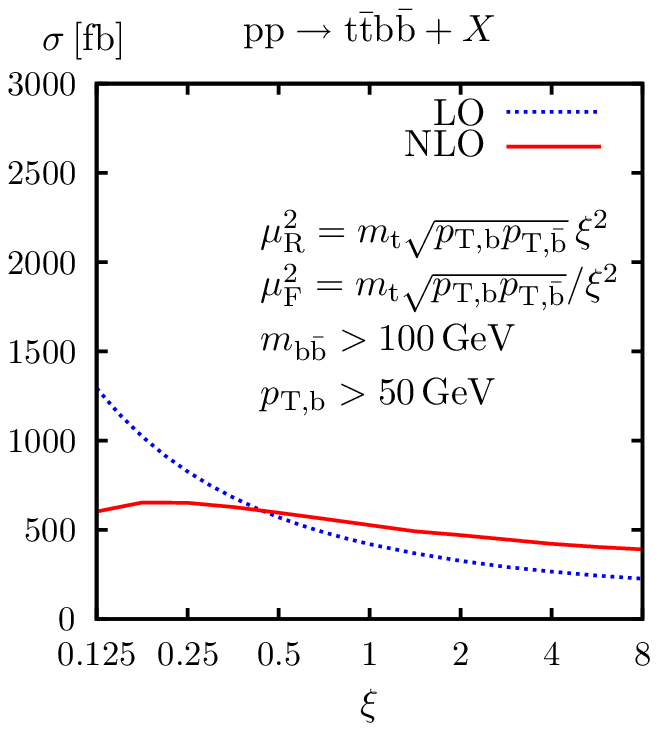}
\vspace*{-.8em}
\caption{Scale dependence of the LO and NLO 
$\Pp\Pp\to\Pt\bar\Pt\Pb\bar\Pb+X$ cross section at \mbox{$\sqrt{s}=14\TeV$}
in setup~IV.
The left and the right plots describe  
uniform ($\xi_{\mathrm{R}}=\xi_{\mathrm{F}}=\xi$)
and antipodal ($\xi_{\mathrm{R}}=\xi_{\mathrm{F}}^{-1}=\xi$)
scale variations, respectively. 
}
\label{fig:sigma_tot_4}
\end{figure}

\subsection{Setup IV}
As observed in \reffi{fig:pTbsoft_dist_1}, the $\Pt\bar\Pt\Pb\bar\Pb$
cross section is dominated by relatively soft b jets, which tend to
saturate the $p_{\rT,\Pb}>20\GeV$ cut.  It is thus interesting to
investigate the influence of this cut on the $\Pt\bar\Pt\Pb\bar\Pb$
background and, in particular, on the NLO corrections.  To this end,
we have studied a variation of setup I where the $p_{\rT,\Pb}$ cut is
increased from 20 to 50 GeV.
The LO and NLO cross sections and their scale dependence are shown in
\reffi{fig:sigma_tot_4}.  At the central scale we obtain
$\sigma_{\mathrm{LO}}=419.4(1)\fb$ and
$\sigma_{\mathrm{NLO}}=526(2)\fb$, corresponding to a correction
factor $K=1.25$.  The higher $p_{\rT,\Pb}$ cut reduces the NLO cross
section by 46\% as compared to setup I.  The LO and the NLO scale
dependence remain very similar as in setup~I
(cf.~\reffi{fig:sigma_tot_1}): the uncertainty corresponding to
factor-two scale variations amounts to 77\% in LO and 21\% in NLO.

Inspecting various kinematic distributions we find, as in setup~I,
that the NLO corrections have little impact on the shapes. In general
the kinematic dependence of the NLO correction factor is below 10\%.
The largest shape distortion is observed in the $\Delta
R_{\Pb\bar\Pb}$ distribution: similarly as in setup I
(cf.~\reffi{fig:dRbb_dist_1}) we find an increase of the $K$ factor
in the region $0.4< \Delta R_{\Pb\bar\Pb} \lsim 1$ by about 30\%.

\section{Conclusions}
\label{se:conclusion}

The direct production of $\Pt\bar\Pt\Pb\bar\Pb$ final states
represents the major background to the production of Higgs bosons in
association with top--antitop-quark pairs at the LHC,
$\Pp\Pp\to\Pt\bar\Pt\PH$, where the Higgs boson decays into
$\Pb\bar\Pb$ pairs.
This process can lead to a direct measurement of the top-quark Yukawa
coupling.  Apart from improvements in the experimental analysis, a
successful exploitation of this demanding channel requires predictions
for $\Pt\bar\Pt\Pb\bar\Pb$ production at the next-to-leading-order
level in QCD, maybe even further improvements.

Extending our first results on the total cross section
for $\Pp\Pp\to\Pt\bar\Pt\Pb\bar\Pb$
published earlier, we have presented a detailed study of integrated
and differential cross sections at the LHC, discussing in particular a
dynamical scale choice, the influence of various cuts on the outgoing
b~quarks, and the impact of a veto on an additional hard jet.  We
observe that the traditional choice of a constant scale determined by
the energy threshold for the process underestimates
the $\Pt\bar\Pt\Pb\bar\Pb$ background by a factor of two, while an
appropriate dynamical scale, which is tied to the transverse momenta
of the b~quarks, stabilizes the perturbative predictions much better.
Moreover, the $K$ factor is reduced from $1.8$ to $1.2$.  Using the
new scale choice, the corrections have little impact on the shapes of
distributions if standard cuts are applied.  Strengthening the cut on
the transverse momentum of the bottom quarks has no big influence on
the $K$ factor and the effect of the NLO corrections on the shape of
the distributions. On the other hand, imposing a jet veto of $100\GeV$
reduces the $K$ factor to $0.9$ and enhances the impact of the NLO
corrections on the shapes of distributions.
In the regime of highly boosted Higgs bosons, which offers better
perspectives to observe the $\Pt\bar\Pt\PH$ signal, we find
significant distortions of the kinematic distributions, while the $K$
factor is $1.3$.

Our calculation builds on the Feynman-diagrammatic approach, \ie on an
algorithmic reduction of each Feynman diagram to a canonical standard
form, which is automatically processed to {\sc Fortran} output, and on
a numerical reduction of tensor loop integrals to an appropriate set
of scalar master integrals.  A key feature of the diagram-by-diagram
approach is that colour sums can be preformed very efficiently.
Helicity summation is simplified by introducing a basis of
$\mathcal{O}(1000)$ basic structures. The reduction to these
structures can be performed in a process-independent way.  The
numerical tensor-integral reduction employs dedicated methods that
have been developed to treat the numerically delicate phase-space
regions where small Gram determinants appear in denominators during
the traditional tensor reduction.  Real corrections are integrated
using well-known dipole subtraction methods.  We find that our
Feynman-diagrammatic approach provides very high numerical stability
and CPU efficiency, a result that is very encouraging in view of
future challenging next-to-leading-order calculations for important
multiparticle processes.

\section*{Acknowledgments}
We thank Thomas Hahn for technical help in structuring the very long
source code, as well as L.~Dixon, M.~Mangano and C.~Papadopoulos for
discussions.  This work is supported in part by the European
Community's Marie-Curie Research Training Network under contract
MRTN-CT-2006-035505 ``Tools and Precision Calculations for Physics
Discoveries at Colliders'' and the Japan Society for the Promotion of
Science.

\appendix
\section*{Appendix}

\section{Some details on the reduction of standard matrix elements}
\label{app:smes} 

In \refse{se:virtcor} we have briefly described our prodecure to
reduce the numerous helicity structures for
$\Pg\Pg\to\Pt\bar\Pt\Pb\bar\Pb$ to a standard form. We proceed in two
steps, the first step employing only identities that hold in arbitrary
$D\ne4$ space--time dimensions, followed by the second step
(after cancelling UV divergences) that builds on four-dimensional
identities.  As also anticipated in \refse{se:virtcor}, we support two
different versions of step~2. The first variant only uses the
reduction \refeq{fivegammaid} of products of five Dirac matrices to
products of three or only single matrices, leading to 970 SMEs. In
this approach no chiral projectors or $\ga_5$ factors are introduced.
The second variant employs more four-dimensional identities, such as
\refeq{eq:chisholm1}, which are derived from Chisholm's identity in
\citere{Bredenstein:2008zb,Denner:2005fg}, and leads to 502 SMEs,
which however involve the $\ga_5$ matrix.  

In the following we describe the second procedure for the crossed
process 
\beq
\Pg(p_1)\,\Pg(p_2)\,\bar\Pt(p_3)\,\Pt(p_4)\,\bar\Pb(p_5)\,\Pb(p_6)\to
0,\eeq 
where the incoming momenta of the corresponding incoming particles
are given in parentheses. The individual steps to reduce the structures
\beq
[\bar{v}(p_3) \Gamma_{a} \omega_\sigma u(p_4)] [\bar{v}(p_5)
\Gamma_{b} \omega_\tau u(p_6)] \equiv [\Gamma_{a}]^\sigma_{34}
[\Gamma_{b}]^\tau_{56}
\eeq 
are as follows:
\begin{enumerate}
\item
First, we eliminate multiple contractions of Dirac matrices between
the two Dirac chains%
\footnote{In this context we recall that multiple contractions of
  Dirac matrices inside a single Dirac chain have already been
  eliminated in the first ($D$-dimensional!) step of the algebraic
  reduction.}  by identities like
\beqar\label{eq:chisholm2}
{\gamma^\mu}
{\gamma^\alpha}
{\gamma^\nu}
\omega_\pm
\otimes
{\gamma_{\mu}}
{\gamma^\beta}
{\gamma_\nu}
=
4g^{\alpha\beta}
{\gamma^\mu}
\omega_\pm
\otimes
\gamma_\mu
\omega_\pm
+
4
\gamma^\beta
\omega_\pm
\otimes
\gamma^\alpha
\omega_\mp
,\nl
{\gamma^\mu}
{\gamma^\alpha}
{\gamma^\nu}
\omega_\pm
\otimes
{\gamma_{\nu}}
{\gamma^\beta}
{\gamma_\mu}
=
4g^{\alpha\beta}
{\gamma^\mu}
\omega_\pm
\otimes
\gamma_\mu
\omega_\mp
+
4
\gamma^\beta
\omega_\pm
\otimes
\gamma^\alpha
\omega_\pm
.
\eeqar
\item Next we shorten strings of five or more Dirac matrices using
  \refeq{fivegammaid} in the massless bottom-quark chain,
  $[\bar{v}(p_5) \Gamma_{b} \omega_\tau u(p_6)]$, leaving three or
  only one Dirac matrix in this chain.  There is at most one
  contraction of Dirac matrices between the two chains.
\item Now we simplify structures that involve bottom-quark Dirac
  chains with three slashed vectors using the trick described in
  (3.40) and (3.41) of \citere{Denner:2005fg}.  In more detail, this
  manipulation replaces the product of two Dirac chains of the form
  $[\dsl{a} \dsl{b} \dsl{c} \dots ]^\si_{34} [\dsl{d} \dsl{e} \dsl{f}
  ]^\tau_{56}$ by products in which either the top chain
  $[\dots]^\sigma_{34}$ has two Dirac matrices less or in which the
  bottom chain contains only one slashed vector.  Using this procedure
  recursively we obviously achieve that bottom chains with three
  slashed vectors are multiplied by top chains containing only a
  single Dirac matrix, i.e.\ they appear in the form
  $[\dsl{a}]^\si_{34} [\dsl{d} \dsl{e} \dsl{f}]^\tau_{56}$.
\item In a further step it is possible to eliminate all products of
  three Dirac matrices in the bottom chain, so that this massless
  Dirac chain contains exactly one Dirac matrix.  If there is a
  contraction between the top and bottom chain, two Dirac matrices can
  easily be shifted from the bottom to the top chain using
  \refeq{eq:chisholm1}. If there is no contraction the product of
  chains looks like $[\dsl{a}]^\si_{34} [\dsl{d} \dsl{e}
  \dsl{f}]^\tau_{56}$ owing to the previous step.  If the set
  $\{d,e,f\}$ contains at most one polarization vector $\veps_i$
  $(i=1,2)$ of the two incoming gluons, we can assume that $\{d,e,f\}$
  contains either the momentum $p_3$ or $p_4$. If this is not the
  case, it can be achieved upon using momentum conservation and the
  Dirac equation in the bottom chain. The factors $\dsl{p}_i$
  $(i=3,4)$ in the bottom chain easily allow for shifting all but one
  Dirac matrix to the top chain by the manipulations,
\beqar
\Bigl[\dsl{a}\Bigr]^\pm_{34} \Bigl[\dsl{p}_3 \dsl{e} \dsl{f}\Bigr]^\pm_{56} &=&
\frac{1}{2}
\left( \Bigl[\ga_\mu\dsl{p}_3\dsl{a}\Bigr]^\pm_{34}-m_3
\Bigl[\ga_\mu\dsl{a}\Bigr]^\pm_{34} \right)
\Bigl[\ga^\mu \dsl{e} \dsl{f}\Bigr]^\pm_{56}
\nn\\
&=&
\frac{1}{2}
\left( \Bigl[ \ga_\mu\dsl{f} \dsl{e} \dsl{p}_3 \dsl{a} \Bigr]^\pm_{34}
-m_3 \Bigl[ \dsl{e}\dsl{f}\ga_\mu \dsl{a} \Bigr]^\pm_{34} \right) \Bigl[\ga^\mu\Bigr]^\pm_{56},
\nn\\
\Bigl[\dsl{a}\Bigr]^\pm_{34} \Bigl[\dsl{p}_3 \dsl{e} \dsl{f}\Bigr]^\mp_{56} &=&
\dots =
\frac{1}{2}
\left( \Bigl[ \dsl{e}\dsl{f} \ga_\mu \dsl{p}_3 \dsl{a} \Bigr]^\pm_{34}
-m_3 \Bigl[ \ga_\mu\dsl{f}\dsl{e}\dsl{a} \Bigr]^\pm_{34} \right)
\Bigl[\ga^\mu\Bigr]^\mp_{56},
\nn\\
\Bigl[\dsl{a}\Bigr]^\pm_{34} \Bigl[\dsl{p}_4 \dsl{e} \dsl{f}\Bigr]^\pm_{56} &=&
\dots =
\frac{1}{2} \left( \Bigl[\dsl{a}\dsl{p}_4\ga_\mu\dsl{f} \dsl{e}\Bigr]^\pm_{34}
+m_4 \Bigl[\dsl{a}\dsl{e} \dsl{f}\ga_\mu\Bigr]^\mp_{34} \right) \Bigl[\ga^\mu\Bigr]^\pm_{56},
\nn\\
\Bigl[\dsl{a}\Bigr]^\pm_{34} \Bigl[\dsl{p}_4 \dsl{e} \dsl{f}\Bigr]^\mp_{56} &=&
\dots =
\frac{1}{2}
\left( \Bigl[\dsl{a}\dsl{p}_4 \dsl{e}\dsl{f} \ga_\mu \Bigr]^\pm_{34}
+m_4 \Bigl[\dsl{a}\ga_\mu\dsl{f}\dsl{e}\Bigr]^\mp_{34} \right)
\Bigl[\ga^\mu\Bigr]^\mp_{56},
\eeqar
where we used $p_i^\mu=\{\dsl{p}_i,\gamma^\mu\}/2$ and the Dirac
equation in the 34-chain and \refeq{eq:chisholm1}.  We recall that we
consistently take momenta as incoming.  The case that no factor
$\dsl{p}_{3/4}$ appears in the bottom chain can only occur if the
other two slashed vectors in the bottom chain belong to polarization
vectors. Thus, we can assume that the vector $a$ is a momentum out of
$\{p_1,p_2,p_5,p_6\}$ and that $\{d,e,f\}$ contains $p_1$ or $p_2$. We
can even omit $p_2$ on either side if we eliminate it via momentum
conservation, so that the considered bottom chains all contain a
factor $\dsl{p}_1$.  Two Dirac matrices can now be shifted from the
bottom to the top chain using the relations
\beqar
\Bigl[\dsl{p}_1\Bigr]^\pm_{34} \Bigl[\dsl{p}_1 \dsl{e} \dsl{f}\Bigr]^\pm_{56} &=&
\frac{1}{2}
\left( \Bigl[\dsl{p}_1\ga_\mu\dsl{p}_1\Bigr]^\pm_{34}+p_1^2\Bigl[\ga_\mu\Bigr]^\pm_{34} \right)
\Bigl[\ga^\mu \dsl{e} \dsl{f}\Bigr]^\pm_{56}
\nn\\
&=&
\frac{1}{2}
\left( \Bigl[\dsl{p}_1\dsl{e} \dsl{f}\ga_\mu\dsl{p}_1\Bigr]^\pm_{34}
+p_1^2\Bigl[\ga_\mu\dsl{f} \dsl{e}\Bigr]^\pm_{34} \right)
\Bigl[\ga^\mu\Bigr]^\pm_{56},
\nn\\
\Bigl[\dsl{p}_1\Bigr]^\pm_{34} \Bigl[\dsl{p}_1 \dsl{e} \dsl{f}\Bigr]^\mp_{56} &=&
\dots =
\frac{1}{2} \left( \Bigl[\dsl{p}_1\ga_\mu \dsl{f}\dsl{e} \dsl{p}_1\Bigr]^\pm_{34}
+p_1^2\Bigl[\dsl{e}\dsl{f} \ga_\mu \Bigr]^\pm_{34} \right)
\Bigl[\ga^\mu\Bigr]^\mp_{56},
\nn\\
\Bigl[\dsl{p}_5\Bigr]^\pm_{34} \Bigl[\dsl{p}_1 \dsl{e} \dsl{f}\Bigr]^\pm_{56} &=&
\frac{1}{2} \Bigl[\ga_\mu\Bigr]^\pm_{34}
\Bigl[\ga^\mu\dsl{p}_5\dsl{p}_1 \dsl{e} \dsl{f} \Bigr]^\pm_{56} 
=
\frac{1}{2} \Bigl[ \ga_\mu\dsl{p}_1\dsl{p}_5 \Bigr]^\pm_{34}
\Bigl[\ga^\mu\dsl{e} \dsl{f} \Bigr]^\pm_{56}
\nn\\
&=&
\frac{1}{2} \Bigl[ \ga_\mu \dsl{f} \dsl{e} \dsl{p}_1\dsl{p}_5 \Bigr]^\pm_{34}
\Bigl[\ga^\mu \Bigr]^\pm_{56},
\nn\\
\Bigl[\dsl{p}_5\Bigr]^\pm_{34} \Bigl[\dsl{p}_1 \dsl{e} \dsl{f}\Bigr]^\mp_{56} &=&
\dots =
\frac{1}{2} \Bigl[ \dsl{p}_5 \dsl{p}_1 \dsl{e}  \dsl{f}\ga_\mu \Bigr]^\pm_{34}
\Bigl[\ga^\mu \Bigr]^\mp_{56},
\nn\\
\Bigl[\dsl{p}_6\Bigr]^\pm_{34} \Bigl[\dsl{p}_1 \dsl{e} \dsl{f}\Bigr]^\pm_{56} &=&
\dots =
\frac{1}{2} \Bigl[\dsl{p}_6\dsl{f} \dsl{e}\dsl{p}_1 \ga_\mu\Bigr]^\pm_{34}
\Bigl[\ga^\mu\Bigr]^\pm_{56},
\nn\\
\Bigl[\dsl{p}_6\Bigr]^\pm_{34} \Bigl[\dsl{p}_1 \dsl{e} \dsl{f}\Bigr]^\mp_{56} &=&
\dots =
\frac{1}{2} \Bigl[ \ga_\mu \dsl{p}_1 \dsl{e} \dsl{f} \dsl{p}_6 \Bigr]^\pm_{34}
\Bigl[\ga^\mu\Bigr]^\mp_{56},
\eeqar
where we have used that the bottom mass is set to zero, i.e.\ 
$m_5=m_6=0$.  Note that we did not use $p_1^2=0$ in the first
relation, in order to indicate that the whole reduction procedure
described here does not only apply to gluons but also to the case
where gluons are replaced by massive vector bosons.
\item
After having reduced all bottom chains to contain only one Dirac 
matrix we apply \refeq{fivegammaid} to the top chain recursively as far as possible.
In this way the top chains contain only up to four Dirac matrices.
\item Next we reduce products of the form
  $[\dsl{a}\dsl{b}\dsl{c}\dsl{d}]^\si_{34} [\dsl{e}]^\tau_{56}$, which
  do not involve a contraction. After eliminating $p_2$ via momentum
  conservation, the product $\dsl{a}\dsl{b}\dsl{c}\dsl{d}$ contains
  one of the factors $\dsl{p}_1\dsl{p}_5$, $\dsl{p}_1\dsl{p}_6$, or
  $\dsl{p}_5\dsl{p}_6$, because at most two of the slashed vectors can
  be polarization vectors.  The majority of such cases can be reduced
  with the following relations,
\beqar
\Bigl[\Ga\dsl{a}\dsl{p}_5\Bigr]^\pm_{34} \Bigl[\dsl{e}\Bigr]^\pm_{56} &=&
\frac{1}{2} \Bigl[\Ga\dsl{a}\ga_\mu\Bigr]^\pm_{34} \Bigl[\ga^\mu \dsl{p}_5\dsl{e}\Bigr]^\pm_{56}  =
\frac{1}{2} \Bigl[\Ga\dsl{a}\ga_\mu\dsl{e}\dsl{p}_5\Bigr]^\pm_{34} \Bigl[\ga^\mu\Bigr]^\pm_{56}  
\nn\\
&=&
-\frac{1}{2} \Bigl[\Ga\dsl{a}\ga_\mu\dsl{p}_5\dsl{e}\Bigr]^\pm_{34} \Bigl[\ga^\mu\Bigr]^\pm_{56}  
+(ep_5) \Bigl[\Ga\dsl{a}\ga_\mu\Bigr]^\pm_{34} \Bigl[\ga^\mu\Bigr]^\pm_{56}
\nn\\
&=&
-\frac{1}{2} \Bigl[\Ga\ga_\mu\dsl{e}\Bigr]^\pm_{34} \Bigl[\dsl{a}\ga^\mu\dsl{p}_5\Bigr]^\pm_{56}  
+(ep_5) \Bigl[\Ga\dsl{a}\ga_\mu\Bigr]^\pm_{34} \Bigl[\ga^\mu\Bigr]^\pm_{56}
\nn\\
&=&
- \Bigl[\Ga\dsl{p}_5\dsl{e}\Bigr]^\pm_{34} \Bigl[\dsl{a}\Bigr]^\pm_{56}  
+ (ap_5) \Bigl[\Ga\ga_\mu\dsl{e}\Bigr]^\pm_{34} \Bigl[\ga^\mu\Bigr]^\pm_{56}  
+(ep_5) \Bigl[\Ga\dsl{a}\ga_\mu\Bigr]^\pm_{34} \Bigl[\ga^\mu\Bigr]^\pm_{56}
\nn\\
&=&
\Bigl[\Ga\dsl{e}\dsl{p}_5\Bigr]^\pm_{34} \Bigl[\dsl{a}\Bigr]^\pm_{56} 
+(ap_5) \Bigl[\Ga\ga_\mu\dsl{e}\Bigr]^\pm_{34} \Bigl[\ga^\mu\Bigr]^\pm_{56}
-(ep_5) \Bigl[\Ga\ga_\mu\dsl{a}\Bigr]^\pm_{34} \Bigl[\ga^\mu\Bigr]^\pm_{56},
\hspace{2em}
\nn\\
\Bigl[\Ga\dsl{a}\dsl{p}_5\Bigr]^\pm_{34} \Bigl[\dsl{e}\Bigr]^\mp_{56} &=&
\dots = 
-\Bigl[\Ga\dsl{e}\dsl{p}_5\Bigr]^\pm_{34} \Bigl[\dsl{a}\Bigr]^\mp_{56} 
+(ap_5) \Bigl[\Ga\dsl{e}\ga_\mu\Bigr]^\pm_{34} \Bigl[\ga^\mu\Bigr]^\mp_{56}
\nn\\
&& {}
-(ae) \Bigl[\Ga\dsl{p}_5\ga_\mu\Bigr]^\pm_{34} \Bigl[\ga^\mu\Bigr]^\mp_{56}
+(ep_5) \Bigl[\Ga\dsl{a}\ga_\mu\Bigr]^\pm_{34} \Bigl[\ga^\mu\Bigr]^\mp_{56},
\nn\\
\Bigl[\Ga\dsl{a}\dsl{p}_6\Bigr]^\pm_{34} \Bigl[\dsl{e}\Bigr]^\pm_{56} &=&
\dots = 
-\Bigl[\Ga\dsl{e}\dsl{p}_6\Bigr]^\pm_{34} \Bigl[\dsl{a}\Bigr]^\pm_{56} 
+(ap_6) \Bigl[\Ga\dsl{e}\ga_\mu\Bigr]^\pm_{34} \Bigl[\ga^\mu\Bigr]^\pm_{56}
\nn\\
&& {}
-(ae) \Bigl[\Ga\dsl{p}_6\ga_\mu\Bigr]^\pm_{34} \Bigl[\ga^\mu\Bigr]^\pm_{56}
+(ep_6) \Bigl[\Ga\dsl{a}\ga_\mu\Bigr]^\pm_{34} \Bigl[\ga^\mu\Bigr]^\pm_{56},
\\
\Bigl[\Ga\dsl{a}\dsl{p}_6\Bigr]^\pm_{34} \Bigl[\dsl{e}\Bigr]^\mp_{56} &=&
\dots =
\Bigl[\Ga\dsl{e}\dsl{p}_6\Bigr]^\pm_{34} \Bigl[\dsl{a}\Bigr]^\mp_{56} 
+(ap_6) \Bigl[\Ga\ga_\mu\dsl{e}\Bigr]^\pm_{34} \Bigl[\ga^\mu\Bigr]^\mp_{56}
-(ep_6) \Bigl[\Ga\ga_\mu\dsl{a}\Bigr]^\pm_{34} \Bigl[\ga^\mu\Bigr]^\mp_{56}.
\nn
\eeqar
Here $\Ga$ is any string of Dirac matrices, and again $m_5=m_6=0$ was
used.  If $a\ne e$, these relations interchange $a$ and $e$ (if $a=e$
they are useless).  Since $e$ can only be $p_1$, $p_3$, or $p_4$, we
can use these identities to shift $p_3$ or $p_4$ to the top chain,
where these momenta are eliminated by the Dirac equation, so that we
are left with the cases $e=p_1$.  For
$[\Ga\dsl{p}_5\dsl{p}_6]^\si_{34} [\dsl{p}_1]^\tau_{56}$ this relation
can be used to transfer $p_6$ to the bottom chain, again triggering
the Dirac equation there. We are left with the two cases
$[\Ga\dsl{p}_1\dsl{p}_5]^\si_{34} [\dsl{p}_1]^\tau_{56}$ and
$[\Ga\dsl{p}_1\dsl{p}_6]^\si_{34} [\dsl{p}_1]^\tau_{56}$, which are
reduced according to
\beqar
\Bigl[\Ga\dsl{p}_1\dsl{p}_6\Bigr]^\pm_{34} \Bigl[\dsl{p}_1\Bigr]^\pm_{56} &=&
\frac{1}{2} \Bigl[\Ga\dsl{p}_1\ga_\mu\Bigr]^\pm_{34} \Bigl[\dsl{p}_1\dsl{p}_6\ga^\mu\Bigr]^\pm_{56}
=
\frac{1}{2} \Bigl[\Ga\dsl{p}_1\dsl{p}_6\dsl{p}_1\ga_\mu\Bigr]^\pm_{34} \Bigl[\ga^\mu\Bigr]^\pm_{56}
\nn\\
&=&
(p_1 p_6) \Bigl[\Ga\dsl{p}_1\ga_\mu\Bigr]^\pm_{34} \Bigl[\ga^\mu\Bigr]^\pm_{56} 
-\frac{p_1^2}{2} \Bigl[\Ga\dsl{p}_6\ga_\mu\Bigr]^\pm_{34} \Bigl[\ga^\mu\Bigr]^\pm_{56},
\nn\\
\Bigl[\Ga\dsl{p}_1\dsl{p}_5\Bigr]^\pm_{34} \Bigl[\dsl{p}_1\Bigr]^\mp_{56} &=&
\dots
=
(p_1 p_5) \Bigl[\Ga\dsl{p}_1\ga_\mu\Bigr]^\pm_{34} \Bigl[\ga^\mu\Bigr]^\mp_{56} 
-\frac{p_1^2}{2} \Bigl[\Ga\dsl{p}_5\ga_\mu\Bigr]^\pm_{34} \Bigl[\ga^\mu\Bigr]^\mp_{56},
\nn\\
\Bigl[\Ga\dsl{a}\dsl{p}_1\dsl{p}_5\Bigr]^\pm_{34} \Bigl[\dsl{p}_1\Bigr]^\pm_{56} &=&
\dots
=
 2(ap_1) \Bigl[\Ga\dsl{p}_5\Bigr]^\pm_{34} \Bigl[\dsl{p}_1\Bigr]^\pm_{56} 
-2(ap_5) \Bigl[\Ga\dsl{p}_1\Bigr]^\pm_{34} \Bigl[\dsl{p}_1\Bigr]^\pm_{56}
\nn\\
&& {}
+(p_1 p_5) \Bigl[\Ga\dsl{p}_1\ga_\mu\dsl{a}\Bigr]^\pm_{34} \Bigl[\ga^\mu\Bigr]^\pm_{56}
-\frac{p_1^2}{2} \Bigl[\Ga\dsl{p}_5\ga_\mu\dsl{a}\Bigr]^\pm_{34} \Bigl[\ga^\mu\Bigr]^\pm_{56},
\nn\\
\Bigl[\Ga\dsl{a}\dsl{p}_1\dsl{p}_6\Bigr]^\pm_{34} \Bigl[\dsl{p}_1\Bigr]^\mp_{56} &=&
\dots
=
 2(ap_1) \Bigl[\Ga\dsl{p}_6\Bigr]^\pm_{34} \Bigl[\dsl{p}_1\Bigr]^\mp_{56} 
-2(ap_6) \Bigl[\Ga\dsl{p}_1\Bigr]^\pm_{34} \Bigl[\dsl{p}_1\Bigr]^\mp_{56}
\nn\\
&& {}
+(p_1 p_6) \Bigl[\Ga\dsl{p}_1\ga_\mu\dsl{a}\Bigr]^\pm_{34} \Bigl[\ga^\mu\Bigr]^\mp_{56}
-\frac{p_1^2}{2} \Bigl[\Ga\dsl{p}_6\ga_\mu\dsl{a}\Bigr]^\pm_{34} \Bigl[\ga^\mu\Bigr]^\mp_{56},
\eeqar 
where the Dirac structure $\Ga$ and the vector $a$ can be chosen arbitrarily.
\label{item-abcd}
\item Now we reduce products of Dirac chains of the type
  $[\ga_\mu\dsl{a}\dsl{b} \dsl{c}]^\si_{34} [\ga^\mu]^\tau_{56}$,
  i.e.\ four Dirac matrices in the top chain with a contraction to the
  bottom chain.  Among the vectors $a$, $b$, $c$, there is at least
  one of the momenta $p_1$, $p_5$, $p_6$.  The momenta $p_5$, $p_6$
  can be shifted to the bottom chain via \refeq{eq:chisholm1} and
  subsequently eliminated with the Dirac equation. If only $p_1$ (but
  not $p_5$, $p_6$) occurs in the top chain, the full Dirac structure
  can only be $[\ga_\mu\dsl{p}_1\dsl{\veps}_1\dsl{\veps}_2]^\si_{34}
  [\ga^\mu]^\tau_{56}$, which is kept as a standard structure.
\label{item-gabc}
\item Now we repeat steps \ref{item-abcd} and \ref{item-gabc} for top
  chains involving three Dirac matrices, in order to eliminate such
  structures as far as possible, and subsequently again for top chains
  with two matrices.
\item The remaining products of the form $[\dsl{p}_i]^\si_{34}
  [\dsl{p}_j]^\tau_{56}$ are reduced as far as possible as described
  in (B.4) of \citere{Bredenstein:2008zb}.
\item Finally, it is convenient to replace all chirality projectors
  $\omega_\pm=(1\pm\ga_5)/2$ in terms of vector (no $\ga_5$) and
  axial-vector ($\propto\ga_5$) structures in the two Dirac chains,
  because only the combinations vector$\otimes$vector and
  axial-vector$\otimes$axial-vector contribute owing to the parity
  symmetry of the process in NLO QCD.
\end{enumerate}

\section{Benchmark numbers for the virtual corrections}
\label{app:benchmark}

In order to facilitate a comparison to our calculation, in this
appendix we provide explicit numbers on the 
squared LO amplitude and the corresponding virtual correction for a
single non-exceptional phase-space point. The set of momenta for
the partonic reaction $\Pg\Pg \to \Pt\bar\Pt\Pb\bar\Pb$ is chosen as
\beqar
p_\Pg^\mu &=& \scriptstyle (500,0,0,500),
\nn\\
p_\Pg^\mu &=& \scriptstyle (500,0,0,-500),
\nn\\
p_{\Pt}^\mu &=& \scriptstyle
(327.5045589027869,
 107.1276753641986,
-107.9290580423663,
-233.1168284428635),
\nn\\
p_{\bar\Pt}^\mu &=& \scriptstyle
(276.6425142763093,
-107.4949148022111,
 153.8289259355409,
-107.3397668261919),
\nn\\
p_{\Pb}^\mu &=& \scriptstyle
(233.9459027189062,
  82.55875671042013,
 -77.70592645955253,
 204.6375480757531),
\nn\\
p_{\bar\Pb}^\mu &=& \scriptstyle
(161.9070241019976,
 -82.19151727240762,
  31.80605856637796,
 135.8190471933023),
\label{eq:PSpoint}
\eeqar
which defines the same phase-space point as already chosen for
$q\bar q\to\Pt\bar\Pt\Pb\bar\Pb$ in \citere{Bredenstein:2008zb}.
The components are given in GeV and $\Mt=172.6\GeV$.  
For the spin- and colour-averaged squared LO amplitude we obtain
\beqar
|\M^{\LO}|^2/g_{\mathrm{s}}^8 &=&  
5.437061775267626
\cdot10^{-9}\GeV^{-4},
\nn\\
|\M^{\LO}|^2/g_{\mathrm{s}}^8\Big|_{\mathrm{Madgraph}} 
&=&  
5.437061775267649
\cdot10^{-9}\GeV^{-4},
\eeqar
where we divided out the strong coupling constant $g_{\mathrm{s}}$.  

We express NLO contributions in the $2\to 4$ phase space
as Laurent series in $\epsilon=(4-D)/2$,
\beqar
|\M|^2&=&
\left(1+c_\Gamma\sum_{k=0}^2 \de_\NLO^{(k)}\epsilon^{-k}\right)
|\M^{\LO}|^2,
\eeqar
where we factor out the LO term and the normalization factor
\beqar
c_\Gamma&=&
\frac{(4\pi)^\epsilon \Gamma(1+\epsilon)\Gamma^2(1-\epsilon)}
{\Gamma(1-2\epsilon)}
=\frac{(4\pi)^\epsilon}
{\Gamma(1-\epsilon)}+\mathcal{O}(\epsilon^3)
\nn\\
&=&(4\pi)^\epsilon
\Gamma(1+\epsilon)
-\frac{\pi^2}{6}\epsilon^2
+\mathcal{O}(\epsilon^3).
\eeqar

We split the result into the two parts,
\newcommand{\loops}{\mathrm{loops}}
\beqar
\de_\NLO^{(k)}&=&
\de_\loops^{(k)}+
\de_\rI^{(k)},
\eeqar
which correspond to the contributions of renormalized loop diagrams
(loops) and the $I$ operator of the dipole subtraction function as
defined in \citere{Catani:2002hc}.
The numbers in \refta{Tab:benchmark} have been obtained in the
't~Hooft--Feynman gauge using the 't Hooft--Veltman variant of
dimensional regularization (four-dimensional external partons).  We
set the scale of dimensional regularization and the renormalization
scale as $\mu=\mu_{\mathrm{R}}=\Mt$.  The corresponding values of the
strong coupling constant in the renormalization scheme described in
\refse{se:numres} are
\beq
\alpha_{\mathrm{s}}(\Mt)|_{\mathrm{LO}} =
0.1178730139006150, \qquad
\alpha_{\mathrm{s}}(\Mt)|_{\mathrm{NLO}}=
0.1076396017050965.
\eeq
The agreement between our two independent versions of the virtual
corrections is typically about 10~digits at regular phase-space points.
\begin{table}
$$
\begin{array}{|c|c|c|c|c|}
\hline
& & \de^{(2)} & \de^{(1)} & \de^{(0)} \\
\hline
\hline
\loops & \mathrm{version~1} & -0.1484719139263099 &  \phantom{-}0.0986990876957258&  \phantom{-}0.4123948722195028\\
       & \mathrm{version~2} & -0.1484719139260971 &  \phantom{-}0.0986990876958276&  \phantom{-}0.4123948722188028\\
\hline
\rI  & \mathrm{version~1} &\phantom{-}0.1484719139263437 & -0.0986990876957289&-0.2125642646365643\\
     & \mathrm{version~2} &\phantom{-}0.1484719139263439 & -0.0986990876957291&-0.2125642646365644\\
\hline\hline
\NLO  & \mathrm{version~1} & \phantom{-}0.0000000000000338 &          {-}0.0000000000000031& \phantom{-}0.1998306075828831\\
      & \mathrm{version~2} & \phantom{-}0.0000000000002467 &  \phantom{-}0.0000000000000985& \phantom{-}0.1998306075822384\\
\hline
\end{array}
$$
\caption{Various contributions to the virtual NLO corrections
to $\Pg\Pg \to \Pt\bar\Pt\Pb\bar\Pb$ at the phase-space point \refeq{eq:PSpoint}.}
\label{Tab:benchmark}
\end{table}

Another benchmark result for the virtual corrections to
$\Pg\Pg\to\Pt\bar\Pt\Pb\bar\Pb$ can be found in
\citere{vanHameren:2009dr}.  The quantity presented there and denoted
as `{\tt HELAC-1L}' corresponds to the unrenormalized virtual one-loop
correction plus the mass-renormalization contribution (without
wave-function and coupling constant renormalization). We find
agreement at the 10-digit level.

\end{document}